\documentclass[11pt,letterpaper,fleqn]{article}
\usepackage[letterpaper,margin=1in]{geometry}

\usepackage[utf8]{inputenc}
\usepackage[T1]{fontenc}
\usepackage{mathtools}
\usepackage{amsmath,amssymb,amsthm}
\usepackage{paralist}
\usepackage{hyperref}
\usepackage{enumitem}
\usepackage{subfiles}
\usepackage{xcolor}
\usepackage{tikz}
\usepackage{float}
\usepackage{tabularx}
\usepackage{booktabs}
\usepackage{multirow}
\usepackage{verbatim}
\usepackage{lineno}
\usepackage[numbers,sort]{natbib}
\usepackage{microtype}
\usetikzlibrary{calc,shapes,graphs}
\usetikzlibrary{patterns.meta,positioning}
\usetikzlibrary{arrows.meta,patterns.meta,positioning}
\usepackage{forest}

\usepackage[normalem]{ulem}

\newtheorem{theorem}{Theorem}[section]
\newtheorem{definition}[theorem]{Definition}

\newtheorem{observation}[theorem]{Observation}
\newtheorem{corollary}[theorem]{Corollary}
\newtheorem{lemma}[theorem]{Lemma}
\newtheorem{claim}[theorem]{Claim}
\newtheorem{remark}[theorem]{Remark}
\newtheorem{conjecture}[theorem]{Conjecture}
\usepackage[ruled,lined,noend,linesnumbered]{algorithm2e}
\usepackage{cleveref}

\newcommand{\ignore}[1]{}

\newcommand{\cB}{\mathcal{B}}

\newcommand{\cD}{\mathcal{D}}

\newcommand{\cF}{\mathcal{F}}

\newcommand{\cP}{\mathcal{P}}
\newcommand{\cQ}{\mathcal{Q}}

\newcommand{\NN}{\mathbb{N}}

\newcommand{\RR}{\mathbb{R}}

\newcommand{\eps}{\varepsilon}

\newcommand{\probsize}[1]{\lVert #1 \rVert}

\newcommand{\Daniel}[1]{{\color{olive} Daniel: #1}}

\newcommand{\Marco}[1]{{\color{green!70!black} Marco: #1}}
\newcommand{\Giacomo}[1]{{\color{red!70!black} Giacomo: #1}}

\DeclareMathOperator{\poly}{poly}

\newcommand{\primitive}[1]{\ensuremath{\mathsf{#1}}}
\newcommand{\algo}[1]{\ensuremath{\mathsf{#1}}}

\newcommand{\typemask}{\primitive{TypeMask}}

\newcommand{\PartialMatch}{\primitive{PartialMatch}}
\newcommand{\Eval}{\primitive{Sat}}
\newcommand{\ESat}{\primitive{Edge}\Eval}
\newcommand{\VSat}{\primitive{Vertex}\Eval}
\newcommand{\PartialMatchPre}{\primitive{PartialMatch\text{-}Preprocess}}
\newcommand{\NC}{\primitive{Neigh}}
\newcommand{\Neigh}{\primitive{Neigh}}
\newcommand{\NeighInt}{\primitive{NeighINT}}
\newcommand{\Endpoints}{\algo{Endpoints}}
\newcommand{\ORQCount}{\algo{ORQ\text{-}Count}}
\newcommand{\ORQSample}{\algo{ORQ\text{-}Sample}}
\newcommand{\ORQList}{\algo{ORQ\text{-}Report}}
\newcommand{\ORQPre}{\algo{ORQ\text{-}Preprocess}}

\newcommand{\ADJ}{\primitive{Adj}}
\newcommand{\VENN}{\primitive{Venn}}
\newcommand{\ISH}{\primitive{SubH}}

\newcommand{\CONTAIN}{\primitive{Contain}}

\newcommand{\SP}{\primitive{Superset}}

\newcommand{\UNION}{\primitive{Union}}
\newcommand{\DIFF}{\primitive{Diff}}
\newcommand{\INT}{\primitive{Int}}

\newcommand{\BoundedEdgePartialMatchPre}{\algo{EPartialMatch}\ensuremath{^t}\text{-}\algo{Preprocess}}
\newcommand{\BoundedEdgePartialMatch}{\algo{EPartialMatch}\ensuremath{^t}}
\newcommand{\EdgePartialMatch}{\primitive{Edge}\PartialMatch}
\newcommand{\EdgePartialMatchPre}{\primitive{Edge}\PartialMatchPre}
\newcommand{\VertexPartialMatch}{\primitive{Vertex}\PartialMatch}

\newcommand{\EdgeNeigh}{\primitive{Edge}\Neigh}
\newcommand{\VertexNeigh}{\primitive{Vertex}\Neigh}
\newcommand{\VertexNeighPre}{\VertexNeigh\text{-}\primitive{Preprocess}}

\newcommand{\BuildRectangles}{\primitive{N1}\text{-}\primitive{Preprocess}}

\newcommand{\bE}{\pmb{E}}
\newcommand{\I}{\mathcal{I}}
\newcommand{\R}{\mathcal{R}}

\newcommand{\Pcal}{\mathcal{P}}

\newcommand{\bS}{\pmb{S}}

\newcommand{\Can}{\mathrm{Can}}

\newcommand{\bu}{\pmb{u}}

\newcommand{\B}{\mathcal{B}}
\newcommand{\size}[1]{\ensuremath{\lVert{#1}\rVert}}

\newcommand{\bV}{\pmb{V}}
\newcommand{\IV}{\mathcal{I}}
\newcommand{\IE}{\mathcal{J}}

\newcommand{\NVV}{N_{VV}(v)}
\newcommand{\NVE}{N_{VE}(v)}
\newcommand{\NEE}{N_{EE}(v)}
\newcommand{\NEV}{N_{EV}(v)}

\newcommand{\intStart}{\mathrm{start}}
\newcommand{\intEnd}{\mathrm{end}}

\newcommand{\ve}{\pmb{\vec e}}
\newcommand{\vv}{\pmb{\vec v}}
\newcommand{\vu}{\pmb{\vec u}}
\newcommand{\cT}{\mathcal{T}}
\newcommand{\bp}{\pmb{p}}
\newcommand{\bq}{\pmb{q}}

\newcommand{\vdegen}{\kappa_v}
\newcommand{\edegen}{\kappa_e}
\newcommand{\degen}{\kappa}

\newcommand{\Sec}[1]{\hyperref[sec:#1]{\S\ref*{sec:#1}}} 
\newcommand{\Eqn}[1]{\hyperref[eq:#1]{(\ref*{eq:#1})}} 
\newcommand{\Fig}[1]{\hyperref[fig:#1]{Fig.\,\ref*{fig:#1}}} 
\newcommand{\Tab}[1]{\hyperref[tab:#1]{Tab.\,\ref*{tab:#1}}} 
\newcommand{\Thm}[1]{\hyperref[thm:#1]{Theorem\,\ref*{thm:#1}}} 
\newcommand{\Fact}[1]{\hyperref[fact:#1]{Fact\,\ref*{fact:#1}}} 
\newcommand{\Lem}[1]{\hyperref[lem:#1]{Lemma\,\ref*{lem:#1}}} 
\newcommand{\Prop}[1]{\hyperref[prop:#1]{Prop.~\ref*{prop:#1}}} 
\newcommand{\Cor}[1]{\hyperref[cor:#1]{Cor.~\ref*{cor:#1}}} 
\newcommand{\Conj}[1]{\hyperref[conj:#1]{Conjecture~\ref*{conj:#1}}} 
\newcommand{\Def}[1]{\hyperref[def:#1]{Definition~\ref*{def:#1}}} 
\newcommand{\Alg}[1]{\hyperref[alg:#1]{Alg.~\ref*{alg:#1}}} 
\newcommand{\Clm}[1]{\hyperref[clm:#1]{Clm.~\ref*{clm:#1}}} 
\newcommand{\Step}[1]{\hyperref[step:#1]{Step~\ref*{step:#1}}} 
\newcommand{\Exa}[1]{\hyperref[exa:#1]{Example~\ref*{exa:#1}}} 
\newcommand{\Obs}[1]{\hyperref[obs:#1]{Obs~\ref*{obs:#1}}} 

\newcommand{\PTR}{\mathsf{PTR}}
\newcommand{\nil}{\textsc{nil}}
\newcommand{\arrV}{\mathsf{V}}

\newcommand{\degHeap}{\ensuremath{\mathsf{DegHeap}}}
\newcommand{\sizeHeap}{\ensuremath{\mathsf{SizeHeap}}}
\newcommand{\type}{\ensuremath{\mathsf{TYPE}}}
\newcommand{\hash}{\ensuremath{\mathsf{HASH}}}
\newcommand{\twins}{\ensuremath{\mathsf{TWINS}}}
\newcommand{\node}{\ensuremath{\mathsf{NODE}}}

\newcommand{\transp}[1]{\ensuremath{#1}^{\top}}

\newcommand{\nIE}{\bar{\IE}}

\newcommand{\nJ}{\bar{J}}

\newcommand{\nnz}{\ensuremath{\operatorname{nnz}}}

\title{The Power of Bidegeneracy for Hypergraph Data Structures}
\date{}

\author{Marco Bressan\\
University of Milan\\
{\tt marco.bressan@unimi.it}
\and
Stefano Clemente\\
University of Milan\\
{\tt stefano.clemente@unimi.it}
\and
Giacomo Fumagalli\\
University of Milan\\
{\tt giacomo.fumagalli@unimi.it}
\and
Daniel Paul-Pena\thanks{Work partially completed while the author was at the University of California, Santa Cruz}\\
The University of Sydney\\
{\tt daniel.paulpena@sydney.edu.au}
\and
C. Seshadhri\thanks{Supported by NSF DMS-2023495, CCF-1740850, 2402572.} \\
University of California, Santa Cruz\\
{\tt sesh@ucsc.edu}
}

\begin{document}

\maketitle
\thispagestyle{empty}

\begin{abstract} A hypergraph $H = (V,E)$ describes a set system over a vertex universe $V$. It is increasingly popular in the network science community to store data as hypergraphs, instead of graphs, as hypergraphs can represent interactions between subsets of vertices. A natural challenge is 
to design a data structure for $H$ to quickly answer basic questions about it. For example, one may ask how many neighbors a vertex $v$ has, whether two hyperedges $e$ and $f$ intersect, or whether there exists a hyperedge containing a vertex subset $S$. The equivalent questions are trivial for graphs and can be answered by inspecting the adjacency list. The gold standard, from a data structures perspective, would be near-linear time preprocessing and query answering in (poly)logarithmic time. Since this is unlikely for arbitrary $H$, we ask for well-motivated hypergraph properties that can make such a goal possible.

We introduce \emph{bidegeneracy}, a measure of hypergraph sparsity that strictly generalizes many other measures, including bounded degree, bounded rank, and bounded $\infty$-degeneracy.
While these parameters were exploited to accelerate specific tasks such as subgraph counting, we show that bounded bidegeneracy leads to fast hypergraph \emph{data structures}.
For a large set of fundamental queries like the ones above, we get near-linear time preprocessing and query answering in (poly)logarithmic time, even for the listing and sampling versions.
Moreover, direct reductions from assumptions like the Orthogonal Vectors Conjecture suggest that these data structures require an almost-quadratic preprocessing or an almost-linear query time if one makes no assumptions on the input. Even if these conjectures may be false, it would only lead to tiny
polynomial improvements over these bounds. 

On the technical level, we use bidegeneracy to compute a compressed representation of $H$ and then perform careful reductions to stabbing and searching problems in $\RR^d$.
Our framework significantly extends the well-studied ``partial matching'' problem of finding entries in database using a wildcard search.
It also uncovers an intriguing connection between hypergraph sparsity, compression, and the computational complexity of hypergraph primitives.
We also find that hypergraph datasets from the literature have remarkably low bidegeneracy, thus our results might have practical consequences.
\end{abstract}

\clearpage
\setcounter{tocdepth}{2}
{
\tableofcontents
}


\newpage

\pagenumbering{arabic}

\section{Introduction} \label{sec:intro}

{\em \underline{AI Methodology:} All the main ideas were developed by the authors without the use of AI. AI was used to discover some basic lower bound arguments, verify the definitions, review existing proofs, and search for related literature. The authors also used AI for fixing typos and notation.  All the writing was done solely by the authors, who stand by every line of this document.}

\medskip

A hypergraph $H = (V,E)$ is a representation of a set system, where each hyperedge $e \in E$ is an arbitrary subset of the vertex set $V$.
While set systems are studied extensively throughout combinatorics, there is an increasing trend in network science and data mining towards thinking of data in terms of hypergraphs~\cite{VeBeKl20,VeBeKl22,AnCoPo+23,CaDeFa+23} (the recent survey~\cite{LeBuEl+25} has hundreds of papers on this topic).
By allowing for arbitrary hyperedge sizes, we can represent interactions beyond pairs. Many instances of real-world graphs are actually hypergraphs, where hyperedges are replaced by cliques. For example, while a coauthor network is represented as a graph, the raw data is a hypergraph. The vertex set consists of researchers and a hyperedge corresponds to the authors of a publication.
Hypergraphs are a faithful model of the original data and practitioners suggest they should be analyzed directly. There are numerous software packages developed to process hypergraphs~\cite{Sh20,hypnx,hypgraphx}.

It is natural to view hypergraphs from a data structures perspective. Given a hypergraph $H$, we wish to preprocess and store $H$ to answer basic \emph{hypergraph primitives}. Consider the following natural primitives, many
of which are used in practical algorithms.

\smallskip
\begin{asparaitem}
\item \ADJ: given vertices $u,v$, decide whether there exists a hyperedge containing both $u$ and $v$. The dual version, \INT, asks to decide whether two given hyperedges $e, f$ intersect. 
\item \CONTAIN: given hyperedges $e, f \in E$, decide whether $e \subseteq f$. (We think of $e, f$ as identifiers/pointers for the corresponding sets.) 
\item \EdgeNeigh: given hyperedge $e$, return the number of hyperedges intersecting it. The sampling version returns a uniform random hyperedge intersecting $e$ and is a core primitive for hypergraph random walks~\cite{AJM+20,LFA+22,CLSJ24}.
\item \SP: given a subset $S$ of vertices, count the number of hyperedges $f \supseteq S$. In hypergraph mining, it has been used to show that the hypergraph structure can improve link prediction~\cite{BAS+18}. 
\item \ISH: return the subhypergraph induced by $S$. Specifically, list out all distinct, non-empty sets $e \cap S$ (over hyperedges $e$). Such operations are common in subhypergraph counting~\cite{PaSe26,BrBrDe+26}.
\item \UNION: given a subset $F$ of hyperedges, compute the union size $|\bigcup_{f \in F} f|$. This quantity is used to measure the ``density'' of a set of hyperedges $F$~\cite{LCS21}. 
\item \VENN: given three hyperedges, compute the ``Venn diagram'' of all intersection patterns between them. This specific structure is called a ``hypertriangle'' and is a core operation in practical algorithms for hypergraph motifs~\cite{LeJiSh20,YiWaZh+24,paul-pena_counting_2026}.
\item \VertexNeigh: given a vertex $v$, return the number of its neighbors. The listing version can be used to obtain the Gaifman graph of $H$, which is a key preprocessing step in many hypergraph algorithms.
\item \EdgePartialMatch: given a set $P$ of ``positive'' vertices and a set $N$ of ``negative'' vertices, count the number of hyperedges that contain $P$ and have no vertex of $N$. This classic problem was first posed $50$ years ago in the context of database search~\cite{Ri76} and has been heavily studied since~\cite{MiNi+95,ChInPa02,CoGoLe04,JaKh+04,Pa10,Pa11,AnJi+25}.
\end{asparaitem}

\smallskip



We note that some of these operations are related and could be considered the dual of another one; many primitives are known under different names, such as set disjointness or subset containment.
Nevertheless, we state them separately to see the rich landscape of basic primitives on set systems and their utility in practice. For graphs, these primitives are either trivial or vacuous. For bounded rank hypergraphs (hyperedges have bounded size), these primitives can be solved by indexing each hyperedge $e$ over all $S \subseteq e$. 
At the other end of the spectrum, for arbitrary hypergraphs, fine-grained complexity conjectures suggest that we can only get slight improvements over trivial algorithms. The trivial algorithms either (i) precompute and store solutions, or (ii) for each query, compute the answer by a full scan over the vertices/hyperedges.
Even if these complexity conjectures are falsified~\cite{AlWi26,openai-eth}, it would only suggest small polynomial improvements over the trivial algorithms. The gold standard for such data structures is to achieve near-linear preprocessing time and (poly)logarithmic time to answer queries.


Are there natural conditions on hypergraphs and a data structure framework that allows for quickly answering all these hypergraph primitives?
These natural conditions should contain graphs and bounded rank hypergraphs and, if possible, contain properties currently used to accelerate hypergraph algorithms (such as the various hypergraph degeneracies used to list and count sub-hypergraphs).
We would like to discover practically motivated classes of hypergraphs that contain non-trivial hypergraphs with unbounded rank.


\subsection{Results} \label{sec:results}

\paragraph{A new sparsity measure.} One of our primary contributions is the notion of hypergraph \emph{bidegeneracy}. We give a short overview of this concept and expand on it in \Sec{interval}.
We can represent a hypergraph $H = (V,E)$ as a Boolean \emph{incidence matrix} $M_H \in \{0,1\}^{n \times m}$ indexed by $V$ and $E$, where $M_H[u,e] = 1$ iff $u \in e$. The size of $H$, denoted $\size{H}$, is the number of non-zeroes of $M_H$.

\begin{figure}\centering
\begin{tikzpicture}[
  scale=.7,
  font={\sffamily\small},
  node distance=1cm,
  every node/.style={align=center},
  arr/.style={-latex, line width=.8pt, shorten <=2pt, shorten >=2pt},
]
  \node (graphs) at (-2,0)  {Graphs};
  \node (degree) at (3.5,0) {Bounded degree};
  \node (rank)   at (-2,-2) {Bounded rank};
  \node (hdegen)  at (3.5,-2) {Bounded\\ hyperedge degeneracy};
  \node (nice)   at (-2,-4) {Bounded niceness};
  \node (ldegen)  at (3.5,-4) {Bounded $\infty$-degeneracy};
  \node (kq) at (9, -4) {Bounded $(k,q)$-cores};
  \node (bipinci) at (9, -2) {Bounded degeneracy\\ of incidence graph};
  \node (bideg)  at (3.5,-6.5) {Bounded bidegeneracy};

  \draw[arr] (graphs) -- (rank);
  \draw[arr] (rank)   -- (nice);
  \draw[arr] (degree) -- (nice);
  \draw[arr] (degree) -- (hdegen);
  \draw[arr] (nice)   -- (bideg);
  \draw[arr] (hdegen)  -- (ldegen);
  \draw[arr] (ldegen)  -- (bideg);
  \draw[arr] (kq) -- (bideg);
  \draw[arr] (bipinci) -- (kq);
\end{tikzpicture}
\caption{We give strict implications between various notions of sparsity defined in the literature. Bounded bidegeneracy, introduced in this paper, is implied by all of them. See~\Cref{apx:implications}.}
\label{fig:implications}
\end{figure}
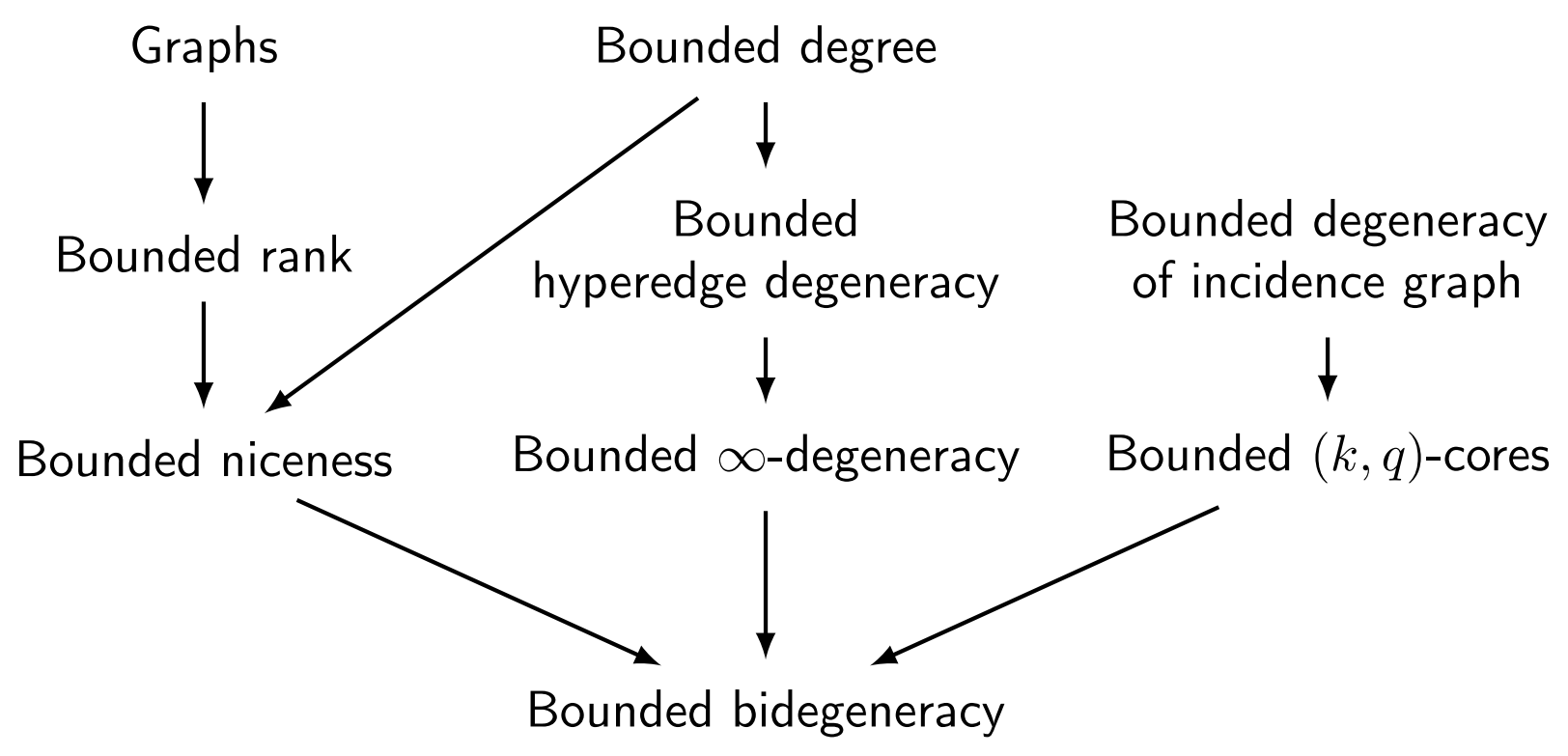

\begin{definition}[Simplified version of~\Def{bidegen}] \label{def:bidegen-intro}
    A hypergraph $H$ is \emph{$\degen$-bidegenerate} if every submatrix of $M_H$ with \uline{unique} rows and columns contains a row or a column of Hamming weight at most $\degen$.
\end{definition}
The inspiration for bidegeneracy comes from the classic notion of graph degeneracy. The  degeneracy of a graph is the largest minimum degree of any subgraph and is a common measure of sparsity~\cite{Na64,MaBe83,ChNi85}.
However, bounded bidegeneracy is a more general notion, as it looks at both rows and columns of $M_H$. 
Every graph is $2$-bidegenerate, and in general every rank $r$ (the maximum cardinality of a hyperedge) hypergraph is $r$-bidegenerate.
(Note that graphs can have unbounded degeneracy.)
Several notions of hypergraph degeneracy have recently been proposed, including hypergraph $l$-degeneracy~\cite{PaSe26}, hyperedge degeneracy~\cite{paul-pena_counting_2026}, and $(k,q)$-cores~\cite{LeGo+23}.
As shown by the implication graph of \Fig{implications}, all these notions are subsumed by \Def{bidegen-intro}.
Since their parameters are typically small for real-world datasets~\cite{bressan2026countinghypergraphletscolorcoding}, so is bidegeneracy (as we indeed find, see the remark below~\Thm{nc-main}).

One distinguishing feature of~\Def{bidegen-intro} is that seemingly ``dense'' hypergraphs can have bounded bidegeneracy.
Consider the hypergraph $H$ with hyperedges $e_1 \subset e_2 \subset \ldots \subset e_n$, where $e_i=\{1,\ldots,i\}$. 
The incidence matrix has $\Omega(n^2)$ nonzeros, so $\size{H} = \Omega(n^2)$.
But $H$ is $1$-bidegenerate, since in every submatrix with unique rows and columns, the row of largest id has weight $1$.
All definitions of degeneracy mentioned above, instead, basically bound the sparsity of submatrices of $M_H$, which is (mostly) equivalent to bounding the Hamming weight of rows/columns in \emph{all} submatrices of $M_H$.
Another notable feature of bidegeneracy is that it is dual-invariant: if $H$ is $\degen$-bidegenerate, then so is $\transp{H}$, the hypergraph obtained by transposing $M_H$.
This means that running time bounds for a primitive, say \ADJ, transfer automatically to its dual, \INT.
For many previous notions of hypergraph degeneracy, this does not hold.

\paragraph{Solving the primitives.}
We define two core primitives, called \ESat\ and \VertexNeigh. 
We show in \Sec{primitives} that many common queries, including all those mentioned so far, can be easily solved by these core primitives and their duals.
These primitives are described as a Boolean condition over hyperedges (or vertices). 
The output is the number of solutions, that is, the number of hyperedges (or vertices) that satisfy the condition.
We also consider the stronger version of weighted counting, sampling, and constant-delay enumeration.
The weighted counting problem is to output the sum of weights of the solutions. 
The sampling problem is to output a solution with probability proportional to its weight. The constant-delay enumeration problem requires listing all distinct solutions with constant delay, after some setup computation (abusing notation, we simply say that the query takes time $T$, if the setup takes time $T$).

We show that both \ESat\ and \VertexNeigh\ admit fast algorithms over bidegenerate hypergraphs.
This implies fast algorithms for all the primitives mentioned above.
For bounded bidegeneracy hypergraphs, we can answer (weighted) \ADJ\ queries in logarithmic time after a $\tilde{O}(\size{H})$ preprocessing.
We give complementary conditional lower bounds suggesting that our dependence on the parameters (the bidegeneracy of $H$ and the size of the query) cannot be reduced without blowing up the running time from near-linear to quadratic, or the query time from polylogarithmic to linear.

\medskip\noindent
\uline{The \ESat\ primitive.}
A \emph{vertex formula} is defined as a Boolean formula $\varphi$ over variables $\{x_v : v \in V\}$; we denote by $|\varphi|$ the size of $\varphi$.
A hyperedge $e$ can be represented by an incidence vector where the $v$-th coordinate is $1$ iff $v \in V(e)$. A hyperedge \emph{satisfies} $\varphi$ if the corresponding incidence vector satisfies the formula.
Let $\varphi(H)$ denote the set of satisfying edges. 
The primitive \ESat\ takes in input a vertex formula $\varphi$ and outputs the size of $\varphi(H)$.
The dual \VSat\ primitive is defined by \emph{edge formulas} that are satisfied by vertices.
This primitive generalizes the well-studied ``partial match'' problem.
%
Our main result for \ESat\ follows.
\begin{theorem} \label{thm:main-eval-bounded}
For every $t \ge 1$ there is an algorithm for \ESat\ with the following preprocessing and query time guarantees. Given an input a $\degen$-bidegenerate hypergraph $H$:
\begin{asparaitem}
    \item Preprocessing: The algorithm constructs a data structure in time $O(\size{H} \degen^{t-1} \log^{t+2}\size{H})$.
    \item Query time: Subsequently, the counting, sampling, and constant-delay enumeration of any \ESat\ query on $t$ variables can be answered in time $O(\degen \log^{t+2} \size{H})$.
\end{asparaitem}
\end{theorem}
\noindent
Note that the dependence on $\degen$ of \Cref{thm:main-eval-bounded} is only polynomial.
Thus, as long as $\degen=\size{H}^{o(1)}$, we have preprocessing time $\size{H}^{1+o(1)}$ and query time $\size{H}^{o(1)}$.
Therefore we can expect the algorithm to be effective in practice (it also does not hide large hidden factors).

By some further optimization of~\Cref{thm:main-eval-bounded}, we can solve many basic primitives even faster.
For example, we can solve the counting version of \ADJ$(u,v)$, \INT$(e,f)$, \DIFF$(e,f)$, \UNION$(e,f)$ in $O\big(\size{H} (\degen + \sqrt{\log \size{H}})\log\size{H}\big)$ preprocessing time and $O(\degen \log \size{H})$ query time. Bounds only slightly higher hold for $\VENN(e,f,g)$. See~\Cref{tab:faster_queries} for a more complete list.

We give also a second result for \ESat. The strength of this result is that the  data structure can now answer \ESat\ queries \emph{of any length}.
The price to pay is a significantly worse dependence on $\degen$, which makes the algorithm useful only as long as $\degen = o(\log \size{H} / \log\log \size{H})$.
\begin{theorem}[Simplified version of~\Thm{Eval}] \label{thm:main-eval} There is an algorithm for \ESat\ with the following preprocessing and query time guarantees. Given an input a $\degen$-bidegenerate~hypergraph~$H$:
    \begin{asparaitem}
        \item Preprocessing: The algorithm constructs a data structure in time $\size{H} \log^{O(\degen)}\size{H}$. 
        \item Query time: Subsequently, the counting, sampling, and constant delay enumeration of any \ESat\ query $\varphi$ can be answered in time $\degen^{O(|\varphi|)}\log^{O(\degen)}\size{H}$.
    \end{asparaitem}
\end{theorem}

\medskip\noindent\uline{The \VertexNeigh\ primitive.}
The neighborhood of a vertex $v$ is $N(v)=\{u \ | E(u) \cap E(v) \neq \emptyset\}$, where $E(u)$ is the set of edges incident with $u$.
The primitive \VertexNeigh\ asks, given $v$, to compute $|N(v)|$. 
For hypergraphs, this problem is substantially harder than computing the degree (which is the number of incident hyperedges). For graphs, this problem is trivial. The dual query, $\EdgeNeigh$, asks for the number of hyperedges that intersect with a given $e$.

\begin{theorem}[Simplified version of~\Thm{NC}] \label{thm:nc-main} There is an algorithm for \VertexNeigh\ with the following preprocessing and query time guarantees. Given a $\degen$-bidegenerate hypergraph $H$:
    \begin{asparaitem}
        \item Preprocessing: The algorithm constructs a data structure in time $\size{H} \log^{O(\degen)}\size{H}$. 
        \item Query time: Subsequently, the counting, sampling, and $O(\log\size{H})$-delay enumeration of any \VertexNeigh\ query can be answered in time $(\degen \log \size{H})^{O(\degen)}$.
    \end{asparaitem}
\end{theorem}
\noindent \uline{Remark.}
The exponential dependence on $\degen$ of~\Cref{thm:main-eval} and~\Cref{thm:nc-main} may appear excessive for any real application.
However, this is not the case.
The full definition of bidegeneracy,~\Def{bidegen}, actually uses a \emph{pair} of parameters $(\vdegen,\edegen)$, one for vertices and one for edges; and the full versions of our theorems,~\Cref{thm:Eval} and~\Cref{thm:NC}, depend exponentially in only \emph{one} of them.
For example, the bounds for \VertexNeigh\ are exponential in $\vdegen$, but only \emph{linear} in $\edegen$ (and only in the preprocessing!).
We ran some preliminary experiments on the six real-world datasets used in~\cite{bressan2026countinghypergraphletscolorcoding}, one of which has $\size{H} \simeq 70$M, maximum degree $\simeq 30$K, and maximum edge size $\simeq 10$K, and we found that in no case $\vdegen$ or $\edegen$ exceeds $1477$ when the other is $0$.
This suggests that the algorithms above can be fast in practice too.

\paragraph{Lower bounds.}
Assuming the Orthogonal Vectors Conjecture, the bounds of \Thm{main-eval-bounded}, \Thm{main-eval} and \Thm{nc-main} are close to optimal: algorithms for these primitives have an exponential dependence on the bidegeneracy, or quadratic preprocessing time, or linear query time. Even if the conjecture fails (so there are, say $O(n^{2-\alpha})$ time algorithms for Orthogonal Vectors), it would only mean that the lower bounds are worse by an $n^\alpha$ factor.

\begin{theorem} \label{thm:main-lb}
 If for some $\eps, \delta > 0$ there exists an algorithm for \ESat\ or \VertexNeigh\ with $2^{o(\degen)} \size{H}^{2-\delta}$ preprocessing time and $2^{o(\degen + t)} \size{H}^{1-\eps}$ query time, where $t=|\varphi|$ for \ESat\ and $t=0$ for \VertexNeigh, then the Orthogonal Vectors Conjecture fails.
\end{theorem}
\noindent In particular, for \VertexNeigh\ the dependence on $\degen$ cannot be improved beyond exponential (assuming truly subquadratic preprocessing and sublinear query time). The upper bounds for \VertexNeigh\ cannot be extended to formulas, such as computing $N(u) \cap N(v)$, since by computing neighborhood intersections on graphs one can solve Set Disjointness; see~\Cref{sec:LB}.
Thus, \VertexNeigh\ is substantially harder than \ESat.

\paragraph{Further applications.}
We give some concrete applications of our results.

\medskip\noindent\uline{The partial match problem:} As mentioned earlier, the \EdgePartialMatch\ problem has been studied in the context of database search~\cite{Ri76,MiNi+95,ChInPa02,CoGoLe04,JaKh+04,Pa10,Pa11,AnJi+25}. We think of the $m$ hyperedges as vectors in $\{0,1\}^n$ that form a database. The queries are vectors in $\{0,1,*\}^n$ (where $*$ are wildcards) and our aim is to find a database vector matching the query. The vertices/dimensions labeled by $1$ (resp. $0$) form the positive set $P$ (resp. negative set $N$). Somewhat confusingly for hypergraph notation, in the partial match literature, our $m$ is called $n$, while our $n$ is called $d$. There is a rich literature of cell probe and communication lower bounds showing that we cannot get significant improvements in the worst-case over the trivial algorithms~\cite{JaKh+04,Pa10,Pa11}. Significant effort has gone into just getting sublinear query time and near-linear storage with restrictions on the query~\cite{Ri76,ChInPa02,CoGoLe04}, and a recent result gives an improvement for the limited wildcard setting~\cite{AnJi+25}. In contrast, if the bidegeneracy is bounded, we get $(\log n)^{O(t)}$ query time for $t$-sized queries with a near-linear sized data structure.

\medskip\noindent\uline{Computing the Gaifman graph.} The Gaifman graph of a hypergraph $H$ (also known as projection graph, primal graph, or clique expansion graph) is the graph $G_H$ obtained by replacing every hyperedge by a clique, so that two vertices are adjacent in $G$ if and only if they are in a common hyperedge of $H$.
Computing $G_H$ is a fundamental step in many hypergraph mining applications.
The naive approach to compute $G_H$ takes time $O(\size{H}^2)$, although it can be accelerated through fast matrix multiplication. 
If $H$ is $O(1)$-bidegenerate, we can use \VertexNeigh\ to list the neighbors of every vertex $v$ and compute the adjacency list representation of $G_H$ in time $\tilde{O}(\size{H}) + \size{G_H}$.
Alternatively, we can use \ADJ\ to compute $G_H$ in time $\tilde{O}(|V|^2)$.
We can also perform other operations on $G_H$, such as power iteration by weighting vertices and collecting neighbor weights using \VertexNeigh.
More in general we can access $G_H$ through primitives such as \ADJ\ and \VertexNeigh, without explicitly computing it.

\medskip\noindent\uline{Hypergraph mining algorithms.} 
Several hypergraph mining algorithms can be immediately rewritten or accelerated using our primitives.
For instance, the hypergraphlet counting algorithm of~\cite{bressan2026countinghypergraphletscolorcoding} was carefully designed to bypass the quadratic lower bounds of Orthogonal Vectors by exploiting the so-called \emph{niceness} of the hypergraph. However, that algorithm can be written entirely in terms of \ADJ, \VertexNeigh, \ISH\ and so on, and bounded niceness implies bounded bidegeneracy.
In the hypertriangle counting algorithms of~\cite{lee_hypergraph_2024,paul-pena_counting_2026}, the expensive check of the intersection pattern of three edges takes $O(\kappa \log^2\size{H})$ time by a call to \VENN.
In hypergraphs with low bidegeneracy, random walks can be simulated efficiently~\cite{AJM+20,LFA+22,CLSJ24}: for example, start from a hyperedge $e_1$, move to an intersecting edge $e_2$ uniformly at random, and so on. The same holds for random walks over vertices. More in general one can efficiently perform walks over small connected subsets of vertices or hyperedges.

\subsection{Related work} \label{sec:related}
Graph degeneracy was first defined in the context of graph coloring and the seminal paper of Chiba-Nishizeki showed the relevance for clique counting~\cite{ChNi85}. There is a long history of exploiting degeneracy for subgraph counting~\cite{ChNi85,Ep94,AhNeRo+15,JhSePi15,PiSeVi17,OrBr17,JaSe17,Br19,PaSe20,Br21,BePaSe21,BeGiLe+22}. Refer to the short survey~\cite{Se23} for details.

There are many notions of graph and hypergraph degeneracy.
Closest to~\Def{bidegen-intro} are the notions of $\ell$-degeneracy~\cite{PaSe26}, hyperedge degeneracy~\cite{paul-pena_counting_2026}, and $(k,q)$-cores~\cite{LeGo+23}.
The term ``bidegeneracy'' has already been used, but only for graphs~\cite{ChenL00021,baudin2024bbksimplerfasteralgorithm}.
All these notions are strictly contained in ours; refer to~\Cref{fig:implications} and \Cref{apx:implications}.
At a first glance, the \emph{twin-width} of a graph may seem related to our peeling procedure via the incidence graph of $H$.
However, the twin-width of the incidence graph of $H=K_n$ is $\Omega(\sqrt{n})$, see~\cite{Jungho_twinwidth}, but the bidegeneracy is just $2$.

There are numerous software packages to process hypergraphs~\cite{Sh20,hypgraphx,hypnx}, underscoring 
the importance of having efficient data structures for storing hypergraphs. Currently, these packages
do not support the rich primitives defined earlier. We believe that it would be quite influential
to have practical implementations of the bidegeneracy ideas and primitives from this paper.

Hypergraphs can be seen as databases, for which there is a large literature on query answering.
On hypergraphs that are \emph{nowhere dense},\footnote{Formally, with an incidence graph coming from a nowhere-dense graph family.} every (Boolean) first-order query can be tested in constant time after a pseudo-linear-time preprocessing~\cite{schweikardt_enumerating_FO}.
Unfortunately, nowhere density rules out even the class of all \emph{graphs}, and  it is too harsh a requirement to be useful in practice.
The same issue is shared by many other measures of sparsity, such as treewidth or expansion~\cite{Torunczyk20_aggregate}.

The interval representation of~\Cref{sec:interval} can be seen as a \emph{labeling scheme} that assigns labels of $O(\degen \log n)$ bits to vertices and edges; one can then decide whether $v$ and $e$ are incident by looking only at their labels.
However, the relation with labeling schemes seems to end there; existing labeling schemes require $H$'s incidence graph to satisfy strong conditions, such as bounded cliquewidth~\cite{courcelle_query_2003}. 


\subsection{Main ideas} \label{sec:ideas}

\uline{Peeling the hypergraph.}
Starting from the seminal work of Matula and Beck~\cite{MaBe83} and Chiba and Nishizeki~\cite{ChNi85}, graph degeneracy and vertex orderings have been shown to be tightly related. The classic algorithm to compute the graph degeneracy produces a ``peeling'' order of the vertices, wherein a vertex of currently minimum degree is repeatedly deleted. This ordering can be used to orient the input graph as a DAG of low outdegree. This DAG is the primary object that is algorithmically exploited for fast algorithms for bounded degeneracy graphs~\cite{Co09,PiSeVi17,Br19,BePaSe21,Se23,PaSe25}.  The central fact is that, at the time of removal, a vertex has degree at most the degeneracy.
There are a multitude of hypergraph degeneracy definitions. But all the corresponding algorithms go via a peeling procedure by minimum degree and use the central degree fact described above.

When one thinks of bidegeneracy as a peeling procedure, things get complicated, because deleting a vertex can produce a hypergraph with multiple copies of the same hyperedge.
To comply with the definition of bidegeneracy, one must now remove those copies, even though they may not have small size.
Moreover, removed copies cannot be just forgotten, as they were distinct hyperedges in the starting hypergraph, and one would throw away information about the vertices they contain.
The same holds for vertices (removing a hyperedge can create more vertices of the same type). These deduplication steps can interleave with the removal of vertices and hyperedges, cascading into a chain of operations.

We handle these issues by constructing two directed binary \emph{auxiliary forests}, one for vertices and one for hyperedges, that record the deletions of copies by the peeling process. The hyperedge forest $\cT_E$ has one leaf $X_e$ for every hyperedge $e$, and the same for the vertex forest $\cT_V$.
To delete from the hypergraph the copy $f$ of a hyperedge $e$, we actually delete both $e,f$ and replace them with a new hyperedge $ef$ containing the same vertices. This ``merge'' operation is then recorded in the hyperedge forest by adding one node $X_{ef}$ with $X_e$ and $X_f$ as child.
When a hyperedge is deleted because it has size at most $\degen$, instead, the corresponding node of $\cT_E$ is frozen---it remains forever the root of its subtree.
The two forests are then used to construct the special representation of the hypergraph that is at the heart of our results.
Note also that the process just described requires, upon any cancellation, to immediately detect whether new duplicated vertices or edges have been produced.
By a careful choice of data structures, we can show that the whole procedure, including the construction of $\cT_V$ and $\cT_E$, can be implemented deterministically in only $O(\size{H} \ln^{1.5} \size{H})$.

\medskip\noindent\uline{Compressing $H$ via interval representations.}
Our main discovery is the notion of \emph{interval representation of hypergraphs}. 
The vertices and hyperedges of the hypergraph $H$ are stored as ordered lists of identifiers, $\bV$ and $\bE$.
Then, we have a function $\IV$ (resp.\ $\IE$) that maps each vertex (resp.\ hyperedge) to a disjoint collection of intervals in $\bE$ (resp.\ $\bV$).
The key property is that, if $e$ and $v$ are incident, then this incidence is certified by either $e$ being in one of the intervals of $\IV(v)$, or vice versa (but not both).
In all this, the bidegeneracy $\degen$ bounds from above the number of intervals in each of the mappings $\IV(v)$ and $\IE(e)$.
We can show that an interval representation of $H$ can be obtained from the forests $\cT_V$ and $\cT_E$. This entails performing a postorder traversal to determine the order of the leaves, which yields precisely $\bV$ and $\bE$.
Afterwards, for every vertex $v$ of the hypergraph we locate at most $\degen$ subtrees in $\cT_E$.
A symmetric procedure yields the intervals for the hyperedges.
\Def{interval_representation} gives the formalism for interval representations and \Sec{interval} gives the details on efficiently computing them.

The matrix viewpoint leads to the following deeper observation. The interval representation implies that, under a suitable permutation of indices, we can express the incidence matrix $M$ as 
the sum of Boolean matrices $M_V + M_E$, such that each row of $M_V$ (resp.\ column of $M_E$) has at most $\degen$ blocks of consecutive ones. Thus, the matrix $M$
can be \emph{compressed}, since we only need to store the starting and ending indices of each block. Hence, the input hypergraph can be stored in $O((m+n)\degen)$ space, which can be much smaller than $\size{H}$. This is a significant departure from all existing notions of (hyper)graph degeneracy. All previous degeneracy notions do not give an asymptotic reduction in the input storage. For example, hyperedge degeneracy can be used to bound the total number of (hyper)edges, so the raw storage of the input is (up to constant factors) the same as storing the degeneracy orientations.

\medskip\noindent\uline{Queries via computational geometry.}
The compression provided by interval representations opens up a new dimension for data structures for hypergraph
processing. Fundamentally, all primitives boil down to understanding whether certain collections of entries $M[v,e]$ have value $1$ (where $v$ and $e$ vary over choices depending on the primitive). That question can be answered by looking at a row of $M_V$ or a column of $M_E$, each of which has a block structure. Building on that intuition, we can represent the relevant incidence information of vertices (and hyperedges) by a collection of $O(\degen)$ integers that indicate the starting and ending index of the various blocks. So this is an $O(\degen)$-dimensional representation of the relevant information. By a careful formalization, we can  convert all primitives to geometric questions in this $O(\degen)$-dimensional space and solve them by using geometric range searching techniques.

Assuming bidegeneracy is bounded, consider the \ESat\ query. We first discuss the approach for the general bound of \Thm{main-eval} that works for queries of any size. 
The properties of the interval representations allow for a reduction of satisfying vertex formulas to problems on intersection patterns of specific intervals. Specifically, we determine if a hyperedge $e$ satisfies formula $\phi$ by inspecting the intervals $\IE(e)$ 
and $\IV(v)$ (for $v$ in the formula $\phi$). Essentially, the question is about intersection patterns of constant size. 
We can represent each hyperedge $e$ as a point in $\RR^d$ (for constant $d$) based on the intervals of $\IE(e)$. Any vertex formula $\phi$ can be converted into a collection of disjoint boxes in $\RR^d$ (again
based on the interval representation) such that $e$ satisfies $\phi$ iff the corresponding point lies in one of these boxes. We can now set up classic orthogonal range searching data structures to solve the (weighted) counting, sampling, and listing versions of \ESat. The dual invariance of the bidegeneracy
definition directly gives algorithms for \VSat\ as well.

In this construction, the dimension $d$ of the points is linear in $\degen$, leading to an exponential dependence in $\degen$ for the query time. This setup is needed for arbitrary queries (indeed, our lower bounds show that exponential dependencies in $\degen$ are unavoidable). When the query size $t$ is bounded (for primitives like $\ADJ$), we can optimize our technique to reduce the dependence on $\degen$ to just polynomial (\Thm{main-eval-bounded}). Since the query only involves $t$ vertices, a careful inspection shows that we only need to discover $t$ intervals (in $\bV$ and $\bE$) to certify that an edge $e$ satisfies a formula. It turns out to be convenient to dualize our geometric viewpoint and convert to a problem of \emph{box stabbing}. 
We represent hyperedges as a collection of boxes in $\RR^d$ (for constant $d$), where $d$ is \emph{independent} of $\degen$ and only depends on $t$. The numbers of boxes in the collection depends only polynomially in $\degen$. Formulas are now represented
as vectors (thus the geometric picture is the dual of the previous approach).
We now set up box stabbing data structures to solve the various versions of \ESat. The exponential dependence is only on $t$ (which needs to be known in advance), since the dimension is independent of $\degen$.

\medskip\noindent\uline{The complications of \VertexNeigh.}
For \VertexNeigh\ we follow the above approach, but with significant complications because of the ``second-order'' nature of the query.
For \ESat, whether an edge $e$ satisfies a query $\varphi$ involving vertices $v_1,\ldots,v_t$ depends only on the column of $e$ and the rows of $v_1,\ldots,v_t$ in the incidence matrix $M_H$.
For \VertexNeigh, instead, whether $u$ is a neighbor of $v$ is a function of \emph{all the columns of $M_H$}.
Thus, for \VertexNeigh, it is not obvious how to set up the geometric problem between the query vertex and other vertices.
We bypass the obstacle by breaking down the neighborhood $N(v)$ of the query vertex $v$ into an ``easy'' subset $N_0(v)$ and a ``hard'' subset $N_1(v)$.
Recall that, in the interval representation of $H$, every vertex holds a collection $\IV(v)$ of at most $\degen$ intervals of hyperedges in the ordering $\bE$.
Then, $N_0(v)$ consists of those neighbors of $v$ whose incidence with $v$ is certified by an edge being in both $\IV(v)$ and $\IV(u)$.
Therefore $N_0(v)$ is easy in the sense above---for any vertex $u \in N_0(v)$ the adjacency with $v$ is certified by the rows of $u$ and $v$ alone.
Exploiting this fact, we show that $N_0(v)$ can be counted using an approach analogous to \VSat\ queries, with the difference that now both vectors and boxes represent vertices.

To deal with $N_1(v)$, we apply a different approach.
We show that the set of pairs of vertices $\{(u,v) \ | \ u \in N_1(v)\}$ can be expressed as a collection of axis-parallel rectangles in $\RR^2$.
The sides of these rectangles are constructed from the intervals in the interval representation of $H$.
The construction also crucially uses an additional property of interval representations, \emph{niceness}, that is given for free by our peeling algorithm.\footnote{Not to be confused with the one of~\cite{bressan2026countinghypergraphletscolorcoding}.}
The problem of computing now $|N_1(v)|$ becomes a 2-d stabbing problem, where the query is the vertical line $(v,\cdot)$, and the answer is the size of the \emph{union} of the $y$-sides of boxes that cross that line.
Existing results for this kind of problem are too abstract and yield only vacuous bounds, thus we adapt them into an efficient algorithm.
We then enrich the data structures in order to support listing and sampling.

\subsection{Manuscript organization}
\Cref{sec:interval} introduces hypergraph bidegeneracy, the interval representation of $H$, and an algorithm for computing it.
\Cref{sec:impl-eval} describes the algorithm that uses the interval representation of $H$ to solve \ESat\ and \VSat\ queries via a reduction to Orthogonal Range Searching, proving~\Cref{thm:main-eval}.
\Cref{sec:bounded_query} deals with the case of bounded-size queries, proving~\Cref{thm:main-eval-bounded}.
\Cref{sec:vneigh} provides the algorithm for \VertexNeigh, proving~\Cref{thm:nc-main}.
Finally,~\Cref{sec:LB} proves the lower bounds of~\Cref{thm:main-lb}.
All omitted material can be found in the appendix.

\section{Hypergraph bidegeneracy and interval representations} \label{sec:interval}

We set up some notation and give a detailed exposition of hypergraph bidegeneracy. We also explain the notion of subhypergraphs, which gives more context for bidegeneracy and the relation to other hypergraph degeneracies.

We will need to distinguish between the \emph{identifier} of a hyperedge and the actual vertices it contains. We use $e \in E$ to denote the identifier
and use $V(e)$ to denote the set of vertices that $e$ represents. Similarly, we use $E(v)$ to denote the set of (identifiers of) hyperedges containing $v$.
This setup allows us to deal with multi-hypergraphs, which is critical for defining induced hypergraphs. When we say that two hyperedges are distinct, we mean that the identifiers are different. They could represent the same set.




There are numerous subtleties in defining \emph{subhypergraphs}. There is an entire hierarchy of definitions of induced subhypergraphs~\cite{PaSe26,paul-pena_counting_2026}. 
Observe that different hyperedges may induce the same hyperedge on $S$,
and definitions vary on whether these duplicates are kept.
This leads to the definition of sub-hypergraph used in this work, which is as follows.

\begin{definition} \label{def:subhyper} Consider a hypergraph $H = (V,E)$  with incidence matrix $M_H$ and consider $V' \subseteq V$ and $E' \subseteq E$. Let $M'$ be the submatrix of $M_H$ induced by $V'$ and $E'$.
The subhypergraph $H'=(V',E')$ of $H$ is the hypergraph with incidence matrix $M'$.
\end{definition}

We come to a crucial point. One may think of the deduplication step (removing ``copies'' of a hyperedge) as a natural operation. However, the dual operation
is to remove vertices that participate in the same hyperedges, which is not a standard operation. 
Any dual invariant definition needs to remove both ``duplicated'' vertices and hyperedges.

\begin{definition} \label{def:twin} 
Two distinct vertices $u,v \in V$ are \emph{twins} if $E(u)=E(v)$, and two distinct hyperedges $e,f$ are twins if $V(e) = V(f)$.
A hypergraph is twin-free if it contains no twins.
\end{definition}

Now for the central definition of our paper. This definition is equivalent to \Def{bidegen-intro}, but expressed in terms in subhypergraphs.
\begin{definition} \label{def:bidegen}
    A hypergraph $H$ is \emph{$(\vdegen,\edegen)$-bidegenerate} if every twin-free subhypergraph of $H$ has a vertex of degree at most $\vdegen$ or an edge of size at most $\edegen$. Equivalently, $H$ is $(\vdegen,\edegen)$-bidegenerate if every submatrix of $M_H$ with unique rows and columns contains a row (column) of Hamming weight at most $\vdegen$ ($\edegen$).
    A hypergraph is \emph{$\degen$-bidegenerate} if it is $(\degen,\degen)$-bidegenerate.
\end{definition}
We note that having bounded bidegeneracy is a weaker property than most existing definitions, see~\Fig{implications}. In Appendix~\ref{apx:implications}, we show that these existing notions are subsumed by bounded bidegeneracy.
%


\subsection{Interval representations}

We come to a key tool behind our algorithms: the \emph{interval representation} of a hypergraph.
This is a data structure that specifies orderings of $V$ and $E$, analogous to the degeneracy ordering for graphs~\cite{MaBe83, Se23}, and uses them to encode the hypergraph's incidence relation in a compact form.
This compact form consists in a short sequence of \emph{intervals} along those orderings, something that is absent from (and significantly more complex than) the standard degeneracy ordering over graphs.
We now give the formal definition of interval representation, and then provide some intuition for it.
A set $X$ is a \emph{$k$-interval subset} of $\{1,\ldots,N\}$ if $X$ can be written as the disjoint union of $k$ intervals $X_1,\ldots,X_k$ over $[N]$.
Clearly, every $k$-interval subset can be stored implicitly by using $k$ pairs of integers.
When there is no risk of ambiguity, we use $X$ to denote both the sequence of intervals $X_1,\ldots,X_k$ and their union $X_1 \uplus \ldots \uplus X_k$.

\begin{definition}[Interval representation of a hypergraph]\label{def:interval_representation}
Let $H=(V,E)$ be a hypergraph.
A $(\vdegen,\edegen)$-\emph{interval representation} of $H$ is a tuple $(\bV, \bE,\IV,\IE)$ with the following properties. First, $\bV = (v_1,\ldots,v_n)$ is an ordering of $V$, and $\bE=(e_1,\ldots,e_m)$ is an ordering of $E$. Second, $\IV$ maps every $v \in V$ to a $k_v$-interval subset $\IV(v)=I_1(v)\uplus \ldots\uplus I_{k_v}(v)$ of $\bE$, with $k_v \le \vdegen$; and $\IE$ maps every $e \in E$ to a $k_e$-interval subset $\IE(e)=J_1(e)\uplus \ldots\uplus J_{k_e}(e)$ of $\bV$, with $k_e \le \edegen$.
Finally, $\IV$ and $\IE$ satisfy:
\begin{enumerate}
    \item \textbf{Soundness:} letting $\IE^{-1}(v) = \{e \in E : v \in \IE(e)\}$ and $\IV^{-1}(e) = \{v \in V : e \in \IV(v)\}$,
        \begin{align} 
            E(v) &= \IV(v) \; \uplus \; \IE^{-1}(v), \quad \forall v \in V \,,\label{eq:IV} \\
            V(e) &= \IE(e) \; \uplus \; \IV^{-1}(e), \quad \forall e \in E \,.\label{eq:IE}
        \end{align}
    \item \textbf{Niceness:}
        \begin{align}
            \forall &v\in V, \forall i\in[k_v], \quad e,f\in I_i(v) \Rightarrow \IE(e)=\IE(f)\,, \\
            \forall &e\in E, \forall j\in[k_e], \quad u,v\in J_j(e) \Rightarrow \IV(u)=\IV(v)\,.
        \end{align}
\end{enumerate}
\end{definition}

Let us give some intuition about Definition~\ref{def:interval_representation}.
Ignore for the moment the niceness property (it is a technical condition given for free by our preprocessing, and exploited by the algorithms).
Soundness says that, if $v$ is incident with $e$, then exactly one of two cases holds: either $e \in \IV(v)$, or $v \in \IE(e)$.
Vice versa, if $v$ is not incident with $e$, then neither $e \in \IV(v)$ nor $v \in \IE(e)$.
Thus, under suitable permutations of $V$ and $E$, every incidence between a vertex $v$ and an edge $e$ can be verified by looking at two short lists.
For example, if $H$ has rank at most $k$, which includes the case of graphs, then we can trivially set $\IV(v)=\emptyset$ for all $v$, and set $\IE(e)$ to be the union of at most $|V(e)|\le k$ intervals, one per vertex, regardless of $\bV$ and $\bE$. 
Thus, hypergraphs of rank at most $k$ admit a $(0,k)$-representation, and in particular graphs admit a $(0,2)$-representation.
Symmetrically, hypergraphs of degree at most $k$ admit a $(k,0)$-representation.

In matrix language, Definition~\ref{def:interval_representation} says that the rows and columns of the incidence matrix $M$ of $H$ can be permuted so that $M = M_V + M_E$, where $M_V$ ($M_E$) is a Boolean matrix where every row (column) consists of at most $\vdegen$ ($\edegen$) blocks of consecutive ones.
However, clearly, storing the matrix would be too expensive.
An interval representation, instead, takes space only $O(|V|\cdot\vdegen + |E| \cdot \edegen)$, since each interval can be represented by its endpoints, which take one word of space each. See \Fig{example_intervals} for an example.

\begin{figure}[ht]
\centering
\resizebox{.9\textwidth}{!}{%
\begin{tikzpicture}[
    vtx/.style={circle, fill=black, inner sep=1.8pt},
    edge/.style={line width=1.1pt},
    vlab/.style={text=red!70!black, font=\Large},
    elab/.style={text=teal, font=\Large}
]

\draw[edge] (-1.645, 0.95) ellipse [x radius=3.4, y radius=1.9, rotate=150];
\draw[edge] ( 1.645, 0.95) ellipse [x radius=3.4, y radius=1.9, rotate=30];
\draw[edge] ( 0,    -1.9 ) ellipse [x radius=3.4, y radius=1.9, rotate=270];

\draw[edge] (-3.464, 2) ellipse [x radius=0.9, y radius=1.2, rotate=150];
\draw[edge] ( 3.464, 2) ellipse [x radius=0.9, y radius=1.2, rotate=30];
\draw[edge] ( 0,    -4) ellipse [x radius=0.9, y radius=1.2, rotate=270];

\node[vtx, label={[vlab]-30:$v_1$}] at (-3.689, 1.610) {};
\node[vtx, label={[vlab]-30:$v_2$}] at (-3.239, 2.390) {};
\node[vtx, label={[vlab]210:$v_3$}] at ( 3.239, 2.390) {};
\node[vtx, label={[vlab]210:$v_4$}] at ( 3.689, 1.610) {};
\node[vtx, label={[vlab]90:$v_5$}]  at (-0.45, -4) {};
\node[vtx, label={[vlab]90:$v_6$}]  at ( 0.45, -4) {};
\node[vtx, label={[vlab]90:$v_7$}]  at ( 0,      0.55 ) {};
\node[vtx, label={[vlab]210:$v_8$}] at (-0.476, -0.275) {};
\node[vtx, label={[vlab]330:$v_9$}] at ( 0.476, -0.275) {};

\node[elab] at (-3.940,  2.275) {$e_1$};
\node[elab] at ( 3.940,  2.275) {$e_2$};
\node[elab] at ( 0,     -4.55 ) {$e_3$};
\node[elab] at (-1.992,  1.15 ) {$e_4$};
\node[elab] at ( 1.992,  1.15 ) {$e_5$};
\node[elab] at ( 0,     -2.3  ) {$e_6$};

\node[anchor=west, font=\Large] (lists) at (5.2, -0.3) {$
\renewcommand{\arraystretch}{1.2}
\begin{array}{ll}
\multicolumn{2}{l}{\hspace{1.6em}{\color{red!70!black}\mathbf{V}} = ({\color{red!70!black}v_{1}}, {\color{red!70!black}v_{2}}, \dots, {\color{red!70!black}v_{9}})} \\
\multicolumn{2}{l}{\hspace{1.6em}{\color{teal}\mathbf{E}} = ({\color{teal}e_{1}}, {\color{teal}e_{2}}, \dots, {\color{teal}e_{6}})} \\[10pt]
\mathcal{I} \left\{
\begin{array}{l}
{\color{red!70!black}v_{1}} : [{\color{teal}e_{4},e_{4}}] \\
{\color{red!70!black}v_{2}} : [{\color{teal}e_{4},e_{4}}] \\
{\color{red!70!black}v_{3}} : [{\color{teal}e_{5},e_{5}}] \\
{\color{red!70!black}v_{4}} : [{\color{teal}e_{5},e_{5}}] \\
{\color{red!70!black}v_{5}} : [{\color{teal}e_{6},e_{6}}] \\
{\color{red!70!black}v_{6}} : [{\color{teal}e_{6},e_{6}}] \\
{\color{red!70!black}v_{7}} : [{\color{teal}e_{4}}, {\color{teal}e_{6}}] \\
{\color{red!70!black}v_{8}} : [{\color{teal}e_{4}}, {\color{teal}e_{6}}] \\
{\color{red!70!black}v_{9}} : [{\color{teal}e_{4}}, {\color{teal}e_{6}}]
\end{array}\right.
&
\quad \mathcal{J} \left\{
\begin{array}{l}
{\color{teal}e_{1}} : [{\color{red!70!black}v_{1}}, {\color{red!70!black}v_{2}}] \\
{\color{teal}e_{2}} : [{\color{red!70!black}v_{3}}, {\color{red!70!black}v_{4}}] \\
{\color{teal}e_{3}} : [{\color{red!70!black}v_{5}}, {\color{red!70!black}v_{6}}] \\
{\color{teal}e_{4}} : \emptyset \\
{\color{teal}e_{5}} : \emptyset \\
{\color{teal}e_{6}} : \emptyset
\end{array}\right.
\end{array}
$};

\node[anchor=west, font=\Large] at ([xshift=0.3cm]lists.east) {$
\renewcommand{\arraystretch}{1.2}
\begin{array}{c|cccccc}
       & {\color{teal}e_{1}} & {\color{teal}e_{2}} & {\color{teal}e_{3}} & {\color{teal}e_{4}} & {\color{teal}e_{5}} & {\color{teal}e_{6}} \\ \hline
{\color{red!70!black}v_{1}} & {\color{teal}1} & 0  & 0  & {\color{red!70!black}1} & 0  & 0  \\
{\color{red!70!black}v_{2}} & {\color{teal}1} & 0  & 0  & {\color{red!70!black}1} & 0  & 0  \\
{\color{red!70!black}v_{3}} & 0  & {\color{teal}1} & 0  & 0  & {\color{red!70!black}1} & 0  \\
{\color{red!70!black}v_{4}} & 0  & {\color{teal}1} & 0  & 0  & {\color{red!70!black}1} & 0  \\
{\color{red!70!black}v_{5}} & 0  & 0  & {\color{teal}1} & 0  & 0  & {\color{red!70!black}1} \\
{\color{red!70!black}v_{6}} & 0  & 0  & {\color{teal}1} & 0  & 0  & {\color{red!70!black}1} \\
{\color{red!70!black}v_{7}} & 0  & 0  & 0  & {\color{red!70!black}1} & {\color{red!70!black}1} & {\color{red!70!black}1} \\
{\color{red!70!black}v_{8}} & 0  & 0  & 0  & {\color{red!70!black}1} & {\color{red!70!black}1} & {\color{red!70!black}1} \\
{\color{red!70!black}v_{9}} & 0  & 0  & 0  & {\color{red!70!black}1} & {\color{red!70!black}1} & {\color{red!70!black}1}
\end{array}
$};

\end{tikzpicture}%
}
\caption{An example of a $(1,1)$-bidegenerate hypergraph and a valid $(1,1)$-interval representation $(\bV,\bE,\IV,\IE)$. This representation is \emph{sound}, as every vertex and hyperedge gets defined from the intervals $\IV$ and $\IE$, for example $E(v_1) = \IV(v_1) \uplus \IE^{-1}(v_1)=\{e_1,e_4\}$, as $e_4 \in \IV(v_1)$ and $v_1 \in \IE(e_1)$ (hence $e_1 \in \IE^{-1}(v_1)$). The representation is also \emph{nice}, as the edges $e_4,e_5,e_6$ have the same intervals in $\IE$ and each of $(v_1,v_2)$, $(v_3,v_4)$ and $(v_5,v_6)$ have the same intervals in $\IV$. \\On the right the incidence matrix of the hypergraph, in red the entries corresponding to $M_V$ and in blue the entries corresponding to $M_E$, matching the assignments of $\IV$ and $\IE$. The red entries in every row and the blue entries in each column form a single block of 1's.}
\label{fig:example_intervals}
\end{figure}



\subsection{From bidegeneracy to interval representations}

It is easy to see that a hypergraph is $(\vdegen,\edegen)$-bidegenerate if and only if its incidence matrix can be reduced to the empty matrix by a sequence of deletions of rows of (Hamming) weight at most $\vdegen$, and columns of (Hamming) weight at most $\edegen$ (provided that identical rows and columns are merged whenever they appear). However, this characterization is not very useful. Here, we show that such a deletion process can be carried out efficiently while storing information about merges, so as to obtain a $(\vdegen,\edegen)$-interval representation.

\begin{theorem} \label{thm:interval}
Every $(\vdegen,\edegen)$-bidegenerate hypergraph $H$ has a $(\vdegen,\edegen)$-interval representation that, given $H$ and one of $\vdegen$ and $\edegen$, can be computed deterministically in time $O(\size{H}\ln^{1.5}\size{H})$.
\end{theorem}

Finding an interval representation with optimal parameters is hard in general.
Even deciding whether a hypergraph admits a $(0,k)$-interval representation is equivalent to testing whether its incidence matrix satisfies the $k$-consecutive-ones property, which is NP-complete for every $k\ge 2$~\cite{Goldberg95}.
The algorithm of~\Cref{thm:interval} certifies the bidegeneracy of $H$: given $\vdegen$, it finds the smallest $\edegen$ such that $H$ is $(\vdegen,\edegen)$-bidegenerate. In this sense the bidegeneracy of $H$ is efficiently computable.


The rest of this section proves \Cref{thm:interval} by giving an algorithm that computes the interval representation.
In particular, \Cref{sub:peeling_overview} gives a high-level description of the algorithm, \Cref{sub:peeling_correct} proves its correctness, and \Cref{sub:peeling_time} describes a careful implementation that brings the algorithm in the time prescribed by \Cref{thm:interval}.
The main idea is akin to the algorithm for obtaining the degeneracy ordering in a graph, where we peel the lowest degree vertex at each time. In our case, we peel either a vertex of small degree or a hyperedge of small size. This is complicated by the fact that we need to efficiently detect twins on the fly and remove them. The interval representation arises from constructing two auxiliary forests that track the peeling process, and then computing certain intervals over the leaves of those forests.

\subsubsection{The algorithm}\label{sub:peeling_overview}
We start by describing the algorithm in terms of a \emph{peeling process}.
Let $H=(V,E)$, and recall that deleting $v \in V$ yields the hypergraph whose incidence matrix is obtained from the one of $H$ by removing the row of $v$ (similarly, deleting $e \in E$ amounts to removing from the matrix the column of $e$).
Until $H$ has no vertices and no edges, we repeat the following step.
\begin{itemize}
    \item 
If $H$ contains twin vertices $u,v$, then delete $u$ and $v$ and add a new vertex $x_{uv}$ of type $E(u)=E(v)$.
If $H$ contains twin edges $e,f$, then delete $e$ and $f$ and add a new edge $x_{ef}$ of the type $V(e)=V(f)$.
We refer to this operation as a \emph{merge}.
    \item 
If $H$ is twin-free, then delete a vertex of degree at most $\vdegen$ if one exists, otherwise delete an edge of size at most $\edegen$.
We refer to this operation as a \emph{deletion}.
\end{itemize}

Throughout the process we construct two data structures, the \emph{vertex forest} $\cT_V$ and the \emph{edge forest} $\cT_E$.
At the beginning, $\cT_V$ contains one isolated node $X_v$ per vertex $v \in V$; similarly, $\cT_E$ contains one isolated node $X_e$ per edge of $E$.
We then update the forests as follows, maintaining the invariant that every vertex (edge) of $H$ created along the process, or originally in $V$ ($E$), is associated to a unique node in $\cT_V$ ($\cT_E$).
\begin{itemize}
    \item 
When we delete a vertex $v$, we store in $X_v$ the pointers to all nodes $X_e$ of $\cT_E$ such that $e$ contains $v$.
Denote this set of pointers by $\PTR(X_v)$.
Note that, by construction,  $|\PTR(X_v)| \le \vdegen$.
Similarly, when we delete an edge $e$, we store in $X_e$ the set $\PTR(X_e)$ of pointers to all nodes $X_v$ of $\cT_V$ such that $v$ is in $e$, and $|\PTR(X_e)| \le \edegen$.
    \item
When we merge two vertices $u,v$, we create in $\cT_V$ a new node $X_{uv}$ associated to the new vertex $x_{uv}$, and set $X_u$ and $X_v$ as its children, thus merging their subtrees into a new tree rooted at $X_{uv}$.
Similarly, when we merge edges $e,f$, we create in $\cT_E$ a new node $X_{ef}$ with $X_e$ and $X_f$ as children.
\end{itemize}

\noindent At the end of the process we perform a postorder traversal of $\cT_V$, and record:
\begin{itemize}
    \item The order in which the leaf nodes $\{X_v: v \in V\}$ are visited. This gives the total order $\bV=(v_1,\ldots,v_n)$ over $V$.
    \item For every node $X$ in the forest, the interval $J(X)$ formed in $\bV$ by the leaves of the subtree rooted at $X$, encoded as the pair of indices $j,j'$ such that $J(X)=\{v_j,\ldots,v_{j'}\}$.
\end{itemize}
Do the same for $\cT_E$ to obtain the total order $\bE$ and the intervals $I(\cdot)$.
For every node $X$ in a forest, let $r(X)$ be the root of the subtree it lies in.
Finally, define:
\begin{align}
    \IV(v) &= \{I(X) : X \in \PTR(r(X_v)) \}\, : \forall v \in V\\
    \IE(e) &= \{J(X) : X \in \PTR(r(X_e)) \}\, : \forall e \in E
\end{align}

\begin{figure}[!htbp]
\centering
\begin{minipage}{0.499\textwidth}
\centering
Initial Hypergraph\\[4pt]
\resizebox{\textwidth}{!}{%
\begin{tikzpicture}[
    tnode/.style={circle, draw, line width=1.5pt, minimum size=1.20cm, inner sep=1pt, font=\Huge},
    vtx/.style={circle, fill=black, inner sep=3pt},
    edge/.style={line width=2pt},
    vlab/.style={text=red!70!black, fill=white, fill opacity=0.85, text opacity=1, inner sep=0.5pt, rounded corners=2pt, font=\Huge},
    elab/.style={text=teal, fill=white, fill opacity=0.85, text opacity=1, inner sep=0.5pt, rounded corners=2pt, font=\Huge}
]

\draw[edge] (-1.645, 0.95) ellipse [x radius=3.4, y radius=1.9, rotate=150];
\draw[edge] ( 1.645, 0.95) ellipse [x radius=3.4, y radius=1.9, rotate=30];
\draw[edge] ( 0,    -1.9 ) ellipse [x radius=3.4, y radius=1.9, rotate=270];

\draw[edge] (-3.464, 2) ellipse [x radius=0.9, y radius=1.2, rotate=150];
\draw[edge] ( 3.464, 2) ellipse [x radius=0.9, y radius=1.2, rotate=30];
\draw[edge] ( 0,    -4) ellipse [x radius=0.9, y radius=1.2, rotate=270];

\node[vtx, label={[vlab]-30:$v_1$}] at (-3.689, 1.610) {};
\node[vtx, label={[vlab]-30:$v_2$}] at (-3.239, 2.390) {};
\node[vtx, label={[vlab]210:$v_3$}] at ( 3.239, 2.390) {};
\node[vtx, label={[vlab]210:$v_4$}] at ( 3.689, 1.610) {};
\node[vtx, label={[vlab]90:$v_5$}]  at (-0.45, -4) {};
\node[vtx, label={[vlab]90:$v_6$}]  at ( 0.45, -4) {};
\node[vtx, label={[vlab]90:$v_7$}]  at ( 0,      0.55 ) {};
\node[vtx, label={[vlab]210:$v_8$}] at (-0.476, -0.275) {};
\node[vtx, label={[vlab]330:$v_9$}] at ( 0.476, -0.275) {};

\node[elab] at (-3.940,  2.275) {$e_1$};
\node[elab] at ( 3.940,  2.275) {$e_2$};
\node[elab] at ( 0,     -4.55 ) {$e_3$};
\node[elab] at (-1.992,  1.15 ) {$e_4$};
\node[elab] at ( 1.992,  1.15 ) {$e_5$};
\node[elab] at ( 0,     -2.3  ) {$e_6$};

\begin{scope}[xshift=0cm, yshift=1cm]
\node[font=\Huge, fill=white, inner sep=1pt] at (10.10, -2.3) {$\mathcal{T}_V$};
\node[tnode] (tv1) at (5.10, -0.975) {${\color{red!70!black}v_{1}}$};
\node[tnode] (tv2) at (6.35, -0.975) {${\color{red!70!black}v_{2}}$};
\node[tnode] (tv3) at (7.60, -0.975) {${\color{red!70!black}v_{3}}$};
\node[tnode] (tv4) at (8.85, -0.975) {${\color{red!70!black}v_{4}}$};
\node[tnode] (tv5) at (10.10, -0.975) {${\color{red!70!black}v_{5}}$};
\node[tnode] (tv6) at (11.35, -0.975) {${\color{red!70!black}v_{6}}$};
\node[tnode] (tv7) at (12.60, -0.975) {${\color{red!70!black}v_{7}}$};
\node[tnode] (tv8) at (13.85, -0.975) {${\color{red!70!black}v_{8}}$};
\node[tnode] (tv9) at (15.10, -0.975) {${\color{red!70!black}v_{9}}$};

\node[font=\Huge, fill=white, inner sep=1pt] at (10.10, -7.85) {$\mathcal{T}_E$};
\node[tnode] (te1) at (5.10, -6.55) {${\color{teal}e_{1}}$};
\node[tnode] (te2) at (7.10, -6.55) {${\color{teal}e_{2}}$};
\node[tnode] (te3) at (9.10, -6.55) {${\color{teal}e_{3}}$};
\node[tnode] (te4) at (11.10, -6.55) {${\color{teal}e_{4}}$};
\node[tnode] (te5) at (13.10, -6.55) {${\color{teal}e_{5}}$};
\node[tnode] (te6) at (15.10, -6.55) {${\color{teal}e_{6}}$};

\end{scope}
\path (-4.8, 3.8) rectangle (15.75, -7.3);
\end{tikzpicture}%
}
\end{minipage}
\hfill
\begin{minipage}{0.499\textwidth}
\centering
Step 1\\[4pt]
\resizebox{\textwidth}{!}{%
\begin{tikzpicture}[
    tnode/.style={circle, draw, line width=1.5pt, minimum size=1.20cm, inner sep=1pt, font=\Huge},
    vtx/.style={circle, fill=black, inner sep=3pt},
    edge/.style={line width=2pt},
    vlab/.style={text=red!70!black, fill=white, fill opacity=0.85, text opacity=1, inner sep=0.5pt, rounded corners=2pt, font=\Huge},
    elab/.style={text=teal, fill=white, fill opacity=0.85, text opacity=1, inner sep=0.5pt, rounded corners=2pt, font=\Huge}
]

\draw[edge] (-1.645, 0.95) ellipse [x radius=3.4, y radius=1.9, rotate=150];
\draw[edge] ( 1.645, 0.95) ellipse [x radius=3.4, y radius=1.9, rotate=30];
\draw[edge] ( 0,    -1.9 ) ellipse [x radius=3.4, y radius=1.9, rotate=270];

\draw[edge] (-3.464, 2) ellipse [x radius=0.9, y radius=1.2, rotate=150];
\draw[edge] ( 3.464, 2) ellipse [x radius=0.9, y radius=1.2, rotate=30];
\draw[edge] ( 0,    -4) ellipse [x radius=0.9, y radius=1.2, rotate=270];

\node[vtx, label={[vlab]60:$v_{12}$}]  at (-3.464,  1.7) {};
\node[vtx, label={[vlab]120:$v_{34}$}]  at ( 3.464,  1.7) {};
\node[vtx, label={[vlab]90:$v_{56}$}]   at ( 0,     -4) {};
\node[vtx, label={[vlab]90:$v_{789}$}]  at ( 0,      0) {};

\node[elab] at (-3.940,  2.275) {$e_1$};
\node[elab] at ( 3.940,  2.275) {$e_2$};
\node[elab] at ( 0,     -4.55 ) {$e_3$};
\node[elab] at (-1.992,  1.15 ) {$e_4$};
\node[elab] at ( 1.992,  1.15 ) {$e_5$};
\node[elab] at ( 0,     -2.3  ) {$e_6$};

\begin{scope}[xshift=0cm, yshift=1cm]
\node[font=\Huge, fill=white, inner sep=1pt] at (10.10, -2.3) {$\mathcal{T}_V$};
\node[tnode] (tv1) at (5.10, -0.975) {${\color{red!70!black}v_{1}}$};
\node[tnode] (tv2) at (6.35, -0.975) {${\color{red!70!black}v_{2}}$};
\node[tnode] (tv3) at (7.60, -0.975) {${\color{red!70!black}v_{3}}$};
\node[tnode] (tv4) at (8.85, -0.975) {${\color{red!70!black}v_{4}}$};
\node[tnode] (tv5) at (10.10, -0.975) {${\color{red!70!black}v_{5}}$};
\node[tnode] (tv6) at (11.35, -0.975) {${\color{red!70!black}v_{6}}$};
\node[tnode] (tv7) at (12.60, -0.975) {${\color{red!70!black}v_{7}}$};
\node[tnode] (tv8) at (13.85, -0.975) {${\color{red!70!black}v_{8}}$};
\node[tnode] (tv9) at (15.10, -0.975) {${\color{red!70!black}v_{9}}$};

\node[font=\Huge, fill=white, inner sep=1pt] at (10.10, -7.85) {$\mathcal{T}_E$};
\node[tnode] (te1) at (5.10, -6.55) {${\color{teal}e_{1}}$};
\node[tnode] (te2) at (7.10, -6.55) {${\color{teal}e_{2}}$};
\node[tnode] (te3) at (9.10, -6.55) {${\color{teal}e_{3}}$};
\node[tnode] (te4) at (11.10, -6.55) {${\color{teal}e_{4}}$};
\node[tnode] (te5) at (13.10, -6.55) {${\color{teal}e_{5}}$};
\node[tnode] (te6) at (15.10, -6.55) {${\color{teal}e_{6}}$};

\node[tnode] (tv12)  at (5.72, 0.725) {${\color{red!70!black}v_{12}}$};
\node[tnode] (tv34)  at (8.22, 0.725) {${\color{red!70!black}v_{34}}$};
\node[tnode] (tv56)  at (10.72, 0.725) {${\color{red!70!black}v_{56}}$};
\node[tnode] (tv78)  at (13.22, 0.725) {${\color{red!70!black}v_{78}}$};
\node[tnode] (tv789) at (14.16, 2.1) {${\color{red!70!black}v_{789}}$};
\draw[thick] (tv12) -- (tv1);  \draw[thick] (tv12) -- (tv2);
\draw[thick] (tv34) -- (tv3);  \draw[thick] (tv34) -- (tv4);
\draw[thick] (tv56) -- (tv5);  \draw[thick] (tv56) -- (tv6);
\draw[thick] (tv78) -- (tv7);  \draw[thick] (tv78) -- (tv8);
\draw[thick] (tv789) -- (tv78); \draw[thick] (tv789) -- (tv9);

\end{scope}
\path (-4.8, 3.8) rectangle (15.75, -7.3);
\end{tikzpicture}%
}
\end{minipage}%

\vspace{1.2em}

\begin{minipage}{0.499\textwidth}
\centering
Step 2\\[4pt]
\resizebox{\textwidth}{!}{%
\begin{tikzpicture}[
    tnode/.style={circle, draw, line width=1.5pt, fill=white, minimum size=1.20cm, inner sep=1pt, font=\Huge},
    vtx/.style={circle, fill=black, inner sep=3pt},
    edge/.style={line width=2pt},
    vlab/.style={text=red!70!black, fill=white, fill opacity=0.85, text opacity=1, inner sep=0.5pt, rounded corners=2pt, font=\Huge},
    elab/.style={text=teal, fill=white, fill opacity=0.85, text opacity=1, inner sep=0.5pt, rounded corners=2pt, font=\Huge}
]

\draw[edge] (-1.645, 0.95) ellipse [x radius=3.4, y radius=1.9, rotate=150];
\draw[edge] ( 1.645, 0.95) ellipse [x radius=3.4, y radius=1.9, rotate=30];
\draw[edge] ( 0,    -1.9 ) ellipse [x radius=3.4, y radius=1.9, rotate=270];

\node[vtx, label={[vlab]60:$v_{12}$}]  at (-3.464,  1.7) {};
\node[vtx, label={[vlab]120:$v_{34}$}]  at ( 3.464,  1.7) {};
\node[vtx, label={[vlab]90:$v_{56}$}]   at ( 0,     -4) {};
\node[vtx, label={[vlab]90:$v_{789}$}]  at ( 0,      0) {};

\node[elab] at (-1.992,  1.15 ) {$e_4$};
\node[elab] at ( 1.992,  1.15 ) {$e_5$};
\node[elab] at ( 0,     -2.3  ) {$e_6$};

\begin{scope}[xshift=0cm, yshift=1cm]
\draw[violet, dashed, very thick, -{Stealth[length=4mm, width=3.2mm]}, shorten <=0.65cm, shorten >=0.65cm] (5.10, -6.55) to[bend left=40] (5.72, 0.725);
\draw[violet, dashed, very thick, -{Stealth[length=4mm, width=3.2mm]}, shorten <=0.65cm, shorten >=0.65cm] (7.10, -6.55) to[bend left=40] (8.22, 0.725);
\draw[violet, dashed, very thick, -{Stealth[length=4mm, width=3.2mm]}, shorten <=0.65cm, shorten >=0.65cm] (9.10, -6.55) to[bend left=40] (10.72, 0.725);

\node[font=\Huge, fill=white, inner sep=1pt] at (10.10, -2.3) {$\mathcal{T}_V$};
\node[tnode] (tv1) at (5.10, -0.975) {${\color{red!70!black}v_{1}}$};
\node[tnode] (tv2) at (6.35, -0.975) {${\color{red!70!black}v_{2}}$};
\node[tnode] (tv3) at (7.60, -0.975) {${\color{red!70!black}v_{3}}$};
\node[tnode] (tv4) at (8.85, -0.975) {${\color{red!70!black}v_{4}}$};
\node[tnode] (tv5) at (10.10, -0.975) {${\color{red!70!black}v_{5}}$};
\node[tnode] (tv6) at (11.35, -0.975) {${\color{red!70!black}v_{6}}$};
\node[tnode] (tv7) at (12.60, -0.975) {${\color{red!70!black}v_{7}}$};
\node[tnode] (tv8) at (13.85, -0.975) {${\color{red!70!black}v_{8}}$};
\node[tnode] (tv9) at (15.10, -0.975) {${\color{red!70!black}v_{9}}$};

\node[font=\Huge, fill=white, inner sep=1pt] at (10.10, -7.85) {$\mathcal{T}_E$};
\node[tnode] (te1) at (5.10, -6.55) {${\color{teal}e_{1}}$};
\node[tnode] (te2) at (7.10, -6.55) {${\color{teal}e_{2}}$};
\node[tnode] (te3) at (9.10, -6.55) {${\color{teal}e_{3}}$};
\node[tnode] (te4) at (11.10, -6.55) {${\color{teal}e_{4}}$};
\node[tnode] (te5) at (13.10, -6.55) {${\color{teal}e_{5}}$};
\node[tnode] (te6) at (15.10, -6.55) {${\color{teal}e_{6}}$};

\node[tnode] (tv12)  at (5.72, 0.725) {${\color{red!70!black}v_{12}}$};
\node[tnode] (tv34)  at (8.22, 0.725) {${\color{red!70!black}v_{34}}$};
\node[tnode] (tv56)  at (10.72, 0.725) {${\color{red!70!black}v_{56}}$};
\node[tnode] (tv78)  at (13.22, 0.725) {${\color{red!70!black}v_{78}}$};
\node[tnode] (tv789) at (14.16, 2.1) {${\color{red!70!black}v_{789}}$};
\draw[thick] (tv12) -- (tv1);  \draw[thick] (tv12) -- (tv2);
\draw[thick] (tv34) -- (tv3);  \draw[thick] (tv34) -- (tv4);
\draw[thick] (tv56) -- (tv5);  \draw[thick] (tv56) -- (tv6);
\draw[thick] (tv78) -- (tv7);  \draw[thick] (tv78) -- (tv8);
\draw[thick] (tv789) -- (tv78); \draw[thick] (tv789) -- (tv9);

\end{scope}
\path (-4.8, 3.8) rectangle (15.75, -7.3);
\end{tikzpicture}%
}
\end{minipage}
\hfill
\begin{minipage}{0.499\textwidth}
\centering
Step 3\\[4pt]
\resizebox{\textwidth}{!}{%
\begin{tikzpicture}[
    tnode/.style={circle, draw, line width=1.5pt, fill=white, minimum size=1.20cm, inner sep=1pt, font=\Huge},
    vtx/.style={circle, fill=black, inner sep=3pt},
    edge/.style={line width=2pt},
    vlab/.style={text=red!70!black, fill=white, fill opacity=0.85, text opacity=1, inner sep=0.5pt, rounded corners=2pt, font=\Huge},
    elab/.style={text=teal, fill=white, fill opacity=0.85, text opacity=1, inner sep=0.5pt, rounded corners=2pt, font=\Huge}
]

\draw[edge] (-1.039, 0.60) ellipse [x radius=2.2, y radius=1.6, rotate=150];
\draw[edge] ( 1.039, 0.60) ellipse [x radius=2.2, y radius=1.6, rotate=30];
\draw[edge] ( 0,    -1.2 ) ellipse [x radius=2.2, y radius=1.6, rotate=270];
\path (-4.75, 3.32) -- (-4.75, -5.3);

\node[vtx, label={[vlab]90:$v_{789}$}]  at ( 0,      0) {};

\node[elab] at (-1.732,  1.00 ) {$e_4$};
\node[elab] at ( 1.732,  1.00 ) {$e_5$};
\node[elab] at ( 0,     -2.0  ) {$e_6$};

\begin{scope}[xshift=0cm, yshift=1cm]
\draw[violet, dashed, very thick, -{Stealth[length=4mm, width=3.2mm]}, shorten <=0.65cm, shorten >=0.65cm] (5.10, -6.55) to[bend left=40] (5.72, 0.725);
\draw[violet, dashed, very thick, -{Stealth[length=4mm, width=3.2mm]}, shorten <=0.65cm, shorten >=0.65cm] (7.10, -6.55) to[bend left=40] (8.22, 0.725);
\draw[violet, dashed, very thick, -{Stealth[length=4mm, width=3.2mm]}, shorten <=0.65cm, shorten >=0.65cm] (9.10, -6.55) to[bend left=40] (10.72, 0.725);

\draw[orange, dashed, very thick, -{Stealth[length=4mm, width=3.2mm]}, shorten <=0.65cm, shorten >=0.65cm] (5.72, 0.725) to[bend left=25] (11.10, -6.55);
\draw[orange, dashed, very thick, -{Stealth[length=4mm, width=3.2mm]}, shorten <=0.65cm, shorten >=0.65cm] (8.22, 0.725) to[bend left=25] (13.10, -6.55);
\draw[orange, dashed, very thick, -{Stealth[length=4mm, width=3.2mm]}, shorten <=0.65cm, shorten >=0.65cm] (10.72, 0.725) to[bend left=25] (15.10, -6.55);

\node[font=\Huge, fill=white, inner sep=1pt] at (10.10, -2.3) {$\mathcal{T}_V$};
\node[tnode] (tv1) at (5.10, -0.975) {${\color{red!70!black}v_{1}}$};
\node[tnode] (tv2) at (6.35, -0.975) {${\color{red!70!black}v_{2}}$};
\node[tnode] (tv3) at (7.60, -0.975) {${\color{red!70!black}v_{3}}$};
\node[tnode] (tv4) at (8.85, -0.975) {${\color{red!70!black}v_{4}}$};
\node[tnode] (tv5) at (10.10, -0.975) {${\color{red!70!black}v_{5}}$};
\node[tnode] (tv6) at (11.35, -0.975) {${\color{red!70!black}v_{6}}$};
\node[tnode] (tv7) at (12.60, -0.975) {${\color{red!70!black}v_{7}}$};
\node[tnode] (tv8) at (13.85, -0.975) {${\color{red!70!black}v_{8}}$};
\node[tnode] (tv9) at (15.10, -0.975) {${\color{red!70!black}v_{9}}$};

\node[font=\Huge, fill=white, inner sep=1pt] at (10.10, -7.85) {$\mathcal{T}_E$};
\node[tnode] (te1) at (5.10, -6.55) {${\color{teal}e_{1}}$};
\node[tnode] (te2) at (7.10, -6.55) {${\color{teal}e_{2}}$};
\node[tnode] (te3) at (9.10, -6.55) {${\color{teal}e_{3}}$};
\node[tnode] (te4) at (11.10, -6.55) {${\color{teal}e_{4}}$};
\node[tnode] (te5) at (13.10, -6.55) {${\color{teal}e_{5}}$};
\node[tnode] (te6) at (15.10, -6.55) {${\color{teal}e_{6}}$};

\node[tnode] (tv12)  at (5.72, 0.725) {${\color{red!70!black}v_{12}}$};
\node[tnode] (tv34)  at (8.22, 0.725) {${\color{red!70!black}v_{34}}$};
\node[tnode] (tv56)  at (10.72, 0.725) {${\color{red!70!black}v_{56}}$};
\node[tnode] (tv78)  at (13.22, 0.725) {${\color{red!70!black}v_{78}}$};
\node[tnode] (tv789) at (14.16, 2.1) {${\color{red!70!black}v_{789}}$};
\draw[thick] (tv12) -- (tv1);  \draw[thick] (tv12) -- (tv2);
\draw[thick] (tv34) -- (tv3);  \draw[thick] (tv34) -- (tv4);
\draw[thick] (tv56) -- (tv5);  \draw[thick] (tv56) -- (tv6);
\draw[thick] (tv78) -- (tv7);  \draw[thick] (tv78) -- (tv8);
\draw[thick] (tv789) -- (tv78); \draw[thick] (tv789) -- (tv9);

\end{scope}
\path (-4.8, 3.8) rectangle (15.75, -7.3);
\end{tikzpicture}%
}
\end{minipage}%

\vspace{1.2em}

\begin{minipage}{0.499\textwidth}
\centering
Step 4\\[4pt]
\resizebox{\textwidth}{!}{%
\begin{tikzpicture}[
    tnode/.style={circle, draw, line width=1.5pt, fill=white, minimum size=1.20cm, inner sep=1pt, font=\Huge},
    vtx/.style={circle, fill=black, inner sep=3pt},
    edge/.style={line width=2pt},
    vlab/.style={text=red!70!black, fill=white, fill opacity=0.85, text opacity=1, inner sep=0.5pt, rounded corners=2pt, font=\Huge},
    elab/.style={text=teal, fill=white, fill opacity=0.85, text opacity=1, inner sep=0.5pt, rounded corners=2pt, font=\Huge}
]

\draw[edge] (0, -0.2) circle (1.5);
\path (-4.75, 3.32) -- (-4.75, -5.3);

\node[vtx, label={[vlab]90:$v_{789}$}]  at ( 0,      0) {};

\node[elab] at (0, -0.9) {$e_{456}$};

\begin{scope}[xshift=0cm, yshift=1cm]
\draw[violet, dashed, very thick, -{Stealth[length=4mm, width=3.2mm]}, shorten <=0.65cm, shorten >=0.65cm] (5.10, -6.55) to[bend left=40] (5.72, 0.725);
\draw[violet, dashed, very thick, -{Stealth[length=4mm, width=3.2mm]}, shorten <=0.65cm, shorten >=0.65cm] (7.10, -6.55) to[bend left=40] (8.22, 0.725);
\draw[violet, dashed, very thick, -{Stealth[length=4mm, width=3.2mm]}, shorten <=0.65cm, shorten >=0.65cm] (9.10, -6.55) to[bend left=40] (10.72, 0.725);

\draw[orange, dashed, very thick, -{Stealth[length=4mm, width=3.2mm]}, shorten <=0.65cm, shorten >=0.65cm] (5.72, 0.725) to[bend left=25] (11.10, -6.55);
\draw[orange, dashed, very thick, -{Stealth[length=4mm, width=3.2mm]}, shorten <=0.65cm, shorten >=0.65cm] (8.22, 0.725) to[bend left=25] (13.10, -6.55);
\draw[orange, dashed, very thick, -{Stealth[length=4mm, width=3.2mm]}, shorten <=0.65cm, shorten >=0.65cm] (10.72, 0.725) to[bend left=25] (15.10, -6.55);

\node[font=\Huge, fill=white, inner sep=1pt] at (10.10, -2.3) {$\mathcal{T}_V$};
\node[tnode] (tv1) at (5.10, -0.975) {${\color{red!70!black}v_{1}}$};
\node[tnode] (tv2) at (6.35, -0.975) {${\color{red!70!black}v_{2}}$};
\node[tnode] (tv3) at (7.60, -0.975) {${\color{red!70!black}v_{3}}$};
\node[tnode] (tv4) at (8.85, -0.975) {${\color{red!70!black}v_{4}}$};
\node[tnode] (tv5) at (10.10, -0.975) {${\color{red!70!black}v_{5}}$};
\node[tnode] (tv6) at (11.35, -0.975) {${\color{red!70!black}v_{6}}$};
\node[tnode] (tv7) at (12.60, -0.975) {${\color{red!70!black}v_{7}}$};
\node[tnode] (tv8) at (13.85, -0.975) {${\color{red!70!black}v_{8}}$};
\node[tnode] (tv9) at (15.10, -0.975) {${\color{red!70!black}v_{9}}$};

\node[font=\Huge, fill=white, inner sep=1pt] at (10.10, -7.85) {$\mathcal{T}_E$};
\node[tnode] (te1) at (5.10, -6.55) {${\color{teal}e_{1}}$};
\node[tnode] (te2) at (7.10, -6.55) {${\color{teal}e_{2}}$};
\node[tnode] (te3) at (9.10, -6.55) {${\color{teal}e_{3}}$};
\node[tnode] (te4) at (11.10, -6.55) {${\color{teal}e_{4}}$};
\node[tnode] (te5) at (13.10, -6.55) {${\color{teal}e_{5}}$};
\node[tnode] (te6) at (15.10, -6.55) {${\color{teal}e_{6}}$};

\node[tnode] (tv12)  at (5.72, 0.725) {${\color{red!70!black}v_{12}}$};
\node[tnode] (tv34)  at (8.22, 0.725) {${\color{red!70!black}v_{34}}$};
\node[tnode] (tv56)  at (10.72, 0.725) {${\color{red!70!black}v_{56}}$};
\node[tnode] (tv78)  at (13.22, 0.725) {${\color{red!70!black}v_{78}}$};
\node[tnode] (tv789) at (14.16, 2.1) {${\color{red!70!black}v_{789}}$};
\draw[thick] (tv12) -- (tv1);  \draw[thick] (tv12) -- (tv2);
\draw[thick] (tv34) -- (tv3);  \draw[thick] (tv34) -- (tv4);
\draw[thick] (tv56) -- (tv5);  \draw[thick] (tv56) -- (tv6);
\draw[thick] (tv78) -- (tv7);  \draw[thick] (tv78) -- (tv8);
\draw[thick] (tv789) -- (tv78); \draw[thick] (tv789) -- (tv9);

\node[tnode] (te45)  at (12.10, -4.95) {${\color{teal}e_{45}}$};
\node[tnode] (te456) at (13.60, -3.35) {${\color{teal}e_{456}}$};
\draw[thick] (te45) -- (te4);   \draw[thick] (te45) -- (te5);
\draw[thick] (te456) -- (te45); \draw[thick] (te456) -- (te6);

\end{scope}
\path (-4.8, 3.8) rectangle (15.75, -7.3);
\end{tikzpicture}%
}
\end{minipage}
\hfill
\begin{minipage}{0.499\textwidth}
\centering
Step 5\\[4pt]
\resizebox{\textwidth}{!}{%
\begin{tikzpicture}[
    tnode/.style={circle, draw, line width=1.5pt, fill=white, minimum size=1.20cm, inner sep=1pt, font=\Huge},
    vtx/.style={circle, fill=black, inner sep=3pt},
    edge/.style={line width=2pt},
    vlab/.style={text=red!70!black, fill=white, fill opacity=0.85, text opacity=1, inner sep=0.5pt, rounded corners=2pt, font=\Huge},
    elab/.style={text=teal, fill=white, fill opacity=0.85, text opacity=1, inner sep=0.5pt, rounded corners=2pt, font=\Huge}
]

\path (-4.75, 3.32) -- (4.75, -5.3);

\begin{scope}[xshift=-3.5cm, yshift=1cm]
\draw[violet, dashed, very thick, -{Stealth[length=4mm, width=3.2mm]}, shorten <=0.65cm, shorten >=0.65cm] (5.10, -6.55) to[bend left=40] (5.72, 0.725);
\draw[violet, dashed, very thick, -{Stealth[length=4mm, width=3.2mm]}, shorten <=0.65cm, shorten >=0.65cm] (7.10, -6.55) to[bend left=40] (8.22, 0.725);
\draw[violet, dashed, very thick, -{Stealth[length=4mm, width=3.2mm]}, shorten <=0.65cm, shorten >=0.65cm] (9.10, -6.55) to[bend left=40] (10.72, 0.725);

\draw[orange, dashed, very thick, -{Stealth[length=4mm, width=3.2mm]}, shorten <=0.65cm, shorten >=0.65cm] (5.72, 0.725) to[bend left=25] (11.10, -6.55);
\draw[orange, dashed, very thick, -{Stealth[length=4mm, width=3.2mm]}, shorten <=0.65cm, shorten >=0.65cm] (8.22, 0.725) to[bend left=25] (13.10, -6.55);
\draw[orange, dashed, very thick, -{Stealth[length=4mm, width=3.2mm]}, shorten <=0.65cm, shorten >=0.65cm] (10.72, 0.725) to[bend left=25] (15.10, -6.55);
\draw[orange, dashed, very thick, -{Stealth[length=4mm, width=3.2mm]}] (14.86, 2.1) .. controls (17.70, 2.6) and (17.70, -3.35) .. (14.32, -3.35);

\node[font=\Huge, fill=white, inner sep=1pt] at (10.10, -2.3) {$\mathcal{T}_V$};
\node[tnode] (tv1) at (5.10, -0.975) {${\color{red!70!black}v_{1}}$};
\node[tnode] (tv2) at (6.35, -0.975) {${\color{red!70!black}v_{2}}$};
\node[tnode] (tv3) at (7.60, -0.975) {${\color{red!70!black}v_{3}}$};
\node[tnode] (tv4) at (8.85, -0.975) {${\color{red!70!black}v_{4}}$};
\node[tnode] (tv5) at (10.10, -0.975) {${\color{red!70!black}v_{5}}$};
\node[tnode] (tv6) at (11.35, -0.975) {${\color{red!70!black}v_{6}}$};
\node[tnode] (tv7) at (12.60, -0.975) {${\color{red!70!black}v_{7}}$};
\node[tnode] (tv8) at (13.85, -0.975) {${\color{red!70!black}v_{8}}$};
\node[tnode] (tv9) at (15.10, -0.975) {${\color{red!70!black}v_{9}}$};

\node[font=\Huge, fill=white, inner sep=1pt] at (10.10, -7.85) {$\mathcal{T}_E$};
\node[tnode] (te1) at (5.10, -6.55) {${\color{teal}e_{1}}$};
\node[tnode] (te2) at (7.10, -6.55) {${\color{teal}e_{2}}$};
\node[tnode] (te3) at (9.10, -6.55) {${\color{teal}e_{3}}$};
\node[tnode] (te4) at (11.10, -6.55) {${\color{teal}e_{4}}$};
\node[tnode] (te5) at (13.10, -6.55) {${\color{teal}e_{5}}$};
\node[tnode] (te6) at (15.10, -6.55) {${\color{teal}e_{6}}$};

\node[tnode] (tv12)  at (5.72, 0.725) {${\color{red!70!black}v_{12}}$};
\node[tnode] (tv34)  at (8.22, 0.725) {${\color{red!70!black}v_{34}}$};
\node[tnode] (tv56)  at (10.72, 0.725) {${\color{red!70!black}v_{56}}$};
\node[tnode] (tv78)  at (13.22, 0.725) {${\color{red!70!black}v_{78}}$};
\node[tnode] (tv789) at (14.16, 2.1) {${\color{red!70!black}v_{789}}$};
\draw[thick] (tv12) -- (tv1);  \draw[thick] (tv12) -- (tv2);
\draw[thick] (tv34) -- (tv3);  \draw[thick] (tv34) -- (tv4);
\draw[thick] (tv56) -- (tv5);  \draw[thick] (tv56) -- (tv6);
\draw[thick] (tv78) -- (tv7);  \draw[thick] (tv78) -- (tv8);
\draw[thick] (tv789) -- (tv78); \draw[thick] (tv789) -- (tv9);

\node[tnode] (te45)  at (12.10, -4.95) {${\color{teal}e_{45}}$};
\node[tnode] (te456) at (13.60, -3.35) {${\color{teal}e_{456}}$};
\draw[thick] (te45) -- (te4);   \draw[thick] (te45) -- (te5);
\draw[thick] (te456) -- (te45); \draw[thick] (te456) -- (te6);

\end{scope}
\path (-4.8, 3.8) rectangle (15.75, -7.3);
\end{tikzpicture}%
}
\end{minipage}%
\caption{An example of the peeling procedure. \textbf{Step 1}: Twin vertices are combined. $v_1$ and $v_2$ get combined into $v_{12}$, $v_{3}$ and $v_{4}$ into $v_{34}$, $v_{5}$ and $v_{6}$ into $v_{56}$ and $v_{7}$ and $v_8$ into $v_{78}$, which is combined with $v_9$ to form $v_{789}$. The new nodes are added to the forest. \textbf{Step 2}: We peel the hyperedges $e_1,e_2,e_3$ (the dashed arrows in the forest represent the sets $\PTR$). \textbf{Step 3}: We peel the vertices $v_{12},v_{34},v_{56}$. \textbf{Step 4}: We combine the duplicated hyperedges $e_4,e_5$ into $e_45$, which combines with $e_6$ into $e_{456}$, both are added to $\cT_E$.\linebreak \textbf{Step 5}: We peel the last vertex, $v_{789}$.}
\label{fig:example_peeling}
\end{figure}
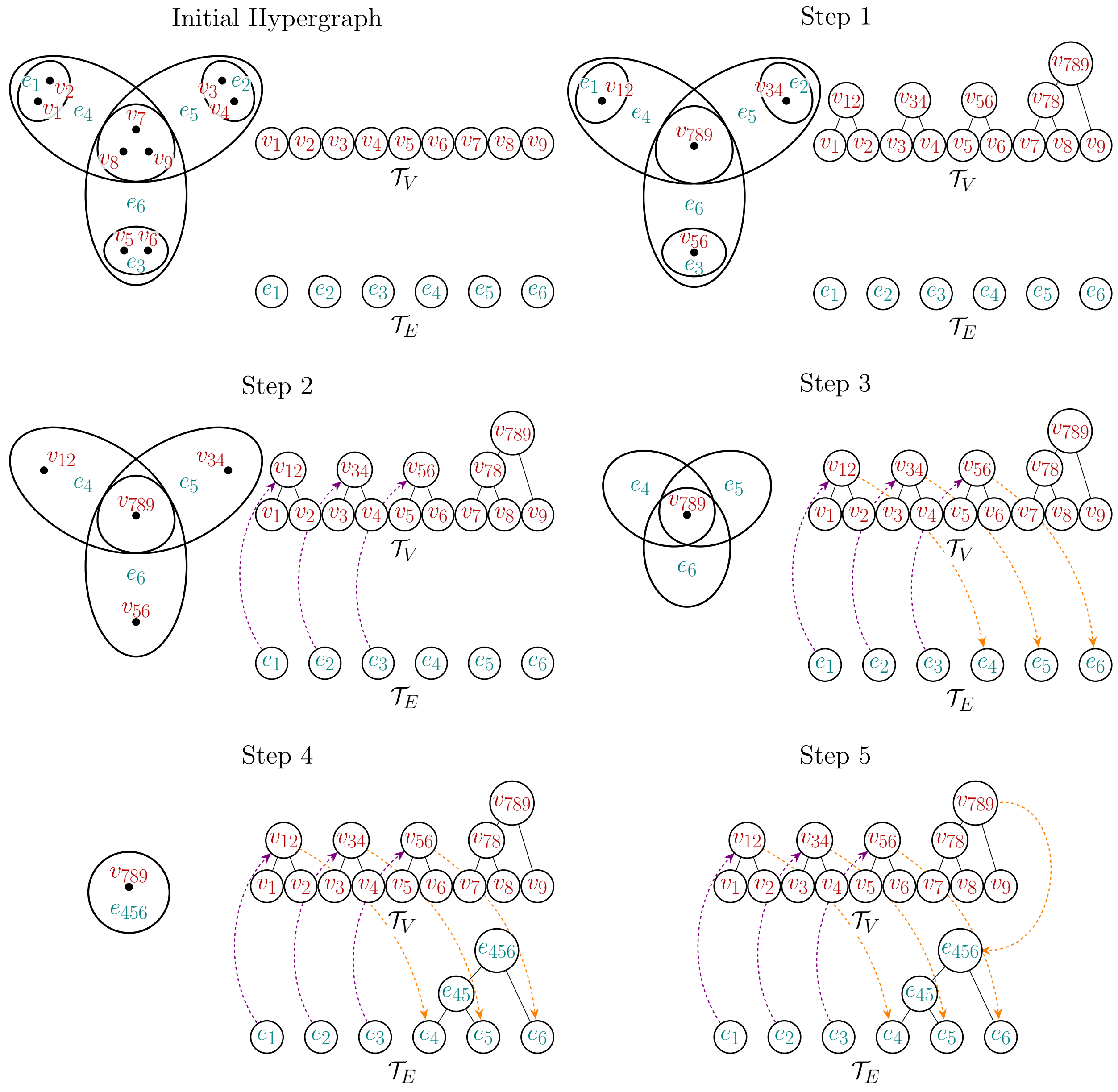

\noindent The pseudocode is given in Algorithm~\ref{alg:representation}. \Fig{example_peeling} gives an example of the execution of the peeling process.
\bigskip

\begin{algorithm}[H]
\caption{\textsc{IntervalRepresentation}}
\label{alg:representation}
\KwIn{A $(\vdegen,\edegen)$-bidegenerate hypergraph $H=(V,E)$ and $\vdegen$ or $\edegen$}
\KwOut{A $(\vdegen,\edegen)$-interval representation $(\bV,\bE,\IV,\IE)$ of $H$}
save $V_0=V$ and $E_0=E$\;\label{line:peel:init}
initialize edgeless forest $\cT_V$ with isolated nodes $\{X_v : v \in V\}$\;
initialize edgeless forest $\cT_E$ with isolated nodes $\{X_e : e \in E\}$\;\label{line:peel:init2}
\While{$H \ne (\emptyset,\emptyset)$}
{\label{line:alg1_ext_loop}
  \uIf{$H$ contains twin vertices $u,v$ (or twin edges $e,f$)\label{line:peel:ifTwin}}{
    update $H$ by merging $u,v$ into a new vertex $uv$ (or $e,f$ into a new edge $ef$)\;
    update $\cT_V$ (or $\cT_E$) by creating a node $X_{uv}$ with $X_u,X_v$ as children (or $X_{ef}$ with $X_e,X_f$ as children)\; \label{line:alg1_mergev}
  }\label{line:alg1_detect_twin}
  \uElseIf{$H$ contains $v$ of degree at most $\vdegen$ (or $e$ of size at most $\edegen$)\label{line:peel:ifSimple}}{
    store $\{X_f : f \in E(v)\}$ as $\PTR(X_v)$ (or $\{X_u : u \in V(e)\}$ as $\PTR(X_e)$)\;
    update $H$ by deleting $v$ (or deleting $e$)\;
  }
}
run a postorder traversal on $\cT_V$ (and $\cT_E$) to compute:\label{line:peel:traversal}\\
\Indp 
    the total order $\bV$ (and $\bE$) given by the visit to the leaves of $\cT_V$ (and of $\cT_E$)\;
    for each node $X$, its root $r(X)$ in the forest\;
    for each node $X$, the interval $J(X) \subseteq \bV$ (and $I(X) \subseteq \bE$) given by its leaves\;
\Indm
let $\IV(v)\gets\{I(X):X\in \PTR(r(X_v))\}$ for each $v \in V_0$\;\label{line:peeling:IX}
let $\IE(e)\gets\{J(X): X\in \PTR(r(X_e))\}$ for each $e \in E_0$\;\label{line:peeling:JX}
\textbf{return} $(\bV,\bE,\IV,\IE)$\;\label{line:peel:return}
\end{algorithm}

\subsubsection{Correctness}\label{sub:peeling_correct}
\begin{lemma}
    \Cref{alg:representation} returns a valid $(\vdegen,\edegen)$-interval representation of $H$.
\end{lemma}
\begin{proof}
    We check the output of \Cref{alg:representation} against Definition~\ref{def:interval_representation}.
    First, observe that the algorithm always terminates, since each iteration of the loop of line~\ref{line:alg1_ext_loop} decreases the total number of vertices and edges by exactly one, either by deleting an element or by merging two twins into one.
    Indeed, at every iteration we are left with a subhypergraph of $H$, and either that subhypergraph contains twins, or it contains a vertex of degree at most $\vdegen$ or an edge of size at most $\edegen$.
    Thus, the construction of both $\cT_V$ and $\cT_E$ takes place as described above.

    Clearly, $\bV$ and $\bE$ are total orderings over respectively $V_0$ and $E_0$.    
    Now, observe that $|\PTR(X)| \le \vdegen$ for every node $X$ in $\cT_V$ and $|\PTR(X)| \le \edegen$ for every node $X$ in $\cT_E$.
    This holds since $\PTR(X)$ is created inside the block of line~\ref{line:peel:ifSimple}, which implies $|E(v)| \le \vdegen$ (or $|V(e)| \le \edegen$).
    Moreover, by construction $I(X)$ and $J(X)$ are indeed intervals.
    We conclude that $\IV(v)$ is a $k_v$-interval subset for all $v \in V_0$, and $\IE(e)$ is a $k_e$-interval subset for all $e \in E_0$, where $k_v \le \vdegen$ and $k_e \le \edegen$.

    \medskip\noindent\underline{Soundness.}
    For every $v \in V_0$ denote by $v^*$ the vertex such that $r(X_v)=X_{v^*}$. Note that $v^*$ can be $v$ itself, or a new vertex created by the algorithm via a merge operation.
    Moreover, let $t(v^*)$ be the time (in the algorithm's execution) at which $v^*$ is deleted; this is well defined, since $X_{v^*}$ is a root if and only if $v^*$ is deleted without being further merged.
    For every $e \in E_0$ define $e^*$ and $t(e^*)$ in the same way.
    Now, for every incident $v \in V_0$ and $e \in E_0$, either $t(v^*) < t(e^*)$ or $t(v^*) > t(e^*)$.
    Suppose $t(v^*) < t(e^*)$ and let $\ell(X)$ be the subtree of $\cT_E$ rooted at $X$.
    The crucial observation is:
    \begin{align}
        \{e \in E_0(v) : t(e^*) > t(v^*)\} = \biguplus_{f \in E(v^*)} \{e : X_e \in \ell(X_f)\} \,.\label{eq:leaves_and_ancestors}
    \end{align}
    To see why,  first note that $\{e : X_e \in \ell(X_f)\}$ is precisely the set of all edges that led to $f$ through some sequence of merge operations; this also implies that the right-hand set is a disjoint union.
    Thus, if $e$ is in the left-hand set, then take $X_f$ to be the highest ancestor of $X_e$ that is created before $t(v^*)$; this implies $v^*$ is incident with $f$ at time $t(v^*)$, that is, $f \in E(v^*)$, and therefore $e$ is in the right-hand set.
    If instead $e$ is in the right-hand set, then observe that if $e \notin E_0(v)$ then either $X_e \notin \ell(X_f)$ or $f \notin E(v^*)$, and the same is implied by $t(e^*)<t(v^*)$.
    Finally, observe that the intervals $\{e : X_e \in \ell(X_f)\}$ for $f \in E(v^*)$ are precisely the intervals $I(X)$ computed by the algorithm at line~\ref{line:peeling:IX} for $X_{v^*}=r(X_v)$.
    Thus, for every $v \in V_0$, if $\IV(v)=I_1(v),\ldots,I_{k_v}(v)$, then
    \begin{align}
        \bigcup_{i=1}^{k_v}I_i(v) = \{e \in E_0(v) : t(e^*) > t(v^*) \}\,.
    \end{align}
    The case $t(v^*) > t(e^*)$ is symmetric, and it yields for every $e \in E_0$:
    \begin{align}
        \bigcup_{j=1}^{k_e}J_j(e) = \{v \in V_0(e) : t(v^*) > t(e^*) \}\,.
    \end{align}
    This proves soundness.

    \medskip\noindent\underline{Niceness.}
    Suppose $e,f \in I_i(v)$ for some $v \in V$ and some $i \in [k_v]$.
    By the construction of the algorithm, this implies $X_e,X_f$ are in the same tree in the forest $\cT_E$.
    This implies $r(X_e)=r(X_f)$, and therefore the algorithm sets $\IE(e)=\IE(f)$, proving niceness. The case of vertices is symmetric.
\end{proof}

\subsubsection{Running time}\label{sub:peeling_time}
\begin{lemma}\label{thm:repr_det_complexity}
    \Cref{alg:representation} can be run in time $O(\size{H} \ln^{1.5}\size{H})$.
\end{lemma}
\begin{proof}
    First, we describe the data structures employed.
    Let $H_0=(V_0,E_0)$ be the input hypergraph.
    We will make sure that, along the algorithm, all vertices are elements of $\{1,\ldots,2|V_0|\}$ and all edges are elements of $\{1,\ldots,2|E_0|\}$.

    For our algorithm to work, we need to efficiently detect when two distinct vertices $u,v$ have identical type, $E(u)=E(v)$, and the same for edges.
    This boils down to maintaining a distinguished id $\eta(X)$ for every set $X$ in a set family under a sequence of updates of the kind ``insert element $x$ in set $i$''.
    As shown in~\cite{TAKWAHLAM199885}, this can be done in update time $O(\log^{1.5} s)$, where $s$ is the sum of the sets' sizes.
    In our case, $s=\size{H}$; thus, upon (say) deleting a vertex from a hyperedge $V(e)$, we can update $\eta(V(e))$ in time $O(\log^{1.5} \size{H})$.
    
    We now describe the data structures for vertices; the same holds for edges. As there is no risk of ambiguity, the data structures have the same name for both.
    Every set, such as $E(v)$, is assumed to be stored in a structure (e.g., a balanced search tree) that supports insertion, deletion, and update in logarithmic time, and querying for the size in constant time.
    We maintain:
    \begin{itemize}
        \item 
        An array \type, indexed by $\{1,\ldots,2|V_0|\}$, such that \type$[v]$ holds $E(v)=\{e \in E: v \in V(e)\}$.
        If $v$ is not currently a vertex of $H$, then \type$[v]=\nil$.
        \item The forest $\cT_V$ using pointers: each node $X=X_v$ stores $v$ as well as the pointers to its children, if any. The same for the forest $\cT_E$.
        \item An array $\node$, indexed by $\{1,\ldots,2|V_0|\}$, such that $\node[v]$ points to $X_v$ if $X_v$ exists in $\cT_V$ and is $\nil$ otherwise.
        \item An array, \hash, such that \hash$[v]=\eta(E(v))$.
        \item A dictionary, \hash$^{-1}$, such that $\hash^{-1}(h)$ holds $V^{-1}(h) := \{v : \eta(E(v))=h\}$.
        \item A set structure, \twins, that holds the set  $\{h : |V^{-1}(h)| \ge 2\}$.
    \end{itemize}
    Finally, we maintain two min-heaps, $\degHeap$ and $\sizeHeap$, that track respectively the smallest degree and the smallest edge size in the current hypergraph.

    \medskip\noindent\uline{Setting up the data structures (lines~\ref{line:peel:init} to~\ref{line:peel:init2}).} Constructing the array \type\ and the dictionaries \type$[v]$ for all $v \in V_0$ takes total time $O(\size{H_0}\ln\size{H_0})$.
    Constructing $\cT_V$ and $\cT_E$ takes time $O(|V_0|+|E_0|)=O(\size{H_0})$, and the same for \node.
    To construct the array \hash\ we compute $\eta(E(v))$ for every $v \in V_0$, which by our assumptions takes total time $O(\size{H_0}\ln^{1.5}(\size{H_0}))$.
    To construct \hash$^{-1}$ we sort and aggregate \hash\ by value, which takes total time $O(|V_0| \ln |V_0|)$.
    Note that this also yields \twins\ for free.
    Constructing $\degHeap$ takes again total time $O(\size{H_0}\ln\size{H_0})$.
    The same applies to the data structures for edges.
    The total time of the setup is therefore $O(\size{H_0}\ln^{1.5}\size{H_0})$.

    \medskip\noindent\uline{Performing a merge (block of line~\ref{line:peel:ifTwin}).} 
    With the data structures set up, we proceed as follows.
    First, observe that by checking whether \twins\ is empty, and by querying \degHeap\ and \sizeHeap, we can perform in time $O(1)$ the tests of lines~\ref{line:peel:ifTwin} and~\ref{line:peel:ifSimple}.
    Suppose then we enter the block of line~\ref{line:peel:ifTwin} as \twins\ contains a hash value $h$; we look $h$ up in \hash$^{-1}$. Suppose we get a set of vertices $V^{-1}(h)$; the case $E^{-1}(h)$ is symmetric.
    Pick any two vertices $u,v \in V^{-1}(h)$.
    Let $z$ be the smallest integer not yet assigned to any vertex along the process.
    (Note that this is just $|V_0|$ plus the number of vertex merges so far, including the current one, so $z$ can be kept as a global counter).
    This ensures $z \in \{1,\ldots,2|V_0|\}$, since the algorithm performs at most $|V_0|$ merges.
    We update all our data structures by replacing $u,v$ with $z$, as follows.
    First, for every $e \in E(v)=E(u)$:
    \begin{itemize}
        \item We update \type$[e]$  by deleting $u$ and $v$ and inserting $z$.
        \item 
        We save $h=\hash[e]$, and we update $\hash[e]$ to its new value $h'=\eta(\type[e])$.
        \item Using $h$ and $h'$ we update \hash$^{-1}$, by removing $e$ from the dictionary \hash$^{-1}(h)$ and adding it to \hash$^{-1}(h')$.
        This entails deleting \hash$^{-1}(h)$ if empty and/or creating \hash$^{-1}(h')$ if not already existing. 
        \item We update \twins\ by inserting $h'$ if $|\hash^{-1}(h')|\ge 2$, and removing $h$ if $|\hash^{-1}(h)|\le 1$.   
        \item We update \sizeHeap\ by decrementing the size of $e$.
    \end{itemize}
    We also update \degHeap\ by replacing $u,v$ with $z$.
    Note that each one of these updates takes time at most $O(\ln^{1.5} \size{H_0})$.
    Thus, the total time spent so far for a merge is $O(\deg(z) \ln^{1.5} \size{H_0})$, where $\deg(z)=|E(z)|$ is the current degree of $z$.
    Finally, from $\arrV$ we retrieve the pointers to $X_u$ and $X_v$, and we create a new node $X_z$ in $\cT_V$ that stores $z$ and the pointers to $X_u$ and $X_v$. We set $\node[z] = X_z$ and $\type[z] = \type[u]$,
    and then set $\type[u] = \type[v] = \nil$.
    This takes just time $O(1)$. We also set \hash$[z]$ = \hash$[u]$, replace $u$ and $v$ with $z$ in \hash$^{-1}($\hash$[z])$ and update {\twins} accordingly. This can be done in $O(\ln \size {H_0})$ time.

    We now show that the total time spent over merging vertices is in $O(\size{H_0}\ln^{1.5}\size{H_0})$; the same holds for edges.
    To this end, recall that by construction each merge corresponds to an internal node $X_z$ of $\cT_V$; as shown above, the time spent for that merge is $O(\deg(z) \ln^{1.5} \size{H_0})$, where $\deg(z)$ is the degree of $z$ in $H$ at merge time.
    Now, for each such $X_z$ choose a distinct leaf node $X_v$ in $\cT_V$ that is a descendant of $X_z$; this is possible as $\cT_V$ is by construction a binary tree.
    Note that $\deg(z) \le \deg_0(v)$, where $\deg_0(v)$ is the degree of $v$ in $H_0$.
    The total cost of vertex merges is thus bounded by:
\begin{align}
\sum_{X_z \in \cT_V \setminus L(\cT_V)} \!\!\!\!\!\!O\big(\deg(z) \ln^{1.5}\size{H}\big)
  &\le \sum_{X_v \in L(\cT_V)} \!\!\!\!O(\deg_0(v) \ln^{1.5}\size{H_0})  
   = O(\size{H_0} \ln^{1.5}\size{H_0})\,.
   \label{eq:sum_cost_merge}
\end{align}

    \medskip\noindent\uline{Performing a deletion (block of line~\ref{line:peel:ifSimple}).}
    Suppose now we enter the block of line~\ref{line:peel:ifSimple} having chosen $v$ with $\deg(v) \le \vdegen$ (the case of $e$ is symmetric). 
    We read $E(v)$ from $\type[v]$, and for every $f \in E(v)$ we add $\node[f]$ to a list.
    Finally, we write the pointer to this list in the entry $\PTR[v]$ of an array $\PTR$ indexed again by $\{1,\ldots,2|V_0|\}$.
    We then update $H$ using the same operations that delete $v$ during a merge.
    First, we read $E(v)$ from $\type[v]$, and for every $e \in E(v)$:
    \begin{itemize}
        \item We update $\type[e]$ by deleting $v$.
        \item We update \sizeHeap\ by decreasing the size of $e$ by one. 
        \item We save $h=\hash[e]$, and we then update $\hash[e]$ to its new value $h'=\eta(\type[e])$.
        \item We remove $e$ from $\hash^{-1}(h)$ and insert it into $\hash^{-1}(h')$.
        As above, we might need to delete or create the corresponding entries. 
        \item We insert $h'$ in \twins\ if $|\hash^{-1}(h')|\geq 2$. (Note that certainly $h \notin \twins$, since $H$ is twin-free if we are doing a deletion, hence no further checks on \twins\ are needed). 
    \end{itemize}
    Every one of these updates takes time at most $O(\ln^{1.5}\size{H})$, for a total of  $O(\deg(v)\ln^{1.5}\size{H})$. 
    Finally, we remove v from $\hash^{-1}(\hash[v])$. We then set $\type[v]=\nil$, and remove $v$ from \degHeap, which cost $O(\ln\size{H})$.

    To bound the total time spent on vertex deletions, we proceed like for merges.
    If a vertex $z$ is deleted, then $X_z$ is the root of a tree in $\cT_V$, and therefore it can be mapped to a distinguished leaf $X_v$ with $v \in V_0$. Moreover, $\deg(z) \le \deg_0(v)$. Adapting~\Cref{eq:sum_cost_merge} yields a total time bound of $O(\size{H}\ln^{1.5}\size{H})$.

    \medskip\noindent\uline{Wrapping up (lines~\ref{line:peel:traversal} to~\ref{line:peel:return}).}
    Performing the postorder traversal of $\cT_V$ and $\cT_E$ and writing $\bV$ and $\bE$ takes time $O(\size{H_0})$, as this is the total size of the forests.
    Note that $r(X)$ and $J(X)$, with $J(X)$ represented by its endpoints, can be computed for free during the traversal.
    Line~\ref{line:peeling:IX} requires, for each $v \in V_0$, to read $X_z=r(v)$, read the nodes in $\PTR[z]$, and for each such node $X$ add $I(X)$ to the list $\IV(v)$.
    This takes time $O(|\PTR[r(v)]|)$ for each $v$.
    By the same charging argument used above, we conclude that line~\ref{line:peeling:IX} takes time $O(\size{H})$.
    The same holds for line~\ref{line:peeling:JX}.

    Summarizing, Algorithm~\ref{alg:representation} can be run in time $O(\size{H}\ln^{1.5}(\size{H}))$, as claimed.
\end{proof}

\section{Answering \ESat\ queries} \label{sec:impl-eval}

In this section, we prove the more general \Thm{main-eval}, showing that we can efficiently preprocess and answer \ESat\ queries on hypergraphs of small bidegeneracy. With the techniques and setup in place, it is easier to explain how to solve \ESat, for fixed length queries (\Thm{main-eval-bounded}). This is done in \Cref{sec:bounded_query}.

The results on \VSat\ follow by taking the dual of $H$.
We actually solve the simpler $\EdgePartialMatch$ primitive; it is rather straightforward to transform every \ESat\ query into a collection of disjoint \EdgePartialMatch\ queries.

Recall that an \EdgePartialMatch\ query takes in a ``positive'' vertex set $P \subseteq V$
and a ``negative'' vertex set $N \subseteq V$. A hyperedge $e$ satisfies this query
if $P \subseteq V(e)$ and $N \cap V(e) = \emptyset$. We will actually solve a stronger
weighted version of this query. Let us assume that there is a weight function $w:E \to \NN$.
In the counting version of this query, we need to output the total weight
of satisfying hyperedges. In the sampling version, we need to sample a satisfying hyperedge
with probability proportional to its weight. And, in the listing version, we need to enumerate all satisfying hyperedges with small delay.

The main result we prove is:
\begin{theorem}\label{thm:PartialMatch}
    There exists an algorithm for $\EdgePartialMatch$ with the following guarantees. Given in input a $(\vdegen,\edegen)$-bidegenerate hypergraph $H = (V,E)$ and an edge weighting $w:E \to \NN$,
    \medskip
    \begin{asparaitem}
        \item First, the algorithm constructs a data structure in time $O\left(2^{O(\edegen)}\size{H} \log^{2\edegen+1}\size{H}\right)$.
        \item Afterwards, the algorithm can answer every query in time $O(\vdegen \,2^{O(\edegen)}\edegen^{O(t)}\log^{2\edegen +1}\size H)$, where $t=|P| + |N|$ is the query size. This holds for counting, sampling, and $O(1)$-delay listing.
    \end{asparaitem}

%
\end{theorem}

\Cref{sec:convert-interval} shows how \EdgePartialMatch\ queries can be expressed as a set of constraints that involve the interval representation of $H$.
\Cref{sec:eval-alg} exploits this connection to describe an algorithm that performs a reduction to orthogonal range searching.
\Cref{sec:proof-partial} wraps up the proof of \Cref{thm:PartialMatch}.
Finally, \Cref{sub:Eval} shows how this yields the results on \Eval\ of \Cref{thm:main-eval}.

\subsection{Converting to a question on intervals} \label{sec:convert-interval}

\Thm{interval} tells us that we can compute a $(\vdegen,\edegen)$-representation $(\bV,\bE,\IV,\IE)$ of $H$ in near-linear time.
Our first step is to convert the question of which edges satisfy an \EdgePartialMatch\ query to a question on the interval representation.
To lighten the notation, we assume that the vertices and the edges of $H$ are relabeled so as to match their rank in $\bV$ and $\bE$.
Thus, $\bV=(v_1,\ldots,v_n)=(1,\ldots,n)$ and $\bE=(e_1,\ldots,e_m)=(1,\ldots,m)$.
Such a relabeling can be obviously computed in linear time.

We introduce some notation about the intervals $\IV$ and $\IE$. Recall from \Def{interval_representation} that every $\IE(e)$ is a set of at most $\edegen$ intervals $J_1(e), \ldots, J_{k_e}(e)$ in $\bV$. For convenience, we add dummy intervals $J_0(e) = (-\infty,0]$ and $J_{k_e+1}(e) = [n+1, \infty)$ that contain no vertices in $\bV$.
For every subinterval $J \subseteq \bV$, denote its endpoints by $\intStart(J)$ and $\intEnd(J)$, so that $J = [\intStart(J), \ldots, \intEnd(J)]$. Without loss of generality, we assume the intervals of $\IE(e)$ to be indexed in increasing order, so that $\intEnd(J_i(e)) < \intStart(J_{i+1}(e))$ for each $i \in [k_e -1]$.

For each $v \in V$, $\IV(v)$ is a collection of $k_v \le \vdegen$ intervals in $\bE$. Let $\IV^\cup(v)$ be the union of these intervals.
For any two subsets $X,Y \subseteq V$, define the union of intervals $\I^\cup(X,Y)$ as follows.
\begin{align}
    \I^\cup(X,Y):=
    \left(\bigcap_{v\in X}\IV^\cup(v)\right)\setminus \left(\bigcup_{u\in Y}\IV^\cup(u)\right) \label{eq:Icup(S)}
\end{align}
The central claim for the reduction is the following:
\begin{claim} \label{clm:eval-interval} A hyperedge $e\in E$ is a match for $\EdgePartialMatch(P,N)$ if and only if there exists a subset $S \subseteq P$ (possibly empty) such that the following claims hold:
    \begin{itemize}
        \item $e \in \I^\cup(P \setminus S, N)$.
        \item There is a mapping $\sigma:S \to [k_e]$ such that, for every $s \in S$,
        $
            s \in [\intStart(J_{\sigma(s)}(e)), \intEnd(J_{\sigma(s)}(e))]
        $

        \item There is a mapping $\tau:N \to \{0\} \cup [k_e]$ such that, for every $t \in N$,
        $
            t \in (\intEnd(J_{\tau(t)}(e)), \intStart(J_{\tau(t)+1}(e)))
        $
    \end{itemize}
    Moreover, if $e$ is a match, then the choice of $S, \sigma, \tau$ is unique.
\end{claim}

\begin{proof} By \Def{interval_representation}, $V(e) = \bigcup_{j=1}^{k_e} J_j(e) \uplus \IV^{-1}(e)$. Take $S$ to be:
\begin{align}
    S = P \cap \bigcup_{j=1}^{k_e} J_j(e)\,.
\end{align}
    Now, suppose $e$ is a match. Then $P \setminus S \subseteq \IV^{-1}(e)$ and $N \cap \IV^{-1}(e) = \emptyset$. Hence, $e \in \IV^\cup(v)$ for all $v \in P \setminus S$, and $v \notin \IV^{-1}(e)$ for all $v \in N$. Hence,
    $e \in \left(\bigcap_{v\in P\setminus S}\IV^\cup(v)\right)\setminus \left(\bigcup_{u\in N}\IV^\cup(u)\right) = \I^\cup(P \setminus S, N)$.
    
    Each $s \in S$ lies in some $J_j(e)$. By \Def{interval_representation}, the intervals $J_j(e)$ are pairwise disjoint, thus $j$ is unique; let then $\sigma(s)=j$.
    Each $t \in N$, instead, does not lie in any $J_j(e)$.
    This means that $t$ lies in the ``spaces'' between these intervals. Formally, there exists a unique $j$ such that $t$ lies strictly between $\intEnd(J_j(e))$ and $\intStart(J_{j+1}(e))$; let then $\tau(t)=j$. That completes the proof that, if $e$ is a match, then the claims of the statement hold.

    Conversely, suppose the three conditions hold. By the first condition, we have $P \setminus S \subseteq \IV^{-1}(e)$ and $N \cap \IV^{-1}(e) = \emptyset$. By the second, every $s \in S$ belongs to some $J_j(e)$, while the third ensures that no $t \in N$ belongs to any $J_j(e)$. The soundness property of the interval representation lets us conclude that $P \subseteq V(e)$ and $N \cap V(e) = \emptyset$, so $e$ is a match.
    Finally, the uniqueness of $S$ follows from soundness, while the uniqueness of $\sigma$ and $\tau$ follows from the disjointness and ordering of the intervals.
\end{proof}

Thus, the problem of determining whether $e$ is a match for $\EdgePartialMatch(P,N)$ is equivalent
to determining the conditions of \Clm{eval-interval}. 

\subsection{The algorithm} \label{sec:eval-alg}

We now prove that the interval problem defined by \Clm{eval-interval} can be solved
by a reduction to  Orthogonal Range Searching.
Given the $(\vdegen,\edegen)$-representation $(\bV,\bE,\IV,\IE)$ of $H$ and the edge weighting $w$, 
the algorithm embeds the edges $e_1,\ldots,e_m$ into a $(2\edegen+2)$-dimensional space based on the interval
representation. The preprocessing step constructs a data structure to answer weighted orthogonal range queries on those points.
To answer a given query $(P,N)$, the algorithm computes a certain set of boxes in the same $(2\edegen+2)$-dimensional space 
and performs an orthogonal range search query for each one of them.
It aggregates all answers, either by summing them (in the counting version) or concatenating them (in the reporting version) or by sampling from them (in the sampling version).

\subsubsection{Preprocessing} \label{sec:eval-prep}

In \Sec{ORQ}, we recall standard results on Orthogonal Range Searching. While there are many such results, we could not find a single one that supports counting, sampling, and listing under weightings; so we provide our own version, \Thm{ORS}, together with a complete proof. Basically, it says one can preprocess a weighted set of $n$ points in $\RR^d$ in $2^{O(d)}n\log^{d-1} n$ time so that, given an axis-aligned box as query, counting/sampling/listing can be done in $2^{O(d)}\log^{d-1} n$ time (for listing, this time is paid only once, and afterwards one can have delay $O(1)$).

We first describe how to convert each hyperedge $e$ into a vector in $\RR^{2\edegen+2}$.
Consider the at most $\edegen$ intervals $J_1(e), J_2(e), \ldots, J_{k_e}(e)$.
The vector associated to $e$ is:
\begin{align}
\ve_e := \big(e,\intStart(J_1(e)),\intEnd(J_1(e)),\dots,\intStart(J_{k_e}(e)),\intEnd(J_{k_e}(e)),n+1,\underbrace{0,\dots,0}_{2(\edegen-k_e)}\big). \label{eq:vec_e}
\end{align}
The extra $n+1$ is just a dummy padding to make the procedure simpler. Technically, we could remove it (and reduce the dimension),
at the cost of extra handling of a corner case.
%
%
Using as input the set of vectors $\{\ve_e : e \in \bE\}$, we construct an orthogonal range query data structure.
The pseudocode is given in Algorithm~\ref{alg:PartialMatch_Preprocess}.

\begin{algorithm}
    \label{alg:PartialMatch_Preprocess}
    \caption{\EdgePartialMatchPre$(\bV,\bE,\IV,\IE,w)$}
    compute the vector $\ve_e$ for every $e \in \bE$\;
    run \ORQPre\ on the set $\{\ve_e : e \in \bE\}$ with weighting $w$\;
\end{algorithm}

\subsubsection{Converting queries to boxes} \label{sec:eval-query}

Each query $(P,N)$ is converted to a collection of axis-parallel boxes,
based on the conditions in \Clm{eval-interval}. We use $t = |P| + |N|$.

Fix a subset $S \subseteq P$. In what follows, we will describe constructions of various intervals for each coordinate that will
be used to construct these axis-parallel boxes. The overall collection of boxes is obtained by varying $S$ and these intervals
of all the coordinates.

\paragraph{The first coordinate.} Recall the definition of $\I^\cup(P \setminus S, N)$ in \Cref{eq:Icup(S)}.
We state an easy claim.

\begin{claim} \label{clm:icups} $\I^\cup(P \setminus S, N)$ is a disjoint union $I_1(S) \uplus \ldots \uplus I_z(S)$ of $z\le\vdegen t + 1$ intervals of $\bE$.
\end{claim}

\begin{proof}
Note that, as $|\IV(v)| \le \vdegen$ for all $v \in V$, then $\bigcap_{v\in P\setminus S}\IV^\cup(v)$ is a disjoint union of at most $|P \setminus S|\vdegen \le |P|\vdegen$ intervals. 
Hence, $\I^\cup(P \setminus S, N)$ is the disjoint union of at most $1+|P|\vdegen + |N|\vdegen = 1+\vdegen t$ intervals.
\end{proof}
Let $\I(P \setminus S, N)=I_1(S), \ldots, I_z(S)$.
Each box constructed from the query $(P,N)$ will have, on its first coordinate, one of the intervals in $\I(P \setminus S, N)$.

\paragraph{The remaining coordinates.} Recalling \Clm{eval-interval}, consider functions $\sigma: S \to [\edegen]$ and $\tau: N \to \{0\} \cup [\edegen]$.
We use a procedure, \Endpoints, to compute a sequence of values $\alpha_j, \beta_j$, for $2 \leq j \leq 2\edegen+2$.
In the procedure, if a set is undefined, the operation involving the $\max$ or $\min$ is a no op.

\begin{algorithm}
    \label{alg:endpoints}
    \caption{\Endpoints$(\sigma,\tau)$}
    \ForEach{$2 \leq j \leq 2\edegen+2$}{
        initialize $\alpha_j = -\infty$ and $\beta_j = \infty$
    }
    set $\alpha_2 = \max\{\tau^{-1}(0)\} + 1/2$\;
    \ForEach{$1 \leq i \leq \edegen$}{
        set $\alpha_{2i+1} = \max\{\sigma^{-1}(i)\}$ and $\beta_{2i} = \min\{\sigma^{-1}(i)\}$\;
        set $\alpha_{2i+2} = \max\{\tau^{-1}(i)\} + 1/2$ and $\beta_{2i+1} = \min\{\tau^{-1}(i)\} - 1/2$
    }
\end{algorithm}

We now state the claim that explicitly converts the matching condition into a geometric condition on boxes.

\begin{claim} \label{clm:matchbox} The hyperedge $e$ is a match for $\EdgePartialMatch(P,N)$ if and only if the following hold. There exist a set $S \subseteq P$, an interval $I \in \I(P \setminus S, N)$, and functions $\sigma: S \to [\edegen]$ and $\tau: N \to \{0\} \cup [\edegen]$ such that, with $\alpha_j, \beta_j$ returned by \Endpoints$(\sigma,\tau)$, we have $\ve_e \in I \times \prod^{2\edegen+2}_{j=2} [\alpha_j, \beta_j]$. Moreover, if $e$ is a match, then the choice of $S, I, \sigma, \tau$ is unique.
\end{claim}

\begin{proof} Suppose $e$ is a match. By \Clm{eval-interval}, there exists $S, \sigma, \tau$ such that
    $e \in \I^\cup(P \setminus S, N)$. (The range of $\sigma, \tau$ are technically $[k_e]$, a subset of $[\edegen]$.)
    Also, for every $s \in S$, $s \in [\intStart(J_{\sigma(s)}(e)), \intEnd(J_{\sigma(s)}(e))]$,
    and for every $t \in N$,  $t \in (\intEnd(J_{\tau(t)}(e)), \intStart(J_{\tau(t)+1}(e)))$. 

    Let us argue that $\ve_e$ is in the desired box, coordinate by coordinate.

    \underline{First coordinate:} Note that $e \in I \in \I^\cup(P \setminus S, N)$. The first coordinate of $\ve_e$ is just $e$,
    so that satisfies the claim.

    \underline{Coordinate $j = 2i$, $1 \leq i \leq k_e$.} The coordinate of $\ve_e$ is $\intStart(J_i(e))$. 
    For each $s \in S$ such that $\sigma(s) = i$, $s \geq \intStart(J_i(e))$. Hence, $\beta_{2i} = \min\{\sigma^{-1}(i)\} \geq \intStart(J_i(e))$.
    For each $t \in N$ such that $\tau(t) = i-1$, $t < \intStart(J_i(e))$ (this also holds for $i=1$, in which case $\tau(t) = 0$).
    Hence, $\max\{\tau^{-1}(i-1)\} < \intStart(J_i(e))$. Since, these are all integers, we can add $1/2$ to left to get $\alpha_{2i} < \intStart(J_i(e))$.
    In summary, the $j$th coordinate of $\ve_e$ lies in $[\alpha_j, \beta_j]$.

    \underline{Coordinate $j = 2i+1$, $1 \leq i \leq k_e$.} The coordinate of $\ve_e$ is $\intEnd(J_i(e))$. 
    For each $s \in S$ such that $\sigma(s) = i$, $s \leq \intEnd(J_i(e))$. Hence, $\alpha_{2i+1} = \max\{\sigma^{-1}(i)\} \leq \intEnd(J_i(e))$.
    For each $t \in N$ such that $\tau(t) = i$, $t > \intEnd(J_i(e))$.
    Hence, $\beta_{2i+1} = \min\{\tau^{-1}(i)\} - 1/2 > \intEnd(J_i(e))$ (using the integrality of the quantities). So, like
    the previous case, the $j$th coordinate of $\ve_e$ lies in $[\alpha_j, \beta_j]$.

    \underline{Coordinate $j \geq 2i+1$, $i > k_e$.} Since the values of $\sigma, \tau$ are at most $k_e$, the values of $\alpha_j, \beta_j$
    maintain their default values of $-\infty, \infty$ in \Endpoints$(\sigma,\tau)$. 

    \underline{Coordinate $j = 2k_e+2$.}
    The coordinate of $\ve_e$ is $n+1$. Since $\sigma \in [k_e]$, we have $\beta_j = \infty$. Moreover, $\alpha_j \leq n+1/2$, since $N \subseteq [n]$. So, the $j$th coordinate of $\ve_e$ lies in $[\alpha_j,\beta_j]$.

    Overall, we conclude that $\ve_e$ is in the desired box. 
    Conversely, suppose $\ve_e$ is in the desired box. The first coordinate implies that we have $e \in \I^\cup(P \setminus S, N)$. For every $s \in S$, the constraints on the corresponding coordinates imply that $s \in J_{\sigma(s)}(e)$. The case for $t \in N$ is analogous. Thus, all the conditions of \Clm{eval-interval} hold, and $e$ is a match.

    If $e$ is a match, the uniqueness holds by the fact that the intervals $I \in \I(P \setminus S, N)$ are disjoint,
    and the uniqueness of $S, \sigma, \tau$ from \Clm{eval-interval}.
\end{proof}

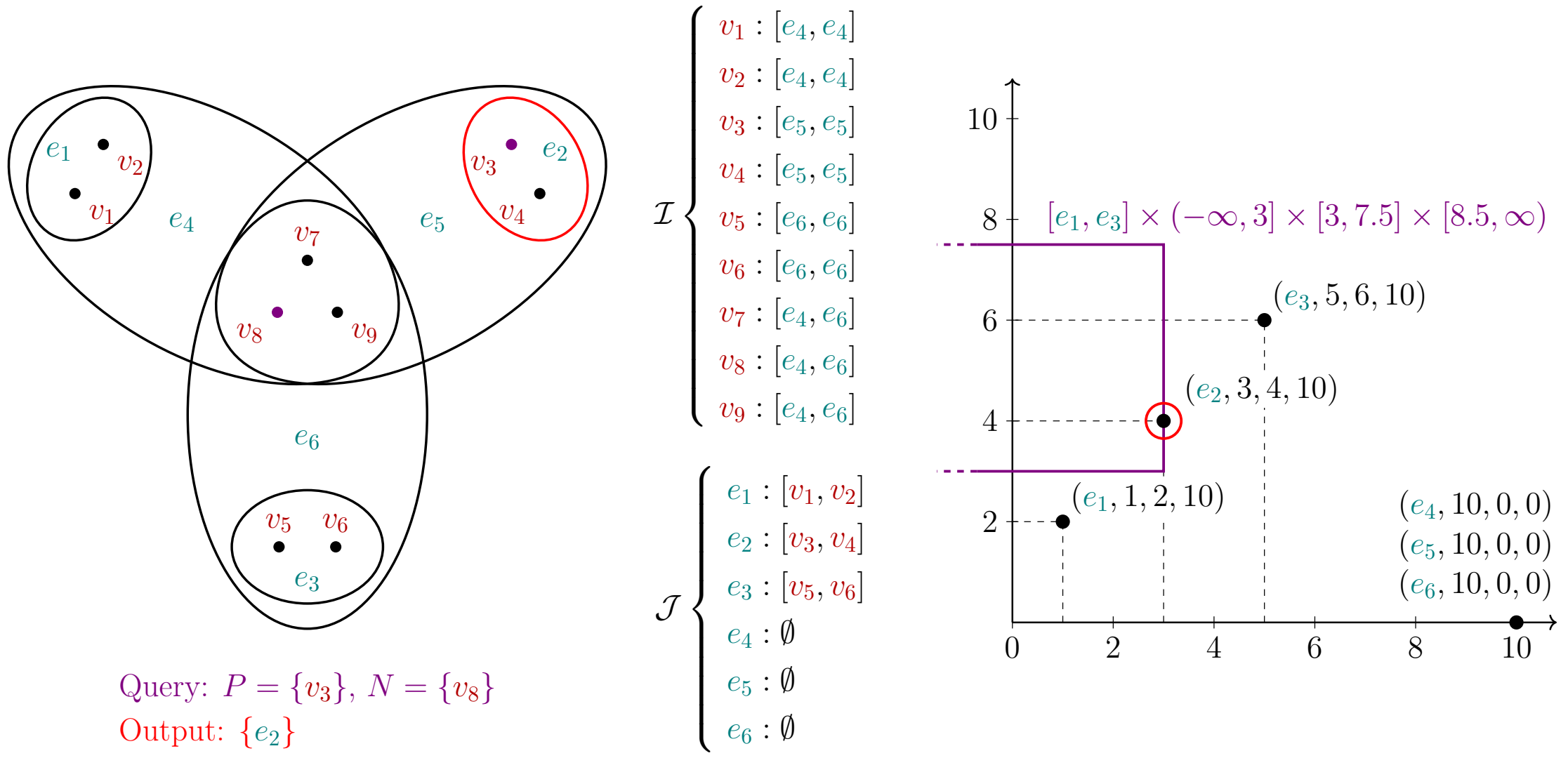
\begin{figure}[ht]
\centering
\resizebox{.9\textwidth}{!}{%
\begin{tikzpicture}[
    vtx/.style={circle, fill=black, inner sep=1.8pt},
    edge/.style={line width=1.1pt},
    vlab/.style={text=red!70!black, font=\Large},
    elab/.style={text=teal, font=\Large}
]

\begin{scope}[yshift=1cm]
\draw[edge] (-1.645, 0.95) ellipse [x radius=3.4, y radius=1.9, rotate=150];
\draw[edge] ( 1.645, 0.95) ellipse [x radius=3.4, y radius=1.9, rotate=30];
\draw[edge] ( 0,    -1.9 ) ellipse [x radius=3.4, y radius=1.9, rotate=270];

\draw[edge] (-3.464, 2) ellipse [x radius=0.9, y radius=1.2, rotate=150];
\draw[edge, red] ( 3.464, 2) ellipse [x radius=0.9, y radius=1.2, rotate=30];
\draw[edge] ( 0,    -4) ellipse [x radius=0.9, y radius=1.2, rotate=270];

\node[vtx, label={[vlab]-30:$v_1$}] at (-3.689, 1.610) {};
\node[vtx, label={[vlab]-30:$v_2$}] at (-3.239, 2.390) {};
\node[vtx, fill=violet, label={[vlab]210:$v_3$}] at ( 3.239, 2.390) {};
\node[vtx, label={[vlab]210:$v_4$}] at ( 3.689, 1.610) {};
\node[vtx, label={[vlab]90:$v_5$}]  at (-0.45, -4) {};
\node[vtx, label={[vlab]90:$v_6$}]  at ( 0.45, -4) {};
\node[vtx, label={[vlab]90:$v_7$}]  at ( 0,      0.55 ) {};
\node[vtx, fill=violet, label={[vlab]210:$v_8$}] at (-0.476, -0.275) {};
\node[vtx, label={[vlab]330:$v_9$}] at ( 0.476, -0.275) {};

\node[elab] at (-3.940,  2.275) {$e_1$};
\node[elab] at ( 3.940,  2.275) {$e_2$};
\node[elab] at ( 0,     -4.55 ) {$e_3$};
\node[elab] at (-1.992,  1.15 ) {$e_4$};
\node[elab] at ( 1.992,  1.15 ) {$e_5$};
\node[elab] at ( 0,     -2.3  ) {$e_6$};

\node[font=\Large, align=left] at (0, -6.6) {{\color{violet}Query: $P = \{{\color{red!70!black}v_{3}}\}$, $N = \{{\color{red!70!black}v_{8}}\}$} \\[2pt] {\color{red}Output: $\{{\color{teal}e_{2}}\}$}};
\end{scope}

\node[anchor=west, font=\Large] (lists) at (5.2, -0.3) {$
\renewcommand{\arraystretch}{1.2}
\begin{array}{l}
\mathcal{I} \left\{
\begin{array}{l}
{\color{red!70!black}v_{1}} : [{\color{teal}e_{4}}, {\color{teal}e_{4}}] \\
{\color{red!70!black}v_{2}} : [{\color{teal}e_{4}}, {\color{teal}e_{4}}] \\
{\color{red!70!black}v_{3}} : [{\color{teal}e_{5}}, {\color{teal}e_{5}}] \\
{\color{red!70!black}v_{4}} : [{\color{teal}e_{5}}, {\color{teal}e_{5}}] \\
{\color{red!70!black}v_{5}} : [{\color{teal}e_{6}}, {\color{teal}e_{6}}] \\
{\color{red!70!black}v_{6}} : [{\color{teal}e_{6}}, {\color{teal}e_{6}}] \\
{\color{red!70!black}v_{7}} : [{\color{teal}e_{4}}, {\color{teal}e_{6}}] \\
{\color{red!70!black}v_{8}} : [{\color{teal}e_{4}}, {\color{teal}e_{6}}] \\
{\color{red!70!black}v_{9}} : [{\color{teal}e_{4}}, {\color{teal}e_{6}}]
\end{array}\right. \\ \\[-6pt]
\mathcal{J} \left\{
\begin{array}{l}
{\color{teal}e_{1}} : [{\color{red!70!black}v_{1}}, {\color{red!70!black}v_{2}}] \\
{\color{teal}e_{2}} : [{\color{red!70!black}v_{3}}, {\color{red!70!black}v_{4}}] \\
{\color{teal}e_{3}} : [{\color{red!70!black}v_{5}}, {\color{red!70!black}v_{6}}] \\
{\color{teal}e_{4}} : \emptyset \\
{\color{teal}e_{5}} : \emptyset \\
{\color{teal}e_{6}} : \emptyset
\end{array}\right.
\end{array}
$};

\begin{scope}[shift={(lists.east)}]
\begin{scope}[xshift=1.8cm, yshift=-3.9cm, scale=0.8]
  \draw[->, thick] (0,0) -- (10.8,0);
  \draw[->, thick] (0,0) -- (0,10.8);
  \foreach \t in {0,2,4,6,8,10} {
    \draw (\t,0.12) -- (\t,-0.12) node[below, font=\Large] {$\t$};
  }
  \foreach \t in {2,4,6,8,10} {
    \draw (0.12,\t) -- (-0.12,\t) node[left, font=\Large] {$\t$};
  }
  \draw[dashed] (1,0) -- (1,2) -- (0,2);
  \draw[dashed] (3,0) -- (3,4) -- (0,4);
  \draw[dashed] (5,0) -- (5,6) -- (0,6);
  \draw[violet, very thick] (-0.7,7.5) -- (3,7.5) -- (3,3) -- (-0.7,3);
  \draw[violet, very thick, dashed] (-0.7,7.5) -- (-1.5,7.5);
  \draw[violet, very thick, dashed] (-0.7,3)   -- (-1.5,3);
  \node[anchor=south east, font=\Large, text=violet] at (10.8,7.55) {$[{\color{teal}e_{1}}, {\color{teal}e_{3}}] \times (-\infty, 3] \times [3, 7.5] \times [8.5, \infty)$};
  \fill (1,2)  circle (4pt);
  \fill (3,4)  circle (4pt);
  \draw[red, very thick] (3,4) circle (0.35);
  \fill (5,6)  circle (4pt);
  \fill (10,0) circle (4pt);
  \node[anchor=south west, font=\Large, fill=white, inner sep=1pt] at (1.1,2.1) {$({\color{teal}e_{1}}, 1, 2, 10)$};
  \node[anchor=south west, font=\Large, fill=white, inner sep=1pt] at (3.35,4.25) {$({\color{teal}e_{2}}, 3, 4, 10)$};
  \node[anchor=south west, font=\Large, fill=white, inner sep=1pt] at (5.1,6.1) {$({\color{teal}e_{3}}, 5, 6, 10)$};
  \node[anchor=south east, font=\Large, align=right] at (10.9,0.25) {$({\color{teal}e_{4}}, 10, 0, 0)$\\$({\color{teal}e_{5}}, 10, 0, 0)$\\$({\color{teal}e_{6}}, 10, 0, 0)$};
\end{scope}
\end{scope}

\end{tikzpicture}%
}
\caption{An example of the construction from \Sec{impl-eval}. Given the same hypergraph as in \Fig{example_intervals}, we preprocess their corresponding four-dimensional points, for which we show their second and third dimension in the plot on the right. We then compute the query $P=\{v_3\}$ and $N=\{v_8\}$ by reducing it to an instance of Orthogonal Range Searching.\\
For $S=\{v_3\}$ we get $\IV^\cup = [e_1,e_3]$ ($S=\emptyset$ leads to an empty interval). For $\sigma: v_3\to 1$ and $\tau: v_8 \to 1$ ($\tau: v_8 \to 0$ leads to a non-valid box) we get the endpoints representing the box in purple. We can check that only $e_2$ is contained in that box and it is the correct output.}
\label{fig:example_match}
\end{figure}

In Algorithm~\ref{alg:PartialMatch}, we give the pseudocode for converting a query into boxes and getting the answer, and in \Fig{example_match} we show an example of the construction.

\begin{algorithm}
    \caption{\EdgePartialMatch$(P,N)$}
    \label{alg:PartialMatch}
    initialize $\B^* = \emptyset$\; \label{line:PartialMatch:B*}
    \ForEach{subset $S \subseteq P$}{ \label{line:PartialMatch:forS}
        construct the family of intervals $\I(P \setminus S, N)$\;            \ForEach{function $\sigma:S \to [\edegen]$}{ \label{line:PartialMatch:forsigma}
                \ForEach{function $\tau:N \to \{0\} \cup [\edegen]$}{ \label{line:PartialMatch:fortau}
                    run \Endpoints$(\sigma,\tau)$ to get $\alpha_j, \beta_j$ values\;
                    \ForEach{interval $I \in \I(P \setminus S, N)$}{
                    add the box $I \times \prod^{2\edegen+2}_{j=2} [\alpha_j, \beta_j]$ to $\B^*$
                    }
                }
            }
        
    }
    for counting,
    
    \Indp
        \Return $\sum_{B^* \in \B^\star} \ORQCount(B^*)$\;\label{line:PartialMatch:ORQ}
    \Indm
    for sampling,
    
    \Indp
        pick $B^* \in \B^\star$ with probability proportional to $\ORQCount(B^*)$\; \label{line:PartialMatch:ORQSample}
        \Return $\ORQSample(B^*)$\;\label{line:PartialMatch:ORQSample1}
    \Indm
    for reporting,
    
    \Indp
        warm up $\ORQList(B^*)$ on every $B^* \in \B^\star$\;\label{line:PartialMatch:ORQList}
        run the rest of $\ORQList(B^*)$ on every $B^* \in \B^\star$\;\label{line:PartialMatch:ORQList2}
    \Indm
\end{algorithm}

\subsection{Proof of \Thm{PartialMatch}} \label{sec:proof-partial}
We wrap up the proof. Given the input hypergraph $H$, we first compute the interval representation 
using \Thm{interval}. We then apply the preprocessing step of Algorithm~\ref{alg:PartialMatch_Preprocess}.
Each query is processed by Algorithm~\ref{alg:PartialMatch}. The correctness
of the output holds by \Clm{matchbox} that correctly maps the query to a collection
of boxes, and by \Thm{ORS} on the correctness of the range searching procedures.

We now give the running time bounds. 
To construct the set $\{\ve : e \in E\}$, we retrieve the family of intervals $\IE(e)=(J_1(e),\ldots,J_{k_e}(e))$ from $\IE$, and we write out $\ve_e$.
Recalling that each interval $J$ is stored precisely as the pair $\intStart(J),\intEnd(J)$, the procedure takes time $O(|E| \cdot \edegen)$.
Finally, by \Thm{ORS}, constructing the data structure for weighted Orthogonal Range Search takes time $O(2^{O(\edegen)}\size H \cdot \log^{2\edegen+1}\size H)$, which dominates $O(|E| \cdot \edegen)$ and gives the bound.

We now analyze the query time. Each query to the range searching data structure takes $O(2^{O(\edegen)}\log^{2\edegen+1}\size H)$ time.
There are $2^{|P|}$ possible subsets $S$. The family of intervals $\I(P\setminus S,N)$ can be computed in $O(\vdegen t \log |E|)$ time. There are $\edegen^{O(t)}$ possible functions $\sigma$ and $\tau$.
Each call to \Endpoints\ takes $O(\edegen + t)$ time (by computing the preimages of $\sigma$ and $\tau$ in a single pass). For fixed $S, \sigma$ and $\tau$, there are at most $O(\vdegen t)$ intervals. Building a single box takes $O(\edegen)$ time. The total query time is therefore $O(2^{|P|}\vdegen\edegen^{O(t)}2^{O(\edegen)}t\log^{2\edegen+1}\size{H}) = O(2^{O(\edegen)}\edegen^{O(t)}\vdegen\log^{2\edegen +1}\size H)$.

\subsection{\Eval\ queries}\label{sub:Eval}

We can answer \Eval\ queries by converting them into a collection of \PartialMatch\ queries.
The following theorem gives time bounds for $\ESat$; as usual, the $\VSat$ version has $\edegen$ and $\vdegen$ switched.
\begin{theorem}\label{thm:Eval}
    There is an algorithm for \ESat\ with the following guarantees.
    Given $\vdegen,\edegen$ and an edge-weighted $(\vdegen,\edegen)$-bidegenerate hypergraph $H$,
    \begin{itemize}
        \item First, the algorithm constructs a data structure in time $O(2^{O(\edegen)}\size{H} \log^{2\edegen+1} \size{H})$.
        \item Subsequently, any counting, sampling, or listing \ESat\ query can be answered in time $\vdegen \, 2^{O(\edegen)}\edegen^{O(t)} \log^{2\edegen+1} \size H$, where $t$ is the query size.
    \end{itemize}
\end{theorem}
\begin{proof}
    First, we run the preprocessing phase of \EdgePartialMatch.
    By \Thm{PartialMatch}, this takes time $O(2^{O(\edegen)}\size{H} \log^{2\edegen+1}\size{H})$.

    Consider now a query $\varphi$. We begin by rewriting $\varphi$ in orthogonal DNF form, $\varphi = \varphi_1 \oplus \ldots \oplus \varphi_N$.
    Every subformula $\varphi_h$ is in the form $\left(\bigwedge_{v \in P_h} x_v\right) \wedge \left(\bigwedge_{v \in N_h} \neg x_v\right)$, and $\varphi_i(H) \cap \varphi_j(H) = \emptyset$ for $i \ne j$.
    This construction can be done in $2^{O(t)}$ time where $t$ is the size of $\varphi$.
    The set $\varphi_i(H)$ is exactly $\EdgePartialMatch(P_i,N_i)$. 
    We can thus answer the different flavors (counting, sampling, listing) of \ESat\ for each $\varphi_i$. Since the matches for all $\varphi_i$ are disjoint, we can combine the answers to get counting/sampling/listing for the original query $\varphi$. 
    Answering each query $\EdgePartialMatch(P_i,N_i)$ takes time
    $\vdegen \, 2^{O(\edegen)}\edegen^{O(t)} \log^{2\edegen+1}\size H$, by \Thm{PartialMatch}.
    As there are at most $2^{O(t)}$ such queries, that is also a bound on the overall running time for an \ESat\ query.
%
\end{proof}

\section{\EdgePartialMatch\ and \ESat\ with fixed query size}\label{sec:bounded_query}
When the query length is known in advance, we can give a faster algorithm for \EdgePartialMatch. For every $t \ge 2$, the algorithm answers queries $(P,N)$ with $|P|+|N|=t$ (note $t=1$ is trivial).

\begin{theorem}\label{thm:EPM_fixed_size}
For every $t \ge 2$ there is an algorithm with the following guarantees. Given a $(\vdegen,\edegen)$-interval representation of a hypergraph $H=(V,E)$ and an edge weighting of $H$,
\begin{itemize}\itemsep0pt
    \item First, the algorithm constructs a data structure in time $O(\size{H}\,\edegen^{t-1} \, \log^{t+2}\size{H})$.
    \item Afterwards, any counting/listing/sampling \EdgePartialMatch\ query $(P,N)$ with $|P|+|N|=t$ can be answered in time $O(\vdegen\,\log^{t+2} \size{H})$.
\end{itemize}
For positive queries ($N=\emptyset$), both exponents are $t$ instead of $t+2$.
Moreover, for the counting version, both exponents can be (further) decreased by $1$.
\end{theorem}
\noindent The algorithm is different than the one behind~\Cref{thm:main-eval} and~\Cref{thm:PartialMatch}, and relies on box stabbing rather than orthogonal range search.

As a first consequence of~\Cref{thm:EPM_fixed_size}, we obtain a faster algorithm for \ESat.
The proof follows immediately from the preprocessing guarantees of~\Cref{thm:interval} and the reduction of \ESat\ to \EdgePartialMatch\ of~\Cref{sub:Eval}.
\begin{corollary}
For every $t \ge 2$ there is an algorithm with the following guarantees. Given $\vdegen \ge 1$ and an edge-weighted $(\vdegen,\edegen)$-bidegenerate hypergraph $H$,
\begin{itemize}\itemsep0pt
    \item First, the algorithm constructs a data structure in time $O\!\left(\size{H} \edegen^{t-1} \log^{t+2} \size{H} \right)$.
    \item Afterwards, any counting/listing/sampling \ESat\ query on $t$ variables can be answered in time $O\!\left(\vdegen \log^{t+2} \size{H}\right)$.
\end{itemize}
\end{corollary}

A second consequence of~\Cref{thm:EPM_fixed_size} is faster algorithms for the counting version of some primitives of interest, see~\Cref{tab:faster_queries}.
Note that $\SP(S)$ is \EdgePartialMatch$(S,\emptyset)$, hence the faster bounds apply.
Moreover, for $\ISH(S)$ we get the same bounds of $\SP(S)$, and for $\VENN(e,f,g)$ the same bounds of $\SP(S)$ for $|S|=3$; this is because we can solve the counting version of \EdgePartialMatch$(P,N)$ by inclusion-exclusion after solving queries \EdgePartialMatch$(S,\emptyset)$, which is $\SP(S)$, for each $S \subseteq P \cup N$.
\begin{table}[h!]\renewcommand{\arraystretch}{1.4}
    \centering
    \begin{tabular}{lll}
    \toprule
        Primitive & Preprocessing time & Query time \\\midrule
        $\ADJ(u,v)$ & $\size{H} \cdot\left(\log^{1.5}\size{H}+ \edegen\log\size{H}\right)$ & $\vdegen \log \size{H}$ \\
        $\INT(e,f)$, $\UNION(e,f)$, $\DIFF(e,f)$ & $\size{H} \cdot\left(\log^{1.5}\size{H}+ \vdegen\log\size{H}\right)$ & $\edegen \log \size{H}$ \\
        $\SP(S)$, $\ISH(S)$ for $|S|=t\ge 2$ & $\size{H} \cdot\big(\log^{1.5}\size{H}+ \edegen^{t-1}\log^{t-1}\size{H}\big)$ & $\vdegen \log^{t-1} \size{H}$\\
        $\VENN(e,f,g)$ & $\size{H} \cdot \big(\vdegen \log\size{H}\big)^2$ & $\edegen \log^2 \size{H}$ \\
        \bottomrule 
     \end{tabular}
    \caption{Guarantees for the counting (and thus decision) version of some primitives of interest. For $\SP$ and $\ISH$, the size of the query $|S|$ must be known before preprocessing.}
    \label{tab:faster_queries}
\end{table}

\subsection{The algorithm}
The basic idea is as follows.
Let $(P,N)$ be the query, and for every $e \in E$ define $S_e = P \cap \IE^\cup(e)$.
Note that, if $e$ matches the query, then:
\begin{align}
    S_e &= \{v \in P : v \in \IE^\cup(e) \} \\
    P \setminus S_e &= \{v \in P : e \in \IV^\cup(v) \}
\end{align}
In words, $S_e$ contains the vertices of $P$ whose incidence with $e$ is certified by the intervals of $e$, while $P \setminus S_e$ contains the vertices of $P$ whose incidence with $e$ is certified by their own intervals.
As a first step, we distinguish edges into ``easy'' and ``hard'', according to $S_e$.

\uline{The easy edges: $S_e \subsetneq P$.}
Suppose $e$ is such that $S_e \subsetneq P$.
Then, $e \in I$ for some $I \in \IV(v)$ for some vertex $v \in P \setminus S_e$.
Intuitively, then, one can find all such edges by looking at each $I$, and for every $e \in I$ checking if $e$ is indeed a match.
Checking all such edges explicitly is clearly infeasible, as they are too many.
However, by the niceness property of Definition~\ref{def:interval_representation}, we need only to check \emph{one} edge $\hat e$ per each interval $I$: indeed, $\hat e$ is a match if and only if all $e \in I$ are.
By guessing every $S \subsetneq P$, we can in this way find all easy edges.

\uline{The hard edges: $S_e = P$.}
Suppose now $S_e = P$.
In this case we cannot search $e$ in the intervals of some $v \in P$, because the incidence between $e$ and $v$ is encoded in $\IE(e)$ instead of $\IV(v)$.
To find these edges we thus perform a reduction to box stabbing.
The basic idea is simple: for every $v \in P$ we know $v \in J_i$ for some $J_i = \IE(e)$, and for every $v \in N$, we know $v \notin J_i$ for all $J_i \in \IE(e)$.
Thus, the vector $(v_1,\ldots,v_t)$ formed by the query vertices lies in a $t$-dimensional box $J_{i_1},\ldots,\bar{J}_{i_t}$, with the ``positive'' sides $J_i$ chosen from $\IE(e)$ and the ``negative'' sides $\bar J_i$ chosen from their complements. Let us pin this down more precisely.

Let $e \in E$, and recall $\IE(e)=J_1(e),\ldots,J_{k_e}(e)$.
Denote by $\nIE(e)=\nJ_0(e),\ldots,\nJ_{k_e}(e)$ the family of intervals that complements $\IE$, so that $\nJ_0(e) \uplus \ldots \uplus \nJ_{k_e}(e) = \bV \setminus \IE^\cup(e).$
Formally,
\begin{align}
    \nJ_j(e) &=[1,\intStart(J_{j+1}(e))-1] && j=0   \\
    \nJ_j(e) &= [\intEnd(J_j(e))+1,\intStart(J_{j+1}(e))-1] && j=1,\ldots,k_e-1 \\
    \nJ_j(e) &= [\intEnd(J_j(e))+1,n] && j=k_e
\end{align}
Note that $\nJ_0(e)$ and $\nJ_{k_e}$ may be empty.
We use the intervals $J_j$ and $\nJ_j$ to transform every edge $e$ into a set of boxes; at query time we then convert the vertices into a vector to stab the boxes.
This is where the preprocessing comes into play, to preprocess the boxes so as to support box stabbing efficiently.

Algorithm~\ref{alg:BoundedPartialMatch_Preprocess} and Algorithm~\ref{alg:BoundedPartialMatch} give respectively the pseudocode of preprocessing and query.
The rest of the section proves the correctness and running time of the algorithms, and wrap up the proof of~\Cref{thm:EPM_fixed_size}.

\begin{algorithm}
    \label{alg:BoundedPartialMatch_Preprocess}
    \caption{\BoundedEdgePartialMatchPre$(\bV,\bE,\IV,\IE,w,\tau)$}
    initialize $\cB_e=\emptyset$\;
    \ForEach{$e \in \bE$}{
        let $\IE(e)=J_1,\ldots,J_{k_e}$ and $\nIE(e)=\nJ_0,\ldots,\nJ_{k_e}$\;
        \ForEach{$(i_1,\ldots,i_{t-\tau}) \in [k_e]^{t-\tau}$ and $(j_1,\ldots,j_{\tau}) \in \{0\} \cup [k_e]^{\tau}$}{
            add to $\cB_e$ the box $[-\infty,e] \times [e,\infty] \times J_{i_1} \times \ldots \times J_{i_{t-\tau}} \times \nJ_{j_1} \times \ldots \times \nJ_{j_{\tau}} \in \RR^{t+2}$\;\label{line:BEPM_add_box}
        }
    }
    preprocess $\bigcup_{e \in E}\cB_e$ for box stabbing\;
\end{algorithm}

\begin{algorithm}
    \caption{\BoundedEdgePartialMatch$(P,N)$}
    \label{alg:BoundedPartialMatch}
    let $P=\{u_1,\ldots,u_{t-\tau}\}$ and $N=\{v_1,\ldots,v_\tau\}$\;
    \ForEach{subset $S \subsetneq P$}{ \label{line:BoundedPartialMatch:forS}
        \ForEach{interval $I \in \I(P \setminus S, N)$}{\label{line:BoundedPartialMatch:forI}
            pick any $\hat e \in I$\;
            \If{$(S \cup N) \cap \IE^\cup(\hat e) = S$}{
                add $I$ to the output set\;\label{line:bounded_partialmatch:addI}
            }
        }
    }    
    \ForEach{interval $I \in \I(\emptyset,N)$\label{line:BoundedPartialMatch:forStab}}{
        perform a box stabbing query on the vector $\bq_I=(\intStart(I),\intEnd(I),u_1,\ldots,u_{t-\tau},v_1,\ldots,v_\tau)$\;\label{line:EPM:stab}
    }
\end{algorithm}

\subsection{Correctness}
The next two claims prove the correctness of our algorithm.

\begin{claim}
The loop of line~\ref{line:BoundedPartialMatch:forS} finds the set of matching edges with $S_e\subsetneq P$.
\end{claim}
\begin{proof}
Suppose $e$ matches the query and $S_e \subsetneq P$.
Since $e \notin \IV^\cup(v)$ for all $v \in N$, it follows that $e \in I$ for some $I \in \I^\cup(P \setminus S_e, N)$.
The algorithm at some point reaches line~\ref{line:BoundedPartialMatch:forI} with $S=S_e$ and the above choice of $I$.
It then picks some $\hat e \in I$, possibly different from $e$.
However, since $P \setminus S \ne \emptyset$, then $e,\hat e \in I$ implies $e,\hat e \in I_i(u)$ for some $u \in P$ and some $i \in [k_u]$; by the niceness property of the interval representation, this implies $\IE(e)=\IE(\hat e)$.
On the one hand this implies $S = P \cap \IE(e)=P \cap \IE(\hat e) \subseteq \IE(\hat e)$; on the other hand, $N \cap \IE(\hat e) = N \cap \IE(e) = \emptyset$.
Therefore $(S \cup N) \cap \IE^\cup(e) = S$, and the algorithm adds $I$, which contains $e$, to the output set.

Now suppose the algorithm adds $e$ to the output set.
Thus, for some $S \subsetneq P$, for some $I \in \I(P \setminus S, N)$, the algorithm added $I$ to the output set, and $e \in I$.
On the one hand, this implies $e \in \I^\cup(P \setminus S, N)$, which means $e \in \IV(v)$ for every $v \in P \setminus S$ and $e \notin \IV(v)$ for every $v \in N$
On the other hand, the algorithm found that $S \subseteq \IE^\cup(\hat e)$ for some $\hat e \in I$.
As noted above, however, $\IE^\cup(e)=\IE^\cup(\hat e)$; therefore $(S \cup N) \cap \IE^\cup(e) = S$, which means $v \in \IE^\cup(e)$ for every $v \in S$ and $v \notin \IE^\cup(e)$ for every $v \in N$.
We conclude that $V(e) \cap (P \cup N) = P$, as claimed.
\end{proof}

\begin{claim}
The loop of line~\ref{line:BoundedPartialMatch:forStab} finds the set of matching edges with $S_e=P$.
\end{claim}
\begin{proof}
Suppose $B \in \cB_e$ is stabbed by some $\bq_I$.
The first two coordinates imply $e \in I$ and thus $e \notin \IV^\cup(v)$ for every $v \in N$; while the last $\tau$ coordinates imply $v \not\in J_i(e)$ for all $i \in [k_e]$ for every $v \in N$.
Thus, $e \notin E(v)$ for every $v \in N$.
Finally, the remaining $t-\tau$ coordinates imply, for every $v \in P$, that $v \in J_i(e)$ for some $i \in [k_e]$.
Thus $S_e=P$, and moreover $e \in E(v)$ for every $v \in P$.

Now suppose $e \in E$ matches the query and $S_e=P$.
We show that, among the boxes stabbed by the vectors $\{\bq_I : I \in \I(\emptyset,N)\}$, exactly one box is in $\cB_e$.
First, $e \in I$ for some unique $I \in \I(\emptyset, N)$.
Thus, for all $I' \in \I(\emptyset, N)$ with $I' \ne I$, the vector $\bq_{I'}$ cannot stab any $B \in \cB_e$.
The vector $\bq_I$ instead stabs every $B \in \cB_e$ in the first two dimensions.
Moreover, since $\IE(e) \cup \nIE(e)$ partitions $\bV$, then every $v \in P \cup N$ falls precisely in one of those intervals, which implies that there exists precisely one $B \in \cB_e$ that is stabbed by $\bq_I$ along the last $t$ dimensions, too.
This concludes the proof.
\end{proof}

\subsection{Running time}
\begin{claim}
Algorithm~\ref{alg:BoundedPartialMatch_Preprocess} can be run in time $O\left(\size{H}\,\edegen^{t-1} \log^{t+2}\size{H}\right)$. The exponent can be decreased from $t+2$ to $t+1$ for the counting version of queries.
\end{claim}
\begin{proof}
For each $e \in E$ the algorithm spends time $O(\edegen)$ to fetch $\IE(e)$ and construct $\nIE(e)$, and creates at most $k_e^{t-\tau}\cdot(k_e+1)^{\tau} \le (k_e+1)^t = O(k_e \cdot \edegen^{t-1})$ boxes, spending time $O(t)=O(1)$ for each of them.
Noting that $\sum_{e \in E}k_e \le \size{H}$, the number of boxes and the running time are both in $O\left(\edegen^{t-1}\sum_{e \in E}k_e \right) = O\left(\edegen^{t-1} \size{H} \right)$.
By~\Cref{thm:BoxStabbing} we can thus preprocess the boxes in time
\begin{align}
    O\left(\edegen^{t-1} \size{H}\right) \cdot O\left(\log \left(\edegen^{t-1} \size{H}\right)\right)^{t+2}
    = O\left(\size{H}\,\edegen^{t-1} \cdot \log^{t+2} \size{H}\right)\,,
\end{align}
which dominates the algorithm. The decrease in the exponent follows again~\Cref{thm:BoxStabbing}.
\end{proof}

\begin{claim}
After \BoundedEdgePartialMatchPre$(\bV,\bE,\IV,\IE,w,\tau)$, Algorithm~\ref{alg:BoundedPartialMatch} can be run in time $O(\vdegen \, \log^{t+2} \size{H})$, and in time $O(\vdegen \log^{t+1} \size{H})$ in the counting version.
\end{claim}
\begin{proof}
The second bound follows again from~\Cref{thm:BoxStabbing}. Let us prove the first one.

\medskip\noindent\uline{First part.}
The algorithm loops over $2^{t-\tau}-1\le 2^t$ subsets $S \subsetneq P$.
For each $S$, computing $\I(P \setminus S,N)$ takes time $O(\vdegen t \log |E|)$, see the proof of \Thm{PartialMatch} in \Cref{sec:proof-partial}.
Checking $(S \cup N) \cap \IE^\cup(\hat e) = S$ requires, for every $v \in S \cup N$, a binary search over the endpoints of $\IE$, which takes time $O(\ln \edegen)$; this is repeated $|\I(P \setminus S,N)| \le \vdegen t +1$ times, see \Clm{icups}.
We maintain the output set as, once again, a disjoint union of intervals.
To add $I$ we perform a binary search, and we update the intervals intersected by $I$.
Every endpoint ever appearing in our list is deleted at most once, and this easily implies a total work of $O(k^* \ln |E|)$ where $k^* \le 2^t (\vdegen t + 1 )$ is the total number of intervals possibly added, and $|E|$ is the maximum size of the list at any time (since the intervals are disjoint and are subsets of $\bE$).
Thus the total cost of updating the output list is $O(2^t \vdegen t^2 \ln |E|)$.
The total time of the first part, as $t=O(1)$ and $|E|+\edegen = O(\size{H})$, is:
\begin{align}
    O\left(2^t \cdot \vdegen t \left( \ln |E| + \ln \edegen \right)\right) + O\left(2^t \vdegen t^2 \ln |E|\right) = O\left(\vdegen\, \log\size{H}\right)\,.
\end{align}

\medskip\noindent\uline{Second part.}
Computing $\I(\emptyset,N)$ can be charged to the first part.
For $|\I(\emptyset,N)| \le \vdegen t +1$ times, the algorithm invokes box stabbing, paying each time $O\!\left(\log \left(\edegen^{t-1} \size{H}\right)\right)^{t+2}$, which for $t=O(1)$ is $O\left(\log^{t+2} \size{H}\right)$.
Thus the second part takes in total $O\left(\vdegen \log^{t+2} \size{H}\right)$, which dominates the first part.
\end{proof}

\subsection{The special case $N=\emptyset$}
For the special case $N=\emptyset$ of~\Cref{thm:EPM_fixed_size}, we edit the pseudocode as follows.
In Algorithm~\ref{alg:BoundedPartialMatch_Preprocess}, the box constructed by line~\ref{line:BEPM_add_box} becomes $J_{i_1} \times \ldots \times J_{i_{t-\tau}} \times \nJ_{j_1} \times \ldots \times \nJ_{j_{\tau}} \in \RR^t$.
In Algorithm~\ref{alg:BoundedPartialMatch}, the loop of line~\ref{line:BoundedPartialMatch:forStab} is replaced by just one stabbing query on the vector $(u_1,\ldots,u_{t-\tau},v_1,\ldots,v_\tau)$.
It is immediate to check that the proofs of correctness carry over with minimal modifications.

The running time analysis changes slightly, as follows.
Algorithm~\ref{alg:BoundedPartialMatch_Preprocess} now uses $t$-dimensional rather than $(t+2)$-dimensional boxes; therefore it takes time $O(|E|\,\edegen^t)  \cdot O\left(t\, \log \size{H}\right)^{t}$.
The second part of Algorithm~\ref{alg:BoundedPartialMatch} now costs only $O\left(t\, \log \size{H}\right)^{t}$, hence the total running time is in $O(\vdegen) \cdot O\left(t\, \log \size{H}\right)^{t}$.
For fixed $t$, both expressions yield the claimed bounds.
Moreover, once again, for the counting version we can decrease the exponents from $t$ to $t-1$.

\section{Answering \VertexNeigh\ queries} \label{sec:vneigh}

This section presents our algorithms for $\VertexNeigh$ and $\EdgeNeigh$. We focus on $\VertexNeigh$, as the results for $\EdgeNeigh$ follow by taking the dual of $H$ and exchanging $\vdegen$ and $\edegen$. Akin to $\PartialMatch$, our algorithm relies on an interval representation of $H$. We solve the weighted version, where the input also includes a vertex weight function $w:V\to\NN$. Our main result is:
\begin{theorem}\label{thm:NC}
There exists an algorithm for \VertexNeigh\ with the following guarantees.
Given a $(\vdegen,\edegen)$-interval representation of a weighted hypergraph $H=(V,E)$,
\begin{itemize}
\item First, the algorithm constructs a data structure in time $\edegen 2^{O(\vdegen)} \probsize H \log^{2\vdegen-1}\probsize H$.
\item Then, (weighted) counting, sampling, and $O(\log \size H)$-delay listing of any \VertexNeigh\ query can be done in time $2^{O(\vdegen\log\vdegen)} \log^{2\vdegen-1}\size H$. 
\end{itemize}
\end{theorem}
\noindent Note that the algorithm is exponential in $\vdegen$, but only polynomial in $\edegen$.
Previous work shows that, in many hypergraph datasets, $\vdegen$ is below $20$ when $\edegen$ is in the hundreds or thousands. In fact, this holds for the stronger parameters $\beta$ and $\alpha$ of~\cite{bressan2026countinghypergraphletscolorcoding}. There are good reasons to think that $\vdegen$ and $\edegen$ are significantly smaller.
Thus,~\Cref{thm:NC} may actually perform well in practice.

Let $(\bV,\bE,\IV,\IE)$ be the $(\vdegen,\edegen)$-interval representation of $H=(V,E)$ given as input. As in \Cref{sec:impl-eval}, we assume $V$ and $E$ are relabeled so that $\bV=(1,\ldots,n)$ and $\bE=(1,\ldots,m)$.
Our algorithm constructs two data structures that, for every vertex $v\in V$, allow us to efficiently access two disjoint sets $N_0(v)$ and $N_1(v)$ whose union is $N(v)$. We define these sets as follows:
\begin{align}
    N_0(v) &:= \left\{ u\in V : \I^\cup(v)\cap\I^\cup(u)\neq\emptyset \right\},\\
    N_1(v) &:= N(v)\setminus N_0(v).
\end{align}
We outline the idea behind the two data structures.
\begin{enumerate}
    \item For $N_0(v)$, we preprocess each vertex $u$ by mapping it to a $2\vdegen$-dimensional vector whose coordinates encode the endpoints of the intervals in $\IV(u)$. At query time, the intervals of $\IV(v)$ and their complements induce a family of disjoint axis-aligned boxes. We select precisely those boxes whose endpoint constraints guarantee that at least one interval of the represented vertices intersects $\IV^\cup(v)$. The selected boxes contain exactly the points corresponding to $N_0(v)$, which can be found using orthogonal range queries.
    \item For $N_1(v)$, we take a complementary approach. Unlike $N_0(v)$, this set cannot be characterized by the intersection of the intervals in $\IV(v)$ and $\IV(u)$, so the previous reduction to orthogonal range queries does not apply. We bypass this problem by reducing to a variant of box stabbing, which we call \emph{union stabbing}, for which we show an efficient algorithm. We construct a family of two-dimensional boxes over $V \times V$ that captures all remaining adjacencies, possibly including some already covered by $N_0$. These boxes are obtained by combining intervals from the $\IE$ side, either with individual vertices or with other intervals. Crucially, the niceness property of~\Def{interval_representation} guarantees that a single representative edge suffices to capture the relevant adjacencies for each interval in $\IV(v)$, and that each box either consists entirely of pairs already captured by $N_0$ or contains none of them. We can therefore discard redundant boxes by testing a single pair per box, and reduce queries for $N_1(v)$ to weighted union stabbing on the remaining family.
\end{enumerate}

\begin{figure}[!htbp]
    \centering
    \resizebox{0.85\textwidth}{!}{%
    \begin{minipage}{\textwidth}
        \centering
        \begin{tikzpicture}[x=1.46cm,y=1.46cm,
      hyperedge/.style={draw=black,line width=1.05pt,
        line cap=round,line join=round},
      peelededge/.style={hyperedge,line width=0.5pt,
        dash pattern=on 3pt off 3pt},
      vertex/.style={circle,fill=black,inner sep=1.7pt},
      peeledvertex/.style={vertex,fill=white,draw=black,
        line width=0.5pt,dash pattern=on 1.5pt off 1pt},
      hvertex/.style={vertex,draw=black,double=white,
        double distance=2pt,line width=0.8pt,inner sep=3pt},
      vlabel/.style={font=\large,fill=white,inner sep=.35pt,text=red!70!black},
      elabel/.style={font=\large,text=teal}]
    
    \def\edgeone{(0.91,2.96)
      .. controls (0.84,3.40) and (1.08,3.72) .. (1.48,3.79)
      .. controls (2.00,3.91) and (2.84,3.16) .. (3.03,2.39)
      .. controls (3.19,1.78) and (2.86,1.04) .. (2.39,0.95)
      .. controls (1.91,0.88) and (1.41,1.52) .. (1.20,2.13)
      .. controls (1.04,2.47) and (0.92,2.75) .. cycle}
    
    \def\edgetwo{(0.92,0.52)
      .. controls (0.83,0.12) and (1.12,-0.20) .. (1.50,-0.12)
      .. controls (2.03,-0.03) and (2.78,0.37) .. (3.02,1.20)
      .. controls (3.28,1.85) and (2.86,2.71) .. (2.37,2.75)
      .. controls (1.88,2.83) and (1.43,2.42) .. (1.20,1.72)
      .. controls (1.06,1.21) and (0.92,0.77) .. cycle}
    
    \def\edgethree{(1.94,2.06)
      .. controls (1.98,2.45) and (2.54,2.84) .. (3.04,2.98)
      .. controls (4.10,3.43) and (5.63,3.12) .. (5.93,2.66)
      .. controls (6.29,2.02) and (6.22,1.04) .. (5.73,0.82)
      .. controls (5.16,0.53) and (4.11,1.26) .. (3.31,1.61)
      .. controls (2.62,1.93) and (2.13,1.78) .. cycle}
    \def\edgefour{(1.97,1.05)
      .. controls (1.98,1.49) and (2.38,1.67) .. (2.80,1.80)
      .. controls (3.92,2.28) and (5.21,2.93) .. (5.92,2.73)
      .. controls (6.33,2.61) and (6.45,1.49) .. (6.07,0.99)
      .. controls (5.62,0.32) and (4.15,0.39) .. (3.17,0.59)
      .. controls (2.44,0.72) and (2.06,0.75) .. cycle}
    
    \draw[hyperedge] \edgeone;
    \draw[hyperedge] \edgetwo;
    \draw[hyperedge] \edgethree;
    \draw[peelededge] \edgefour;
    
    \node[peeledvertex] at (1.35,3.20) {};
    \node[vlabel,above right=-1pt and 4pt] at (1.35,3.20) {$v_1$};
    \node[peeledvertex] at (1.35,0.35) {};
    \node[vlabel,below right=-1pt and 4pt] at (1.35,0.35) {$v_2$};
    \node[peeledvertex] at (5.35,1.85) {};
    \node[vlabel,above right=-1pt and 4pt] at (5.35,1.85) {$v_3$};
    \node[hvertex] at (2.35,1.25) {};
    \node[vlabel,above right=-1pt and 6.5pt,font=\large\bfseries]
          at (2.35,1.25) {$v_4$};
    \node[vertex] at (2.35,2.15) {};
    \node[vlabel,above right=-1pt and 6.5pt] at (2.35,2.15) {$v_5$};
    
    \node[elabel] at (0.94,3.98) {$e_4$};
    \node[elabel] at (1.04,-0.47) {$e_3$};
    \node[elabel] at (4.37,3.41) {$e_2$};
    \node[elabel] at (4.46,0.20) {$e_1$};
    
\end{tikzpicture}

        \vspace*{2em}

        \begin{minipage}[c]{0.47\textwidth}
            \centering
            \resizebox{\linewidth}{!}{%
                \begin{tikzpicture}[x=1.15cm,y=1.12cm,
  axis/.style={line width=.88pt,-{Stealth[length=5.1pt,width=4pt]}},
  boxline/.style={line width=.86pt},
  dot/.style={circle,fill=black,inner sep=1.65pt},
  hdot/.style={dot,draw=black,double=white,
    double distance=1.7pt,line width=.72pt,inner sep=2.7pt},
  pointlabel/.style={
    font=\large,fill=white,inner sep=.35pt,
    text=red!70!black
  },
  axislabel/.style={font=\large,text=black},
  boxlabel/.style={font=\normalsize}]

    \def\Cone{[{\color{teal}e_1},{\color{teal}e_2}]}
    \def\Ctwo{[{\color{teal}e_3},{\color{teal}e_4}]}

    \draw[axis] (-0.14,0) -- (6.40,0);
    \draw[axis] (0,-0.12) -- (0,6.06);
    
    \foreach \xa/\xb/\ya/\yb in {
        .72/2.83/3.36/5.74,
        3.37/5.77/.71/2.82,
        3.37/5.77/3.36/5.74}
      {
        \fill[pattern={
          Lines[angle=43,distance=3pt,line width=.1pt]}]
          (\xa,\ya) rectangle (\xb,\yb);
        \draw[boxline] (\xa,\ya) rectangle (\xb,\yb);
      }
    
    \draw[boxline] (.72,.71) rectangle (2.83,2.82);
    
    \foreach \p/\i in {1.12/1,2.47/2,3.88/3,5.27/4}
      {
        \draw[line width=.83pt] (\p,-.09) -- (\p,.09);
        \node[axislabel,below=3pt]
          at (\p,-.09) {$e_{\i}$};
    
        \draw[line width=.83pt] (-.09,\p) -- (.09,\p);
        \node[axislabel,left=5pt]
          at (-.09,\p) {$e_{\i}$};
      }
    
    \node[dot] at (5.27,5.27) {};
    \node[pointlabel,below left=2pt and 4pt]
      at (5.27,5.27) {$v_1$};
    
    \node[dot] at (3.88,3.88) {};
    \node[pointlabel,above right=2pt and 4pt]
      at (3.88,3.88) {$v_2$};
    
    \node[dot] at (2.47,2.47) {};
    \node[pointlabel,below left=2pt and 4pt]
      at (2.47,2.47) {$v_3$};
    
    \node[hdot] at (3.88,5.27) {};
    \node[pointlabel,below right=2pt and 6pt,
          font=\large\bfseries]
      at (3.88,5.27) {$v_4$};
    
    \node[dot] at (2.47,5.27) {};
    \node[pointlabel,below left=2pt and 4pt]
      at (2.47,5.27) {$v_5$};
    
    \node[boxlabel] at (1.78,.35)
      {$\Cone\times\Cone$};

    \node[boxlabel] at (1.78,6.08)
      {$\Cone\times\Ctwo$};

    \node[boxlabel,anchor=west]
      at (3.4,6.08) {$\Ctwo\times\Ctwo$};

    \node[boxlabel,anchor=west]
      at (3.4,.35) {$\Ctwo\times\Cone$};

\end{tikzpicture}%
            }
            \smallskip
            {\small (a) Retrieving $N_0(v_4)$.}
        \end{minipage}
        \hfill
        \begin{minipage}[c]{0.47\textwidth}
            \centering
            \resizebox{\linewidth}{!}{%
                \begin{tikzpicture}[x=1.15cm,y=1.12cm,
  axis/.style={line width=.88pt,-{Stealth[length=5.1pt,width=4pt]}},
  boxline/.style={line width=.86pt},
  stabbing/.style={
    line width=.83pt,
    dash pattern=on 3.2pt off 3.2pt
  },
  axislabel/.style={font=\large,text=black},
  pointlabel/.style={
    font=\large,fill=white,inner sep=.35pt,
    text=red!70!black
  },
  boxlabel/.style={font=\normalsize}]

\def\J{{[{\color{red!70!black}v_4},{\color{red!70!black}v_4}]}}
\def\Jp{{[{\color{red!70!black}v_3},{\color{red!70!black}v_3}]}}

\fill[white] (3.22,4.43) rectangle (3.94,5.15);
\fill[pattern={
  Lines[angle=43,distance=3pt,line width=.1pt]}]
  (3.22,4.43) rectangle (3.94,5.15);
\draw[boxline]
  (3.22,4.43) rectangle (3.94,5.15);

\fill[white] (4.43,3.22) rectangle (5.15,3.94);
\fill[pattern={
  Lines[angle=43,distance=3pt,line width=.1pt]}]
  (4.43,3.22) rectangle (5.15,3.94);
\draw[boxline]
  (4.43,3.22) rectangle (5.15,3.94);

\draw[boxline] (3.22,3.22) rectangle (3.94,3.94);

\draw[boxline] (4.43,4.43) rectangle (5.15,5.15);

\draw[axis] (-.15,0) -- (7.00,0);
\draw[axis] (0,-.12) -- (0,7.03);

\foreach \p/\i in {
    1.16/1,2.37/2,3.58/3,4.79/4,6.00/5}
  {
    \draw[line width=.83pt]
      (\p,-.09) -- (\p,.09);

    \node[axislabel,below=3pt]
      at (\p,-.09) {$v_{\i}$};

    \draw[line width=.83pt]
      (-.09,\p) -- (.09,\p);

    \node[axislabel,left=5pt]
      at (-.09,\p) {$v_{\i}$};
  }

\draw[stabbing]
  (4.79,.20) -- (4.79,6.77);

\node[boxlabel] at (3.48,5.44)
    {$\Jp\times\J$};

\node[boxlabel] at (4.94,2.98)
    {$\J\times\Jp$};

\end{tikzpicture}%
            }
            \smallskip
            {\small (b) Retrieving $N_1(v_4)$.}
        \end{minipage}
    \end{minipage}%
    }
    \vspace*{2em}
    \caption{
        \textbf{Above:} a hypergraph $H$ with peeling sequence $v_1,v_2,e_1,v_3,v_4,v_5$, shown immediately before peeling the highlighted vertex $v_4$. Dashed outlines indicate previously peeled elements.
        \textbf{Below:} Retrieving the neighborhood of $v_4$.
        \textbf{(a)} The vertex-side intervals are $\IV(v_1) = \{[e_4,e_4]\}$, $\IV(v_2) = \{[e_3,e_3]\}$, $\IV(v_3) = \{[e_2,e_2]\}$, $\IV(v_4) = \{[e_3,e_4]\}$, and $\IV(v_5) = \{[e_2,e_4]\}$. The query interval $\IV(v_4)=\{[e_3,e_4]\}$ induces the partition $C_1=[e_1,e_2]$ and $C_2=[e_3,e_4]$. The shaded boxes represent intervals intersecting $\IV(v_4)$. Orthogonal range queries over these boxes yield $\{v_1,v_2,v_4,v_5\}$.
        \textbf{(b)} The only nonempty edge-side interval family is $\IE(e_1)=\{[v_3,v_3],[v_4,v_4]\}$. Let $J=[v_3,v_3]$ and $J'=[v_4,v_4]$. Combining these intervals produces four rectangles: $J\times J$, $J\times J'$, $J'\times J$, and $J'\times J'$. The shaded rectangles $J\times J'$ and $J'\times J$ are the only ones retained after filtering. Stabbing them at $v_4$ yields $N_1(v_4)=\{v_3\}$.
    }
    \label{fig:nc_retrieval}
\end{figure}
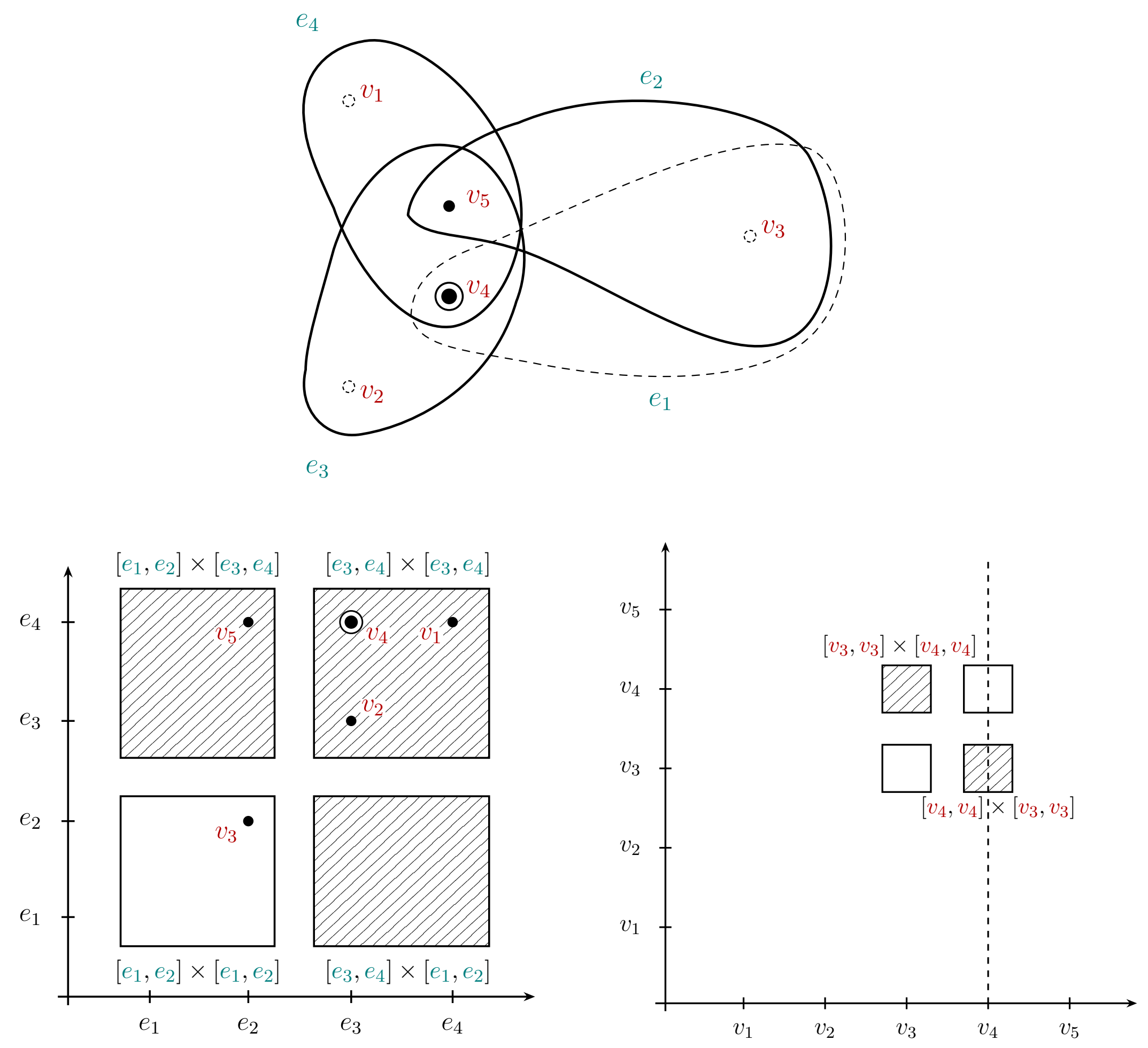

\Cref{fig:nc_retrieval} gives a visual representation of both procedures. The next two subsections formalize this intuition, describing the algorithms for $N_0(v)$ and $N_1(v)$. Since $N_0(v)$ and $N_1(v)$ are disjoint, it is straightforward to combine the two algorithms into a single one.

\subsection{Computing $\mathbf{N_0(v)}$}\label{sec:NV}
This section proves the following result:
\begin{theorem}
    There exists an algorithm that, given a $(\vdegen,\edegen)$-interval representation of a weighted hypergraph $H=(V,E)$, in time $2^{O(\vdegen)}|V|\log^{2\vdegen -1} |V|$ builds a data structure that can answer any counting/listing/sampling query on $N_0(v)$ in time $2^{O(\vdegen \ln \vdegen)} \log^{2\vdegen - 1} |V|$.  
\end{theorem}
The proof performs a reduction to the weighted version of Orthogonal Range Search, similarly to what we did for \PartialMatch.
The idea is to represent every vertex $u \in V$ as a vector $\vu$, and construct on all those vectors a data structure for orthogonal range searching.
Given a query vertex $v$, we will then construct a family of disjoint axis-aligned boxes that contains precisely the vectors of $N_0(v)$.
Throughout this section, we use $\ORQPre, \ORQCount, \ORQList$ and $\ORQSample$ to denote the respective procedures for weighted Orthogonal Range Searching, whose existence is guaranteed by \Cref{thm:ORS}.

\subsubsection{Preprocessing}
For every $u\in V$ with $k_u\ge1$, let $I_i(u)=[l_i,r_i]$ for $i\in[k_u]$, where without loss of generality we assume $r_i < l_{i+1}$.
The vector associated to $u$ is then simply:
\begin{align}
    \vu \coloneq (l_1,r_1,\ldots,l_{k_u},r_{k_u},\underbrace{0, \ldots,0}_{2(\vdegen-k_u)}) \in \{0, \dots,m\}^{2\vdegen}.
\end{align}
Then, construct an orthogonal range counting structure on the indexed family of vectors $\{\vu : u \in \bV\}$. The pseudocode is given by Algorithm~\ref{alg:Neigh_Preprocess}.
\begin{algorithm}
    \label{alg:Neigh_Preprocess}
    \caption{\VertexNeighPre$(\bV,\bE,\IV,\IE,w)$}
    compute the vector $\vu$ for every $u \in \bV$\;
    run \ORQPre\ on the indexed family $\{\vu : u \in \bV\}$ with weighting $w$\;
\end{algorithm}

\subsubsection{Query}
We describe the query phase of \VertexNeigh. The pseudocode is given by Algorithm~\ref{alg:Neigh}.
\begin{algorithm}
    \label{alg:Neigh}
    \caption{\VertexNeigh$(v)$}
    compute the partition $C_1,\ldots,C_{2k_v+1}$ of $\bE$ induced by the intervals $I_1(v),\ldots,I_{k_v}(v)$ of $\I(v)$\; \label{line:Neigh:Cs}
    initialize $\B^* = \emptyset$\; \label{line:Neigh:B*}
    \ForEach{$B \in \cup_{k=1}^{\kappa_v}\B_k(v)$}{ \label{line:Neigh:forB}
        \If{$B$ is a $v$-box}{ \label{line:Neigh:checkVbox}
            add the extended box $B^*= B \times \{0\} \times \ldots \times \{0\}$ to $\B^*$ \label{line:Neigh:extend}
        }
    }
    for counting,
    
    \Indp
        \Return $\sum_{B^* \in \B^\star} \ORQCount(B^*)$\;\label{line:Neigh:ORQ}
    \Indm
    for sampling,
    
    \Indp
        pick $B^* \in \B^\star$ with probability proportional to $\ORQCount(B^*)$\; \label{line:Neigh:ORQSample}
        \Return $\ORQSample(B^*)$\;\label{line:Neigh:ORQSample1}
    \Indm
    for reporting,
    
    \Indp
        warm up $\ORQList(B^*)$ on every $B^* \in \B^\star$\;\label{line:Neigh:ORQList}
        run the rest of $\ORQList(B^*)$ on every $B^* \in \B^\star$\;\label{line:Neigh:ORQList2}
    \Indm
\end{algorithm}

Let $v \in V$ be the query vertex, and let $\I(v)=I_1(v),\ldots,I_{k_v}(v)$, where for every $i = 1,\ldots,k_v$ we have $I_i(v)=[x_i,y_i] \subseteq [m]$.
Again, we assume the intervals are sorted increasingly, so that $y_i < x_{i+1}$.
Define $y_0\coloneq0$ and $x_{k_v+1}\coloneq m+1$.

\paragraph{Boxes.}
Consider then the following set of $c\coloneq2k_v+1$ cells $C_1,\ldots,C_{2k_v+1}$:
\begin{align}
    C_{2j} &\coloneq I_j(v) && j=1,\ldots,k_v\,,\\
    C_{2j+1} &\coloneq [y_{j}+1,\,x_{j+1}-1] && j=0,\ldots,k_v\,.
\end{align}
In words, the even cells $C_2,\ldots,C_{2k_v}$ are the intervals of $\I(v)$, and the odd cells $C_1,\ldots,C_{2k_v+1}$ are the gaps between them.
Note that $C_1$ and $C_{2k_v+1}$ may be empty.

Next, for every $k=1,\ldots,\vdegen$, consider the following grid partition of $[m]^{2k}$ induced by $C_1,\ldots,C_{2k_v+1}$:
\begin{align*}
    \B_k(v) \coloneq \{C_1,\ldots,C_{2k_v+1}\}^{2k}\,.
\end{align*}
Thus, $\B_k(v)$ is a collection of axis-aligned boxes.
For every $B \in \B_k(v)$, the \emph{extension} of $B$ to $2 \vdegen$ dimensions is the box:
\begin{align}
    B^* = B \times \underbrace{\{0\} \times \dots \times \{0\}}_{2(\vdegen-k)}\,.
 \end{align}
For every such $B^*$ let $V(B^*) = \{v \in V : \vv \in B^*\}$.
We now identify the subset of boxes that have the property of containing the vectors of vertices in $N_0(v)$, that is, such that $V(B^*) \subseteq N_0(v)$.
\begin{definition}
Let $k\in\{1,\ldots,\vdegen\}$.
A box $B=C_{a_1}\times C_{b_1}\times\cdots\times C_{a_k}\times C_{b_k}\in\B_k(v)$ is a \emph{$v$-box} if $a_i\le 2j\le b_i$ for some $i\in[k]$ and $j\in[k_v]$. An extended box $B^*$ is a $v$-box whenever $B$ is.
\end{definition}
We let $\B(v) = \B_1(v) \cup \ldots \cup \B_{\kappa_v}(v)$, and we let $\B^*(v)$ be the set of all extensions of $v$-boxes of $\B(v)$.
At query time, the algorithm runs an orthogonal range query on every $v$-box $B^* \in \B^*(v)$, and returns the aggregate size of the output sets.

\subsubsection{Correctness}
The following lemma gives the correctness of the algorithm.
\begin{lemma}\label{lem:NC_VBox_new}
$N_0(v) = \biguplus_{B^* \in \B^*(v)} V(B^*)$, and in particular:
\begin{enumerate}
    \item For every $B^* \in \B^*(v)$ we have $V(B^*) \subseteq N_0(v)$.
    \item For every $u \in N_0(v)$ there exists a unique $B^* \in \B^*(v)$ such that $u \in V(B^*)$.
\end{enumerate}    
\end{lemma}
\begin{proof}

\medskip\noindent\uline{Item 1.}
Let $B^* = B \times \{0\},\dots,\{0\}$ where $B = C_{a_1}\times C_{b_1}\times\cdots\times C_{a_k}\times C_{b_k} \in \B_k(v)$ is a $v$-box.
Since $B$ is a $v$-box, there exist $i \in [k]$ and $j \in [k_v]$ such that $a_i\le 2j\le b_i$, which implies $I_{j}(v) = C_{2j} \in \{C_{a_i},\ldots,C_{b_i}\}$.
Now, suppose $\vu = (l_1,r_1,\ldots,l_{k_u},r_{k_u},0,\ldots,0) \in B^*$ for some $u \in V$.
Since $\vu \in B^*$, then necessarily $k_u=k$, and moreover $l_i \in C_{a_i}$ and $r_i \in C_{b_i}$ for every $i \in [k]$, where $[l_i,r_i]=I_i(u)$.
If $2j=a_i$, this implies $l_i \in C_{a_i}=I_j(v)$, and therefore $I_i(u) \cap I_j(v) \ne \emptyset$; the same conclusion holds if $2j=b_i$.
If instead $a_i<2j<b_i$, then we have $I_j(v) = C_{2j} \subseteq [l_i,r_i] = I_i(u)$, hence again $I_i(u) \cap I_j(v) \ne \emptyset$.
We conclude that $u \in N_0(v)$.

\medskip\noindent\uline{Item 2.} 
Let $u \in N_0(v)$.
Let $\vu=(l_1,r_1,\ldots,l_{k_u},r_{k_u},0,\ldots,0)$ be the vector associated to $u$, and for every $i=1,\ldots,k_u$ let $a_i,b_i \in [2k_v+1]$ be such that $l_i \in C_{a_i}$ and $r_i \in C_{b_i}$.
Note that such $a_i,b_i$ exist and are unique, since $l_i,r_i \in [m]$ and the intervals $C_1,\ldots,C_{2k_v+1}$ form a partition of $[m]$.
Finally, let $B = C_{a_1} \times C_{b_1} \times \ldots \times C_{a_k} \times C_{b_k}$, and let $B^*$ be its extension.
By construction, $B^*$ is unique and $\vu \in B^*$.
It remains to show that $B$ is a $v$-box.
To this end note that, since $u \in N_0(v)$, then there exist $i\in[k_u]$ and $j \in [k_v]$ such that $I_i(u) \cap I_j(v) \ne \emptyset$.
As $l_i \in C_{a_i}$ and $r_i \in C_{b_i}$ this implies $I_j(v) \in \{C_{a_i},\ldots,C_{b_i}\}$, and since $I_j(v)=C_{2j}$, then $a_i \le 2j \le b_i$ too.
\end{proof}

\subsubsection{Running time}
\begin{lemma}\label{lem:Neigh_Preprocess_time}
    Algorithm~\ref{alg:Neigh_Preprocess} can be run in time $2^{O(\vdegen)} |V| \log^{2\vdegen -1} |V|$.
\end{lemma}
\begin{proof}
    Building the vectors takes time $O(\vdegen |V|)$. Since the vectors lie in $\RR^{2\vdegen}$, the bounds follow from \Cref{thm:ORS} with $d = O(\vdegen)$.
\end{proof}

\begin{lemma}\label{lem:Neigh_time}
After executing \VertexNeighPre, Algorithm~\ref{alg:Neigh} can be run:
\begin{itemize}
    \item in the counting and sampling versions, in time $2^{O(\vdegen \ln \vdegen)} \log^{2\vdegen - 1} |V|$.
    \item in the listing version, with $O(1)$-delay per element, after an initial warm-up time of\linebreak $2^{O(\vdegen \ln \vdegen)} \log^{2\vdegen - 1} |V|$.
\end{itemize}
\end{lemma}
\begin{proof}
    The loop of line~\ref{line:Neigh:forB} performs $O\left((2k_v+1)^{2\vdegen}\right) = 2^{O(\vdegen \ln \vdegen)}$ iterations, each of which entails work $O(\vdegen)$ to check if $B$ is a $v$-box, and again work $O(\vdegen)$ to add the extended box $B^*$ to $\B^*$. Then, by \Cref{thm:ORS} with $d = O(\vdegen)$ and $P = V$ all the claimed bounds follow.
\end{proof}

\subsection{Computing $\mathbf{N_1(v)}$}\label{sec:NE}
This section proves the following result:
\begin{theorem}
    There exists an algorithm that, given a $(\vdegen,\edegen)$-interval representation of a weighted hypergraph $H=(V,E)$, in time $O(\edegen \vdegen \probsize H \log \probsize H)$ builds a data structure that, on query $v \in V$, can:
    \begin{itemize}
        \item count $N_1(v)$ in time $O(1)$,
        \item list $N_1(v)$ with $O(\log \probsize H)$ delay per element,
        \item sample from $N_1(v)$ in time $O(\log \probsize H)$.
    \end{itemize}
\end{theorem}

The proof performs a reduction to a particular variant of box stabbing, namely \emph{union stabbing}.
In the classical 1-d formulation, given a set of intervals $X_1,\ldots,X_N$ and a query point $x$, box stabbing asks to retrieve all intervals $X_i$ containing $x$. In union stabbing, instead, each $X_i$ is associated with a second interval $Y_i$, possibly over a different domain. Given $x$, the problem asks now to retrieve $\bigcup_{Y: x \in X} Y$, that is, the \emph{union} of all $Y_i$ whose $X_i$'s are stabbed.
To reduce the problem of computing $N_1(v)$ to union stabbing, define the set of 2-d vectors $N_1 \coloneq \{(v,u) : u \in N_1(v)\}$.
We show that this set can be encoded as the union of a family of axis-aligned rectangles $X_i \times Y_i$ over $V  \times V$.
At this point, $N_1(v)$ will be precisely the result of a union stabbing query on $v$.

Let us give some intuition about the construction of $N_1$.
Observe that every pair $(v,u) \in N_1$ has a witnessing edge in which at least one of the two incidences is represented through $\IE$.
Thus, to capture these adjacencies, we build rectangles by combining all intervals of $\IE$ either with individual vertices or with other intervals of $\IE$. The key observation is that the niceness property of \Def{interval_representation} allows us to capture potentially many adjacencies with a single rectangle. Moreover, the same property guarantees that each rectangle either consists entirely of adjacencies already captured by $N_0$, or contains none of them. Thus, by testing a single pair per rectangle, we can discard the redundant rectangles and obtain a family whose union represents exactly $N_1$. 

The rest of the section is dedicated to formalizing this intuition. 


\subsubsection{Preprocessing}\label{sec:N1_Preprocessing}
Preprocessing is composed of two phases. Let $(\bV, \bE,\IV,\IE)$ be a $(\vdegen, \edegen)$-interval representation of a hypergraph $H = (V,E)$. In the first phase, the procedure builds the family of rectangles representing exactly $N_1$. To this end, it first generates a family $\hat \R$. In general $\hat \R$ could contain pairs that have already been witnessed in $N_0$, thus with an additional scan over $\hat \R$ it computes the family $\R$. As we prove in \Cref{sec:N1_correctness}, for every $v \in V$,  $\R(v) = N_1(v)$. Hence, $\R$ contains all the information we need. In the latter part, the algorithm builds a data structure that supports efficient queries on $\R(v)$ for every $v \in V$. As anticipated in the introduction of this section, we achieve this by reducing the problem to a weighted union stabbing problem, which we formally define as follows. 

\begin{definition}\label{def:union_stabbing}
Let $\Pcal = \{P_1, \dots, P_N\}$ be a family of intervals over $[n]$, let $\cQ = \{Q_1, \dots, Q_N\}$ be a family of intervals over $[m]$, and let $\omega: [m] \to \NN$ be a weighting of $[m]$. For any $A \subseteq [m]$, let $\omega(A) = \sum_{u \in A} \omega(u)$. For $q \in [n]$, define $U(q) = \bigcup_{i: q \in P_i} Q_i$. A \emph{weighted union stabbing} query with $q \in [n]$ asks to:
\begin{itemize}
    \item return $\omega(U(q))$ (counting),
    \item list the elements of $U(q)$ (reporting),
    \item sample from the distribution $\left(\frac{\omega(u)}{\omega(U(q))}\right)_{u \in U(q)}$ (sampling).
\end{itemize}
\end{definition}

Note that union stabbing is a special case of the \emph{stabbing semigroup} problem (See for instance \cite{Agarwal2012AnOD}). In this problem, each interval $P_i$ is assigned to an element of a commutative semigroup $(\bS, +)$, and a query $q \in [n]$ returns the sum of the elements associated to the intervals containing $q$. Observe that, by taking $(\bS, +) = (2^{[m]}, \cup)$ and assigning $Q_i$ to $P_i$ for each $i$, each stabbing-semigroup query $q \in [n]$ returns exactly $U(q)$. However, standard semigroup bounds assume constant-time operations on constant-size elements, whereas representing arbitrary subsets of $[m]$ and computing their unions may require non-constant time and space. Thus, these bounds do not directly yield our counting, reporting, and sampling guarantees.
Nonetheless, we claim that there exists a data structure that supports counting/listing/sampling efficiently on any weighted union stabbing instance. The proof for this can be found in \Cref{sec:UnionStabbingProof}.


\begin{theorem}\label{theorem:UnionStabbing}
There exists a data structure that, given two families of intervals $\Pcal = \{P_1, \dots, P_N\}$, $\cQ = \{Q_1, \dots, Q_N\}$ over $[n]$ and $[m]$ and $\omega : [m] \to \NN$, can be built in time $O(n + m + N\log m)$ and on input $q \in [n]$ supports the following operations:
\begin{itemize}
    \item return $\omega(U(q))$ in time $O(1)$.
    \item list $U(q)$ in total time $O(|U(q)|\log m)$ and with $O(\log m)$ delay.
    \item sample from the distribution $\left(\frac{\omega(u)}{\omega(U(q))}\right)_{u \in U(q)}$ in time $O(\log m)$.
\end{itemize}
\end{theorem}

Now let $\R = \{X_i \times Y_i: i \in [|\R|]\}$ be the family of rectangles built by the algorithm, and let $\R(v) = \bigcup_{i : v \in X_i} Y_i$. Observe that $\R$ can be seen as a weighted union stabbing instance, with $\cP = (X_i)_{i \in [|\R|]}$, $\cQ = (Y_i)_{i \in [|\R|]}$ and weight function $w$. Thus, we apply \Cref{theorem:UnionStabbing} to preprocess this instance, and obtain a data structure $\cD$ capable of efficiently answering queries on $\R(v)$. 

The preprocessing procedure just described is summarized by Algorithm \ref{algo:N1_Preproc}. Similarly to \Cref{sec:NV}, we make use of $\primitive{US}$-$\primitive{Prec}$, $\primitive{US}$-$\primitive{Count}$, $\primitive{US}$-$\primitive{Report}$ and $\primitive{US}$-$\primitive{Sample}$ to refer to the respective union stabbing procedures.


\begin{algorithm}\label{algo:N1_Preproc}
    \caption{\BuildRectangles$(\bV,\bE,\IV,\IE, w)$}
    \label{alg:BuildRectangles}
    $\hat \R \gets \emptyset$\;
    \ForEach{$v \in V$}{
        \ForEach{$I \in \IV(v)$}{
            choose any $e \in I$\;
            \ForEach{$J \in \IE(e)$}{
                $\hat\R \gets \hat\R \cup \{\{v\}\times J, J\times\{v\}\}$\;
            }
        }
    }
    \ForEach{$e \in E$}{
        \ForEach{$J,J' \in \IE(e)$}{
            $\hat \R \gets \hat \R \cup \{ J \times J'\}$\;
        }
    }
    $\R \gets \emptyset$\;
    \ForEach{$R \in \hat \R$}{
        Let $R = X \times Y$, choose any $x \in X, y \in Y$\;
        \If{$\IV^\cup(x) \cap \IV^\cup(y) = \emptyset$}{
            $\R \gets \R \cup \{R\}$
        }
    }

    denote $\R=\{X_i\times Y_i:i\in[N]\}$, then $\cP \gets (X_i)_{i\in[N]}$ and $\cQ \gets (Y_i)_{i\in[N]}$\;
    $\cD \gets \primitive{US}$-$\primitive{Prec}(\cP,\cQ,w)$\;
    \Return $\mathcal D$\;
\end{algorithm}

\subsubsection{Query}
Let $\cD$ be the data structure returned by Algorithm \ref{algo:N1_Preproc}. By \Cref{theorem:UnionStabbing}, this structure supports efficient counting, listing and sampling on $\R(v)$ for every $v \in \bV$. Thus, given $v \in [n]$, we simply perform the corresponding weighted union stabbing query on $\cD$ and return the result. This is formalized in Algorithm \ref{algo:N1_Query}.

\subsubsection{Correctness}\label{sec:N1_correctness}
Let $\R$ be the family of rectangles built by Algorithm \ref{algo:N1_Preproc} on input $(\bV,\bE,\IV,\IE, w)$ and $\cD$ the returned data structure. We prove that $\R(v) = N_1(v) $ for every $v \in V$. Observe that this is sufficient to prove the correctness of all the procedures. Indeed, Algorithm \ref{algo:N1_Preproc} and Algorithm \ref{algo:N1_Query} respectively build and interrogate $\cD$, whose correctness is already guaranteed by \Cref{theorem:UnionStabbing}.

\begin{lemma}
    Let $(\bV, \bE,\IV,\IE)$ be the representation of $H = (V,E)$ given in input to Algorithm \ref{algo:N1_Preproc}. Then, for every $v \in V$, $\R(v) = N_1(v)$. 
\end{lemma}
\begin{proof}
    Let $\hat \R$ be the family built before filtering, let $X \times Y \in \hat \R$, and let $(x,y) \in X \times Y$ be the pair tested at Line 12. We claim that, for every $(x',y') \in X \times Y$,
    \begin{equation}\label{equation:N1_correcteness_lemma}
        y' \in N_0(x') \iff \IV^\cup(x) \cap \IV^\cup(y) \neq \emptyset.
    \end{equation}
    Note that each of $X$ and $Y$ is either a singleton or an interval of $\IE(e)$ for some $e \in E$. In both cases, by the niceness property of the representation, we have $\IV(x) = \IV(x')$ and $\IV(y) = \IV(y')$. Therefore, $\IV^\cup(x') \cap \IV^\cup(y') = \IV^\cup(x) \cap \IV^\cup(y)$, and the claim follows from the definition of $N_0(x')$. We now prove that $\R(v) = N_1(v)$.
    
    \noindent $(\subseteq)$ Let $u \in \R(v)$. By definition of $\R$, there exists $R \in \R$ such that $(v,u) \in R$ and $\IV^\cup(v') \cap \IV^\cup(u') = \emptyset$ for some $(v',u') \in R$. By construction, we have $\R \subseteq \hat \R$. Thus, by \Cref{equation:N1_correcteness_lemma}, we conclude that $u \notin N_0(v)$. Moreover, $u \in N(v)$, since every rectangle of $\R$ is built from incidences with a common edge. Hence, $u \in N(v) \setminus N_0(v) = N_1(v)$.
    
    \noindent $(\supseteq)$ Let $u \in N_1(v)$. Then there must exist an edge $e \in E$ containing $v$ and $u$ such that $v \in \IE^\cup(e)$ or $u \in \IE^\cup(e)$. We distinguish two cases. If both $v \in \IE^\cup(e)$ and $u \in \IE^\cup(e)$, let $J,J' \in \IE(e)$ be two intervals containing $v$ and $u$, respectively. Observe that the algorithm builds the rectangle $R = J \times J'$ at Line 9, which contains $(v,u)$. Otherwise, assume that $e \in \IV^\cup(v)$ and $u \in \IE^\cup(e)$, the symmetric case being analogous. Let $I \in \IV(v)$ be the interval containing $e$ and $J \in \IE(e)$ the one containing $u$. At Line 6, the algorithm selects an edge $e' \in I$ and builds $\{v\} \times J'$ for every $J' \in \IE(e')$. Since $e,e' \in I$, the niceness property guarantees that $\IE(e) = \IE(e')$, meaning that the rectangle $\{v\} \times J$ containing $(v,u)$ is built. Hence, in both cases, $u \in \hat \R(v)$. Moreover, $\IV^\cup(u) \cap \IV^\cup(v) = \emptyset$ since $u \notin N_0(v)$. Thus, by \Cref{equation:N1_correcteness_lemma}, the corresponding rectangle is not discarded, and consequently $u \in \R(v)$.
\end{proof}

\begin{algorithm}
    \caption{\primitive{N1}-\primitive{Query}$(\cD,v)$} \label{algo:N1_Query}
    for counting\;
    \Indp
        \Return $\primitive{US}$-$\primitive{Count}(\cD,v)$\;
    \Indm
    
    for sampling\;
    \Indp
        \Return $\primitive{US}$-$\primitive{Sample}(\cD, v)$\;
    \Indm

    for listing\;
    \Indp
        \Return $\primitive{US}$-$\primitive{Report}(\cD,v)$\;
    \Indm
\end{algorithm}

\subsubsection{Running time}
We first prove that the number of boxes built by Algorithm \ref{algo:N1_Preproc} is bounded by some function on the parameters $\vdegen, \edegen$. Then, we analyze the running time of the three algorithms.

\begin{lemma}\label{lem:NC_NE_boxes}
Let $(\bV, \bE,\IV,\IE,w)$ be the interval representation of $H=(V,E)$ in input to Algorithm \ref{algo:N1_Preproc} and let $\R$ be the family of filtered boxes, then $|\R| = O(\edegen \probsize H)$.
\end{lemma}
\begin{proof}
Let $\hat \R$ be the family of rectangles built at the end of Line 9. Observe that the algorithm builds two different types of rectangles:
\begin{itemize}
    \item for every $v \in V$ and every interval $I \in \IV(v)$ the algorithm picks $e_I \in I$. Then, for every $J \in \IE(e_I)$ it builds $\{v\} \times J$ and $J \times \{v\}$. Recall that, by definition, every $e_I$ has $O(\edegen)$ intervals $J \in \IE(e_I)$. In total, this yields 
    \begin{equation}
        2 \sum_{v \in V} \sum_{I \in \IV(v)} |\IE(e_I)| \leq 2\edegen \sum_{v \in V} |\IV(v)| = O(\edegen \probsize H)
    \end{equation} 
    rectangles.
    \item for every $e \in E$ and every pair of intervals $J,J' \in \IE(e)$, the algorithm builds $J \times J'$. Since every edge has $O(\edegen)$ intervals, a total of 
    \begin{equation}
        \sum_{e \in E} |\IE(e)|^2 \leq \edegen \sum_{e \in E} |\IE(e)| =  O(\edegen \probsize H)
    \end{equation}
    rectangles are built.
\end{itemize}
Since $\R \subseteq \hat \R$, we obtain $|\R| = O(\edegen \probsize H)$, as claimed.
\end{proof}

     

\begin{lemma}
    Let $(\bV, \bE,\IV,\IE,w)$ be the interval representation of $H =(V,E)$ given as input. Then Algorithm \ref{alg:BuildRectangles} can be run in time $O(\edegen \vdegen \probsize H \log \probsize H)$.
\end{lemma}
\begin{proof}
     We first bound the time required to build $\R$. Then, the time to preprocess $\R$ follows directly from \Cref{theorem:UnionStabbing}. Putting these two together proves the lemma.
     
     By \Lem{NC_NE_boxes}, at most $O(\edegen \probsize H)$ rectangles are built, each requiring $O(1)$ time. Let $R \in \hat \R$ and let $(x,y) \in R$ be the pair of elements chosen by the procedure. At Line 13, the algorithm tests whether $\IV^\cup(x) \cap \IV^\cup(y) = \emptyset$. Since $\IV^\cup(x)$ and $\IV^\cup(y)$ are ordered, and each contains $O(\vdegen)$ intervals, this can be done in time $O(\vdegen)$ with a linear scan. Putting everything together, building $\R$ requires time $O(\vdegen \edegen \probsize H)$. Now let $\R=\{X_i\times Y_i:i\in[N]\}$, where $N=|\R|$. The algorithm builds the data structure of \Cref{theorem:UnionStabbing} on the instance $\cP=(X_i)_{i\in[N]}$, $\cQ=(Y_i)_{i\in[N]}$, and weighting $w$. Since both families consist of intervals over $V$, by \Cref{theorem:UnionStabbing} the structure can be built in time $O(|V|+N\log |V|)$. Moreover, by \Lem{NC_NE_boxes}, $N\leq|\hat\R|= O(\edegen \probsize H)$, yielding $O(\edegen \probsize H \log \probsize H)$ time. The claimed bound follows.
\end{proof}


\begin{lemma}
    After executing $\BuildRectangles$, Algorithm \ref{algo:N1_Query} can be run on input $v \in V$
    \begin{enumerate}
        \item in the counting version in time $O(1)$,
        \item in the listing version in total time $O(|\R(v)|\log \probsize H)$, with $O(\log \probsize H)$ delay per element,
        \item in the sampling version in time $O(\log \probsize H)$.
    \end{enumerate}
\end{lemma}
\begin{proof}
    Immediate from \Cref{theorem:UnionStabbing}.
\end{proof}

\section{Lower Bounds}
\label{sec:LB}

In this section, we establish conditional lower bounds for many problems arising naturally in our framework, including $\PartialMatch$ (in both versions) and $\NC$, as well as all the primitives introduced in \Cref{sec:intro}, under the Orthogonal Vectors Conjecture (OVC) and the strong SetDisjointness Conjecture (SDC). We also provide a lower bound for the primitive $\CONTAIN$ based on Boolean Matrix Multiplication (BMM). 

We first recall the formulation of the Orthogonal Vectors Conjecture (OVC) used throughout this section.

\begin{conjecture}[Orthogonal Vectors Conjecture, see~\cite{AbboudWW15}]
There is no $\varepsilon>0$ such that, for all $c\geq 1$, there is an algorithm that, given a set $U\subseteq\{0,1\}^d$ of $n$ Boolean vectors, where $d=c\log n$, determines whether there exist two distinct $a,b\in U$ such that $a\cdot b=0$ in time $O(n^{2-\varepsilon})$.
\end{conjecture}
\noindent In particular, OVC rules out algorithms with running
time $2^{o(d)}n^{2-\varepsilon}$ for any fixed
$\varepsilon>0$.  We refer to the corresponding decision problem as Orthogonal Vectors (OV).

\begin{table}[H]
\centering
\small
\setlength{\tabcolsep}{4pt}
\renewcommand{\arraystretch}{1.45}

\begin{tabular}{|p{.15\linewidth}|p{.30\linewidth}|p{.2\linewidth}|p{.25\linewidth}|}
\hline
\textbf{Hypothesis} &
\textbf{Primitive} &
\textbf{Preprocessing time/space} &
\textbf{Query time} \\
\hline

OVC &
{\raggedright
\EdgePartialMatch/\newline\VertexPartialMatch, \VSat/\ESat, \SP, \ISH,
\NC, \UNION\par} &
$2^{o(\kappa)}\size H^{2-\delta}$ &
$2^{o(\kappa+t)}\lVert H\rVert^{1-\varepsilon}$ \\
\hline

SDC &
\ADJ, \VENN, \NeighInt  &
$\size H ^{1+o(1)}$ &
$\size H^{1/2-\varepsilon}$ \\
\hline

BMM &
\CONTAIN &
$\size H^{\omega_0/2-\delta}$ &
$\size H^{(\omega_0-2)/2-\varepsilon}$ \\
\hline
\end{tabular}

\caption{Summary of conditional lower bounds.
Each row rules out the simultaneous preprocessing
(or space) and query bounds shown, for fixed
$\varepsilon,\delta>0$. Here $\kappa$ denotes
bidegeneracy and $t$ denotes query size.
The SDC lower bound for \textsc{NeighINT} holds
even for constant bidegeneracy, whereas those for
\ADJ\ and \VENN\ hold for general hypergraphs.}
\label{tab:lower-bounds}
\end{table}

\noindent We next recall the SetDisjointness problem and
its associated conjecture:
\begin{definition}[SetDisjointness]
Given a collection of sets $\mathcal{S}=\{S_1,\ldots,S_m\}$ over a universe $U$, the SetDisjointness (SD) problem asks to preprocess $\mathcal{S}$ so that, for any two indices $i,j\in[m]$, one can determine whether $S_i\cap S_j=\emptyset$.
\end{definition}

\noindent The size of a SetDisjointness instance is defined
as $N= |U| +\sum_{i=1}^m |S_i|$.

\begin{conjecture}[Strong SetDisjointness Conjecture, see \citep{GoldsteinKLP17a}]
    Any data structure for the SetDisjointness problem with query time $T$ requires $S = \tilde \Omega(\frac{N^2}{T^2})$ space.
\end{conjecture}

\noindent Our results are summarized in \Cref{tab:lower-bounds}.

\subsection{Reductions from Orthogonal Vectors} \label{sec:lb-partial}

We prove a lower bound for an easier version of \EdgePartialMatch, \SP, where the negative set is empty. Observe that the same lower bound immediately applies to $\ISH$, since $\SP(S)>0$ for some $S \subseteq V$ if and only if $S$ appears in the output of $\ISH(S)$.

\begin{theorem} \label{sec:lb-contain}
Suppose there exist $\varepsilon, \delta>0$ such that for all $r > 0$, there is a data structure solving the decision version of $\SP(S)$ on $\degen$-bidegenerate hypergraphs
with preprocessing time $O(2^{\degen/r} \size{H}^{2-\delta})$ and query time $O(2^{(\degen+|S|)/r} \size{H}^{1-\eps})$.
Then OVC is false.
A similar conclusion holds for $\EdgeNeigh(e)$ and $\UNION(F)$ queries, with query time $O(2^{\degen/r}\size{H}^{1-\eps})$ and $O(2^{(\degen+|F|)/r}\size{H}^{1-\eps})$, respectively.
\end{theorem}

\begin{proof} Fix $c \ge 1$ and consider an OV instance with vectors $x_1,\ldots,x_n\in\{0,1\}^d$, where $d=c\log n$. 
    Let us consider these vectors to be subsets $S_1, S_2, \ldots, S_n \subseteq [d]$. The OV instance is a yes-instance
    iff $\exists i, j$ such that $S_i \subseteq \overline{S_j}$. Let us solve the OV instance by our data structure for
    $\SP$ queries. We first create the hypergraph $H$ corresponding to set system $\overline{S_1}, \overline{S_2}, \ldots, \overline{S_n}$.
    The hypergraph $H$ has $d$ vertices and $n$ hyperedges, and $\size{H} = O(nd)$. It is trivially $d$-bidegenerate and can be
    constructed in $O(nd)$ time. We will choose $r$ to be a sufficiently large constant depending on other (constant) parameters.
    The preprocessing time is $O(2^{d/r} (nd)^{2-\delta})$. To solve the OV instance, we invoke queries $\SP(S_1),\ldots,\SP(S_n)$. The OV instance is a yes-instance iff any of these queries returns a non-zero. The total query
    time is $O(2^{2d/r} (nd)^{2-\eps})$. 

    The overall running time is $O(nd + 2^{r^{-1} c\log n} (n\log n)^{2-\delta} + 2^{2r^{-1}c\log n} (n\log n)^{2-\eps})$. 
    For sufficiently large constant $r$ (depending on $c, \eps, \delta$), the running time is $O(n^{2-\alpha})$, for some constant $\alpha > 0$.
    Hence, OVC is false.

    For $\EdgeNeigh(e)$ queries, we can use a direct reduction. Observe that OV instance is a yes-instance iff there exists
    $i,j$ such that $S_i \cap S_j = \emptyset$. Equivalently, there exists some $i$ such that $\EdgeNeigh(S_i) < n $
    (so there exists some $S_j$ that it does not intersect). We can make $n$ $\EdgeNeigh$ queries to solve the OV
    instance. The running time math of the previous paragraph can be followed to complete the proof.

    For $\UNION$ queries, we use a dual reduction. We start from the same OV instance with vectors $x_1,\ldots,x_n\in\{0,1\}^d$, where $d=c\log n$ and $c \ge 1$. We create the hypergraph $H$ by adding one vertex $v_i$ for each $x_i$, and one hyperedge $e_{\ell} = \{v_i: x_i[\ell] = 1\}$ for each $\ell \in [d]$, and identifying twin edges. $H$ is trivially $d$-bidegenerate and can be constructed in $O(nd)$ time. Consider the set $F_i = \{e_{\ell}: x_i[\ell] = 1\}$. We have $v_j \in \bigcup_{e \in F_i}e \iff \exists \ell : x_i[\ell] = x_j[\ell] = 1 \iff x_i \cdot x_j > 0$, so that $\UNION(F_i) = |\{j \in [n]: x_i \cdot x_j > 0\}|$. To solve the OV instance, we create $n$ $\UNION(F_i)$ queries, where $i \in [n]$. The OV instance is a yes-instance iff $ \UNION(F_i) < n$ for some $i$. Again, the same arguments as the previous paragraphs complete the proof.
\end{proof}

\subsection{Reductions from SetDisjointness}

We first consider a refined version of \NC\ involving the neighborhoods of multiple vertices: given a query set $P\subseteq V$, we want to compute the cardinality of the intersection of their neighborhoods, $|\bigcap_{v\in P}N(v)|$.
We show that, assuming SDC, such queries cannot be supported with polylogarithmic time and near-linear space, even on hypergraphs of constant bidegeneracy. In fact, the lower bound already holds for $|P| = 2$, even when the task is merely to determine whether their neighborhoods intersect.
This shows that the running time bounds we achieve for \NC\ cannot be generalized to a ``neighbor version'' of \PartialMatch.

\begin{definition}[\NeighInt] 
Given a hypergraph $H=(V,E)$, a \NeighInt\ (neighborhood intersection) query takes in input a set of vertices $P\subseteq V$ and asks to compute $|\bigcap_{v\in P}N(v)|$.
\end{definition}

\begin{theorem}
    Assuming SDC, for every $\varepsilon>0$, there is no data structure for \NeighInt\ on $\kappa$-bidegenerate hypergraphs with space $2^{\widetilde O(\kappa)}\size{H}^{1+o(1)}$ and query time $2^{\widetilde O(\kappa)}\size{H}^{\frac{1}{2}-\varepsilon}$.
\end{theorem}

\begin{proof}
    Let $\mathcal S = \{S_1, \dots, S_m\}$ over a universe $U$ be an instance of SD. We construct a bipartite graph $B = (V, E)$ in the following way: for every $v \in U$, we add a vertex $v$ to $V$; for every $S_i \in \mathcal S$, we add a vertex $a_i$ to $V$; then, we add edges $\{u, a_i\}$ to $E$ if and only if $u \in S_i$. The resulting graph is clearly $2$-bidegenerate, has size $\size{B} = O(|U| +\sum_{i=1}^m |S_i|) = O(N)$ and we immediately observe that $N(a_i) \cap N(a_j) = S_i \cap S_j$. So, one single query of $\NeighInt$ with $P=\{a_i, a_j\}$ can determine the size of that intersection, and answer the corresponding SD query. A data structure with space $2^{\widetilde O(\kappa)}\size{H}^{1+o(1)} = O(\size{B}^{1 + o(1)}) = O(N^{1 + o(1)}) = S$ and query time $2^{\widetilde O(\kappa)}O(\size{H}^{\frac{1}{2}-\varepsilon}) = O(\size{B}^{\frac{1}{2}-\varepsilon}) = O(N^{\frac{1}{2} - \varepsilon}) = T$ would be able to answer SD with $ST^2 = O(N^{2-2\varepsilon + o(1)})$, contradicting SDC. 
\end{proof}

\noindent Then, we consider $\ADJ$ and $\VENN$ and show that, assuming SDC, such queries cannot be supported with polylogarithmic time and near-linear space on general hypergraphs.
\begin{theorem}
Assuming SDC, for every $\varepsilon>0$, there is no data structure for $\ADJ$ or $\VENN$ on general hypergraphs with space $\size{H}^{1+o(1)}$ and query time $O(\size{H}^{1/2-\varepsilon})$.
\end{theorem}
\begin{proof}
Let $\mathcal S = \{S_1, \dots, S_m\}$ over a universe $U$ be an instance of SD. 
For $\ADJ$, we construct a hypergraph $H$ with one vertex $v_i$ for each $S_i$ and one hyperedge $e_x$ for each $x\in U$, where $v_i\in e_x$ if and only if $x\in S_i$. Then $\ADJ(v_i,v_j)$ is a yes-instance iff $|S_i\cap S_j| > 0$.

For $\VENN$, we construct the hypergraph $H = (U, \{S_i\}_{i \in [m]})$ and observe that any query $\VENN(S_i, S_j, S_k)$ allows us to compute $|S_i \cap S_j|$ by summing $|(S_i \cap S_j) \setminus S_k|$ and $|S_i \cap S_j \cap S_k|$.

Both constructions take $O(N)$ time and each SD query can be answered with a single query to the corresponding primitive. The claimed bounds follows from SDC.
\end{proof}

\subsection{Reductions from Boolean Matrix Multiplication} \label{sec:bool-matmult}

We let $\omega_0$ denote the Boolean matrix multiplication exponent. Consider answering \CONTAIN\ queries: given hyperedges $e, e'$, determine if $e \subseteq e'$. 

\begin{theorem} \label{thm:lb-contain} Consider a data structure for general hypergraphs for answering \CONTAIN\ queries with preprocessing time
$O(\size{H}^{\alpha})$ and worst-case query time $O(\size{H}^{\delta})$.
Then, either $\alpha \geq \omega_0/2$ or
$\delta \geq (\omega_0-2)/2$.
\end{theorem}

\begin{proof} Consider the problem of multiplying two Boolean $n \times n$ matrices $A$ and $B$. It is convenient to think in terms of $A^T$ and $B$, where the matrix product corresponds to all pairwise dot products of the column vectors.
We create a hypergraph $H$ where the vertex set correspond to the coordinates of the vectors. We define two sets of hyperedges $E_A$ and $E_B$. We treat each column vector of $A^T$ as an incidence vector of a hyperedge of $E_A$. 
We take the \emph{complement} of each column vector of $B$ to get the hyperedges of $E_B$.

Observe that for $e \in E_A$ and $e' \in E_B$, the corresponding matrix product entry is $1$ iff $e \not\subseteq e'$. To compute the matrix product, we construct the hypergraph $H$ and set up the data structure for answering \CONTAIN\ queries. We make query for all pairs of hyperedges $e \in E_A$ and $e' \in E_B$.
The total running time is $O(n^2 + P + n^2 Q)$, where $P$ is the preprocessing time and $Q$ is the query time. Suppose that $P=O(\size{H}^{\alpha})$ and
$Q=O(\size{H}^{\delta})$. Since $\size{H} = O(n^2)$, we get the total running time to be $O(n^2+n^{2\alpha}+n^{2+2\delta})$. We must have
$\max\{2,2\alpha,2+2\delta\}\geq\omega_0$. Hence, either $\alpha \geq \omega_0/2$ or $\delta \geq \frac{\omega_0 - 2}{2}$. 
\end{proof}

\appendix

\section{Auxiliary Proofs}

\subsection{Implications in \Cref{fig:implications}}\label{apx:implications}
In what follows, $M_H \in \{0,1\}^{n \times m}$ denotes the incidence matrix of a hypergraph $H$ with $n$ vertices and $m$ edges.

\medskip\noindent\uline{Niceness~\cite{bressan2026countinghypergraphletscolorcoding}.}
$H=(V,E)$ is $(\alpha,\beta)$-nice if $E = E_1 \uplus E_2$ such that $(V,E_1)$ has rank at most $\alpha$ and $(V,E_2)$ has degree at most $\beta$.
It is immediate to see that hypergraphs of rank at most $\alpha$ and hypergraphs of degree at most $\beta$ are proper special cases of $(\alpha,\beta)$-nice hypergraphs.
Moreover, if $H$ is $(\alpha,\beta)$-nice, then every submatrix of $M_H$ contains a column of weight at most $\alpha$ or a row of weight at most $\beta$; thus $H$ is $(\beta,\alpha)$-bidegenerate.
To see that the implication is strict, take the ``sunflower'' hypergraph on $V=[2n]$, where for every $i \in [n]$ there is a hyperedge $\{i\} \cup \{n+1,\ldots,2n\}$. This hypergraph is not $(\alpha,\beta)$-nice for any $\alpha,\beta < n$, yet it is $1$-bidegenerate: every twin-free submatrix of $M_H$ either contains a row in $[n]$, which has weight $1$, or it has a single row, and therefore a column of weight at most $1$.

\medskip\noindent\uline{$\infty$-degeneracy (\cite{PaSe26}).}
The $\infty$-degeneracy can be defined as the smallest $k$ such that every twin-free submatrix obtained by deleting rows from $M_H$ contains a row of weight at most $k$. Since deleting columns cannot increase the weight of the rows, this implies $(k,0)$-bidegeneracy and thus $k$-bidegeneracy. To see that the containment is strict, the complete graph $K_n$ has $\infty$-degeneracy $n-1$ but bidegeneracy $2$.

\medskip\noindent \uline{Hyperedge degeneracy~\cite{paul-pena_counting_2026}.} This is a stronger notion of $\infty$-degeneracy that keeps duplicates. The hyperedge degeneracy is the smallest $k$ such that every submatrix obtained by deleting rows from $M_H$ contains a row of weight at most $k$. The $\infty$-degeneracy is therefore at most the hyperedge degeneracy. 

\medskip\noindent\uline{$(k,q)$-cores~\cite{LeGo+23}.}
A $(k,q)$-core is the (hypergraph of) a maximal submatrix of $H$ where all rows have weight at least $k$ and all columns have weight at least $q$. Note that this does \emph{not} require the submatrix to be twin-free. Thus, if $H$ has no $(k,q)$-core, then $H$ is $(k,q)$-bidegenerate. To see that the inclusion is strict, just take $H$ for $M_H = \pmb{1}_n \times \transp{\pmb{1}_n}$; it has an $(n,n)$-core but it is $1$-bidegenerate.

\medskip\noindent\uline{Bidegeneracy of the incidence graph~\cite{ChenL00021}}.
The bidegeneracy of a bipartite graph is defined as the maximum, over all subgraphs, of the smallest $(\le 2)$-hop neighborhood of a vertex. In particular, it bounds the minimum degree of every subgraph. Applied to the incidence graph of $H$, it bounds the $(k,k)$-core number of $H$.

\subsection{Implementing primitives} \label{sec:primitives}

We show that list of primitives in the introduction can be implemented
using $\ESat$ or $\VertexNeigh$ queries.

\begin{asparaitem}
\item \ADJ: given $u,v \in V$, decide whether there exist a hyperedge containing both $u$ and $v$. This is the $\ESat$ query $x_u \wedge x_v$.
\item \CONTAIN: given hyperedges $e, e' \in E$, decide whether $e \subseteq e'$.
We can compute the size of $e \setminus e'$ with the dual of an $\ESat$ query,
which is $y_e \wedge \overline{y_{e'}}$.
\item \EdgeNeigh: given $e$, return the number of edges intersecting $e$. This is the dual of a $\VertexNeigh$ query.
\item \SP: given vertex subset $S \subseteq V$, count the number of hyperedges $f \supseteq S$. This is the $\ESat$ query $\bigwedge_{s \in S} x_s$.
\item \ISH: return the subhypergraph induced by $S$. Specifically, list out all distinct, non-empty sets $e \cap S$ (over hyperedges $e$). We make a \SP{} query over all subsets $S' \subseteq S$ and recover the intersection counts by inclusion-exclusion.
\item \UNION: given a subset $F$ of hyperedges, compute the union size $|\bigcup_{f \in F} f|$. This is the dual of an $\ESat$ query $\bigvee_{f \in F} x_f$.
\item \VENN: given three hyperedges $e,f,g$, compute the ``Venn diagram'' of all intersection patterns between these hyperedges. This can be expressed as $7$ different dual $\ESat$ queries, one for each combination of including/excluding each hyperedge (minus the all excluding $\overline{y_e} \wedge \overline{y_f} \wedge \overline{y_g}$).
\item \EdgePartialMatch: Given a set $P$ of "positive vertices" and a set $N$ of "negative vertices", count the number of hyperedges that contain $P$ and have no vertex of $N$. This is again an $\ESat$ query.
\end{asparaitem}

\section{Results on Orthogonal Range Searching}\label{sec:ORQ}
Let $P = (\bp_1,\ldots,\bp_n) \subseteq \mathbb R^d$ be an indexed family of points and $w : P \to \NN$ a weight function. With a slight abuse of notation, we use set notation for indexed families.  For every $X \subseteq P$, denote by $w_X$ the distribution over $X$ that assigns probability $w(x) / \sum_{x' \in X} w(x')$ to $x \in X$. An \emph{orthogonal range search} query is specified by an axis-aligned $d$-dimensional box $B$.
The output of the query comes in three flavors, defined as follows:
\begin{itemize}
    \item In the \emph{counting} version, $w(P \cap B)=\sum_{x \in P \cap B} w(x)$.
    \item In the \emph{sampling} version, (the index of) an element drawn from $w_{P \cap B}$.
    \item In the \emph{reporting} version, (the indices of) the set $P \cap B$ itself.
\end{itemize}
Standard techniques based on range trees~\cite{AgarwalErickson1999RangeSearching, deBerg2008ComputationalGeometry} can be adapted to obtain bounds on all three flavors. The following theorem states these bounds with explicit dependence on the dimension $d$. The classical bounds are immediately recovered when $d$ is constant.
\begin{theorem}[Orthogonal range searching]\label{thm:ORS}
Given an indexed family of points $P \subseteq \mathbb R^d$ and a weight function $w : P \to \NN$, in time $2^{O(d)}|P| \log^{\max(1, d-1)} |P|$ one can construct a data structure that, on input an axis-aligned $d$-dimensional box $B$, supports the following operations:
\begin{itemize}
    \item Returning $w(P \cap B)$ in time $2^{O(d)}\log^{\max(1, d-1)}|P|$.
    \item Sampling from $w_{P \cap B}$ in time $2^{O(d)}\log^{\max(1, d-1)}|P|$.
    \item Listing $P \cap B$ with $O(1)$ delay, after an initial warm-up time of $2^{O(d)}\log^{\max(1, d-1)}|P|$.

\end{itemize}
\end{theorem}

\begin{proof}
We adapt layered range trees (see \cite{deBerg2008ComputationalGeometry}). Let us first recall how queries on range trees work. The case $d=1$ is a simple binary search on an interval, and the bounds follow accordingly. 

Let $d \ge 2$. For a set of $n$ $d$-dimensional points $P$ with coordinates $x_1, \dots, x_d$, a $d$-dimensional range tree is defined recursively. The top level is a balanced BST on the first coordinate $x_1$. Every node $\nu$ represents a canonical subset $P(\nu)$ and stores an associated $(d-1)$-dimensional range tree $T_{d-1}(P(\nu))$ built on coordinates $x_2, \dots, x_d$. Given a query box $B = [a_1, b_1] \times \dots [a_d, b_d]$, the query procedure first decomposes $[a_1, b_1]$ into $O(\log n)$ canonical subsets in the top-level tree. Then it recursively processes the trees associated with these nodes, moving to $x_2$, then to $x_3$, and so on, until it reaches the bottom level, which is a BST ordered according to the last coordinate $x_d$. We note that this last structure can equivalently be represented as a sorted array. At each of the first $d-1$ dimensions, the query interval can be decomposed into at most
$2\lceil\log_2 n\rceil$ canonical subsets.
Consequently, at most
$(2\lceil\log_2 n\rceil)^{d-1} = 2^{O(d)}\log^{d-1}n$
bottom-level arrays are reached, and they
are pairwise disjoint. The last slice $[a_d, b_d]$ locates the endpoints $\ell, r $ of a contiguous sub-interval in each array A (possibly empty), and all the points whose $x_d$ coordinate is in $A[\ell, r]$ are returned. In the counting version, we simply compute the size of each such sub-interval. The recursive procedure visits all these arrays and yields $P \cap B$, or its cardinality $|P \cap B|$. A $d$-dimensional range tree can be constructed in $O(2^{O(d)}n\log^{d-1}n)$ time, and, using fractional cascading, supports unweighted counting queries in $O(2^{O(d)}\log^{d-1}(n))$ time and reporting queries in $O(2^{O(d)}\log^{d-1}n +|P \cap B|)$ time \cite{AgarwalErickson1999RangeSearching, deBerg2008ComputationalGeometry}.
Additionally, we observe that, for every array $A = (p_1, \dots , p_h)$, we can preprocess the prefix sum $S_{A}(j) = \sum_{k=1}^{j} w(p_{k})$ and store it with each entry $A[j]$. Since the total size of all bottom-level arrays is $2^{O(d)}n\log^{d-1}n$, this can be done in $2^{O(d)}n\log^{d-1}n$ time.

Consider now a query box $B = [a_1, b_1] \times \dots [a_d, b_d]$. Let $A_1, \dots, A_s$ be the bottom-level arrays reached by the recursive procedure on input $B$, and, for $i \in [s]$, denote by $(\ell_i, r_i)$ the endpoints located by the slice $[a_d, b_d]$ in the array $A_i$.

A \emph{linked-list summary} of $P \cap B$ is a doubly linked list containing one node for each nonempty sub-interval $A_i[\ell_i,r_i]$. Each node stores four pointers: two pointers $\mathsf{prev}$ and $\mathsf{next}$, respectively to the previous and the next nodes of the list; two pointers $\mathsf{first}$ and $\mathsf{last}$, respectively to $A_i[\ell_i]$ and $A_i[r_i]$. We observe that, since the nonempty sub-intervals are pairwise disjoint, their union is exactly $P \cap B$.

\begin{claim}
In time $2^{O(d)}n\log^{d-1} n$ one can preprocess a range tree so that, on input an axis-aligned $d$-dimensional box $B$, one can return a linked-list summary of $P \cap B$ in time $2^{O(d)}\log^{d-1} n$.
\end{claim}

\begin{proof}
    We run the whole recursive query procedure with $B$ as input. Let $A_1, \dots, A_s$ be the arrays reached at the bottom level. For each $i \in [s]$, $A_i[\ell_i]$ and $A_i[r_i]$ can be located in $O(1)$ additional time through fractional cascading. We do not enumerate the points $P \cap B$. Rather, for each sub-interval, we determine whether it is nonempty in $O(1)$ time and, if so, we add a corresponding node in the linked list. Since at most $2^{O(d)}\log^{d-1} n$ arrays are reached, the whole linked-list summary is constructed in $2^{O(d)}\log^{d-1} n$ time.
\end{proof}

\begin{claim}
Given a linked-list summary of $P \cap B$, one can return $w(P \cap B)$, list $P \cap B$, or sample from $w_{B \cap P}$ with the guarantees of \Cref{thm:ORS}.
\end{claim}

\begin{proof}
    To compute $w(P \cap B)$, we traverse the linked-list summary and, for each $i$, add the sub-interval weight $w(A_i[\ell_i, r_i]) = S_{A_i}(r_i) -S_{A_i}(\ell_i) + w(A_i[\ell_i])$ to an accumulator $W_B$. At the end of the traversal, we have $W_B = w(P \cap B)$.
    
    To list the elements of $P \cap B$, we traverse the linked-list summary and, for each node, scan the associated sub-interval from $A[\ell_i]$ to $A[r_i]$. The delay is $O(1)$ in the worst case: moving to the next element within a sub-interval takes constant time, and when the end of a sub-interval is reached, the pointer to the next node gives direct access to the first element of the next nonempty sub-interval. 
    
    To sample from $w_{P \cap B}$, we traverse the linked-list summary to obtain the weights $W_i :=w(A_i[\ell_i, r_i])$ and $W_B = \sum_i W_i $, as in the counting version. First, we choose a sub-interval proportionally to its weight. Then, we sample a random value in the cumulative-weight range of that sub-interval and find the point $p$ containing that weight through a binary search. Each point of $P \cap B$ is sampled with probability $\Pr(p) = \frac{W_i}{W_B} \cdot \frac{w(p)}{W_i} = \frac{w(p)}{W_B}$. The whole procedure takes time $O(2^{O(d)}\log^{d-1}n + \log n) = O(\log^{\max(1, d-1)}n)$.
\end{proof}
\end{proof}

We observe that the sampling procedure of \Cref{thm:ORS} does not guarantee constant sampling time. We show next that, by paying an additional logarithmic factor in both the preprocessing and the warm-up time, weighted samples from $P \cap B$ can be generated in $O(1)$ time.

\begin{corollary}
    Let $P$ be an indexed family of $n$ points in $\mathbb R^{d \ge 1}$, and let $w: P \rightarrow \mathbb N$ be a weight function. There exists a data structure that can be constructed from $P$ in $2^{O(d)}n \log^{d} n$ time and, on input an axis-aligned $d$-dimensional box $B$, supports weighted sampling from $P \cap B$ in warm-up time $2^{O(d)}\log^d n$ and, thereafter, $O(1)$ time per sample. 
\end{corollary}

\begin{proof}
    We start by constructing a $d-$dimensional range tree on $P$, as in \Cref{thm:ORS}. Let $\mathcal T$ be the set of all trees at the bottom level. For every $T \in \mathcal T$, and for every node $v \in T$, we precompute (bottom-up) $W(v) = \sum_{v' \in P(v)} w(v')$, that is, the sum of the weights of all the points in its canonical subset $P(v)$. For each node such that $W(v) > 0$, we also construct an alias table for the points in $P(v)$ (this can be done in $O(|P(v)|)$ time). Since every point occurs in at most $1 + \lceil\log n\rceil$ canonical subsets at each of the first $d-1$ levels, we have $\sum_{T \in \mathcal T}|T| \leq n(1+ \lceil\log n\rceil)^{d-1} = O(2^{O(d)}n\log^{d-1}n)$; furthermore, within each bottom-level tree, every point occurs in at most $1 + \lceil\log n\rceil$ alias tables. Hence, the total size of all alias tables is $2^{O(d)}n\log^d n$, and the whole preprocessing takes time $2^{O(d)}n\log^d n$.
    
    Consider now a query box $B = [a_1, b_1] \times \dots \times [a_d, b_d]$. We run the standard query procedure on the range tree until the bottom level. At most $2^{O(d)}\log^{d-1} n$ bottom-level trees are reached. In each of them, the interval $[a_d,b_d]$ is decomposed into at most $2\lceil\log n\rceil$ canonical subsets. The total number of canonical subsets in all trees is, therefore, $q := 2^{O(d)}\log^d n$. We observe that they are pairwise disjoint and their union is exactly $P \cap B$. Let $v_1, \dots ,v_q$ be the nodes representing these canonical subsets. In the warm-up phase, we compute $W_B = \sum_{i=1}^q W(v_i)$
    and create a second alias table on the $v_i$, so that $\Pr(v_i) = \frac{W(v_i)}{W_B}$. This can be done in $2^{O(d)}\log^d n$ time. Note that $W_B = \sum_{i=1}^q W(v_i) = \sum_{i=1}^q \sum_{p \in P(v_i)} w(p) = \sum_{p \in P \cap B} w(p)$. 
    
    To sample from $w_{P \cap B}$, we first sample a node $v_i$ with probability proportional to its weight $W(v_i)$ using the second alias table. Then, we sample a point $p$ from $P(v_i)$ using the precomputed alias table. Both operations take $O(1)$ time. It holds $\Pr(p) = \frac{W(v_i)}{W_B}\cdot \frac{w(p)}{W(v_i)} = \frac{w(p)}{W_B}$ for every $p \in P \cap B$. 
\end{proof}

\section{Results on Stabbing}\label{sec:Stabbing}
Let $\B$ be a family of $N$ axis-parallel boxes in $\RR^d$, and let $w:\B\to\NN$ be a weight function. For a point $q\in\RR^d$, define $\B(q)=\{B\in\B:q\in B\}$. For any $\B'\subseteq\B$, let $w(\B')=\sum_{B\in\B'}w(B)$.

\begin{definition}
    Given a family $\B$ of axis-parallel boxes in $\RR^d$ and a weight function $w:\B\to\NN$, a \emph{weighted stabbing} query with $q\in\RR^d$ asks to:
    \begin{itemize}
        \item compute $w(\B(q))$ (counting),
        \item list the elements of $\B(q)$ (reporting),
        \item sample from the distribution
        $\left(\frac{w(B)}{w(\B(q))}\right)_{B\in\B(q)}$
        (sampling).
    \end{itemize}
\end{definition}
It is the weighted version of a classical problem in computational geometry, see for instance \cite[Section 10.4]{deBerg2008ComputationalGeometry} and \cite[Section 1]{Chazelle1983Filtering}. With an approach similar to the one of \Cref{sec:ORQ}, we can adapt classical stabbing techniques to solve these three versions of the weighted variant. Ultimately, we prove the following theorem.

\begin{theorem}\label{thm:BoxStabbing}
    For any family of axis-parallel boxes $\B=\{B_1,\dots,B_N\}$ in $\RR^d$ and weight function $w:\B\to\NN$, in time $2^{O(d)}N\log^d N$ one can build a data structure that, given $q\in\RR^d$, supports the following operations:
    \begin{itemize}
        \item return $w(\B(q))$ in time $2^{O(d)}\log^d N$,
        \item list $\B(q)$ with $O(1)$ delay after an initial warm-up time of $2^{O(d)}\log^d N$,
        \item sample from the distribution $\left(\frac{w(B)}{w(\B(q))}\right)_{B\in\B(q)}$ in time $2^{O(d)}\log^d N$.
    \end{itemize}
    Moreover, for counting only, a $\log N$ factor can be removed from both preprocessing and query.
\end{theorem}

We recall the definition of weighted union stabbing introduced in \Cref{sec:N1_Preprocessing}.

\begin{definition}\label{def:appendix_union_stabbing}
Let $\Pcal$ be a family of intervals over $[n]$, let $\cQ$ be a family of intervals over $[m]$, and let $\omega: [m] \to \NN$ be a weighting of $[m]$. A \emph{weighted union stabbing} query with $q \in [n]$ asks to:
\begin{itemize}
    \item return $\omega(U(q))$ (counting),
    \item list the elements of $U(q)$ (reporting),
    \item sample from the distribution $\left(\frac{\omega(u)}{\omega(U(q))}\right)_{u \in U(q)}$ (sampling).
\end{itemize}
\end{definition}

We prove that there exists a data structure that supports these three operations efficiently on any weighted union stabbing instance.

\begin{theorem}\label{theorem:appendix_UnionStabbing}
For any family of intervals $\Pcal = \{P_1, \dots, P_N\}$, $\cQ = \{Q_1, \dots, Q_N\}$ over $[n]$ and $[m]$ respectively, and $\omega : [m] \to \NN$, there exists a data structure that can be built in time $O(n + m + N\log m)$ and, on input $q \in [n]$, supports the following operations:
\begin{itemize}
    \item return $\omega(U(q))$ in time $O(1)$.
    \item list $U(q)$ in total time $O(|U(q)|\log m)$ and with $O(\log m)$ delay.
    \item sample from the distribution $\left(\frac{\omega(u)}{\omega(U(q))}\right)_{u \in U(q)}$ in time $O(\log m)$.
\end{itemize}
\end{theorem}

The rest of the section is organized as follows. In \Cref{{sec:SegmentTrees}} we recall how a segment tree works, this is the standard structure to answer stabbing queries. Then, in \Cref{sec:BoxStabbingProof} and \Cref{sec:UnionStabbingProof} we show how to modify this structure in order to support efficient weighted stabbing and weighted union stabbing query, effectively proving \Cref{thm:BoxStabbing}  and \Cref{theorem:appendix_UnionStabbing}.



\subsection{Segment tree}\label{sec:SegmentTrees}
In this section, we briefly recall how multidimensional segment trees work.

Let us first recall how one-dimensional segment trees work. Let $\I$ be a family of $N$ intervals over $\RR$. By standard coordinate compression, we reduce it to a problem in the discrete domain $[n]$, with $n = O(N)$, in time $O(N \log N)$. A \emph{segment tree} $T_n$ is a balanced binary tree where each node $x$ is associated with an interval $J_x \subseteq [n]$. The root represents $[n]$. A node $x$ with interval $J_x = [a,b]$ is internal if $a < b$, in which case its left child has interval $[a, \lfloor\frac{a+b}{2}\rfloor]$ and its right child has interval $[\lfloor\frac{a+b}{2}\rfloor + 1, b]$. Each node keeps pointers to its children (if any). Leaves have singleton intervals $\{p\}$, with $p \in [n]$. Hence $T_n$ has $O(n)$ nodes and height $O(\log n)$. Each interval $I \in \I$ is stored at its \emph{canonical nodes}, namely the nodes $x$ such that $J_x \subseteq I$ and either $x$ is the root or $J_{x'} \not\subseteq I$, where $x'$ is the parent of $x$. We denote this set by $\Can(I)$. The intervals associated with the nodes in $\Can(I)$ form a partition of $I$, and $|\Can(I)|=O(\log n)$. Thus, every input interval is stored at $O(\log n)$ nodes. Given $q \in [n]$, the stabbing query follows the root-to-leaf path corresponding to $q$ and returns the intervals stored at the visited nodes. The query thus visits $O(\log n)$ nodes and retrieves $O(\log n)$ pairwise disjoint lists whose union is exactly $\I(q)$. The segment tree, including the lists of stored intervals, can be built in $O(N \log N)$ space and time.

Now let $\B = \{B_1, \dots, B_N\}$ be a family of axis-aligned boxes in $\RR^d$, where $B_i = I_{i,1} \times \cdots \times I_{i,d}$. A $d$-dimensional segment tree storing $\B$ is defined recursively as follows. First, a one-dimensional segment tree over the first coordinate is built, where each box $B_i$ is assigned to $\Can(I_{i,1})$. Recursively, at step $j$, every node $x$ of a segment tree on the $j$-th coordinate stores an associated $(d-j)$-dimensional segment tree built on $B_i' = I_{i,j+1} \times \cdots \times I_{i,d}$ for each box $B_i$ assigned to $x$. Again, by standard coordinate compression, the tree can be built in $2^{O(d)}N\log^d N$ time.
Let $q = (q_1, \dots, q_d) \in \RR^d$ be a query point, the query procedure follows the root-to-leaf path of $q_1$ on the top-level segment tree, and recursively does the same thing on the next coordinate for each tree associated to the visited nodes. Intuitively, after processing the first $j$ coordinates, the visited structures represent exactly the boxes $B_i$ such that $(q_1, \dots, q_j) \in I_{i,1} \times \cdots \times I_{i,j}$. Since a box can appear at most once in every root-to-leaf path, when the procedure reaches the bottom level, the lists contain disjoint elements and their union is exactly $\B(q)$. Since $O(\log n)$ nodes are visited in each path, at most $2^{O(d)}\log^d N$ bottom-level lists are reached. We summarize all these standard facts in the next remark.

\begin{remark}\label{remark:SegmentTree}
    Let $\B = \{B_1, \dots, B_N\}$ be a family of axis-aligned boxes in $\RR^d$. A $d$-dimensional segment tree storing $\B$ can be built in time and space $2^{O(d)}N\log^d N$. Given $q \in \RR^d$, the tree identifies at most $2^{O(d)}\log^d N$ pairwise disjoint lists whose union is exactly $\B(q)$, in time $2^{O(d)}\log^d N$.
\end{remark}

In the next section, we show a simple way to adapt this standard structure in order to support counting, listing and sampling weighted stabbing queries.

\subsection{Proof of \Cref{thm:BoxStabbing}}\label{sec:BoxStabbingProof}
Let $T(\B,\omega)$ be the $d$-dimensional segment tree containing $\B$. Remark~\ref{remark:SegmentTree} reminds us that this can be done in time $2^{O(d)}N\log^d N$. We claim that, by spending additional $2^{O(d)}N\log^d N$ time in the preprocessing phase, $T(\B,\omega)$ is able to answer weighted stabbing queries within the claimed bounds. The proof is organized as follows: we first exhibit the preprocessing procedure and then prove that the newly obtained structure can efficiently answer to stabbing queries.

\medskip\noindent\uline{Preprocessing.} Let $L_x$ be a bottom-level list associated with a node $x$. Recall that, by construction, $L_x$ stores the boxes $B_{i_1}, \dots, B_{i_j}$ assigned to $x$. For each bottom-level list $L_x$, we store $\omega(L_x) = \sum_{B \in L_x} \omega(B)$ and its prefix sums. This can be done in $2^{O(d)}N\log^d N$ time. Indeed, by Remark~\ref{remark:SegmentTree}, the entire structure, including all bottom-level lists, has size $2^{O(d)}N\log^d N$. Thus, a single traversal of these lists is sufficient to compute both their total weights and prefix sums within the claimed time.

\medskip\noindent\uline{Query.} We prove the claimed bounds for each query type.
\begin{itemize}
    \item counting: query $T(\B,\omega)$ on $q$ and retrieve lists $L_1, \dots, L_k$. Computing and returning $\sum_{i=1}^k \omega(L_i)$ is sufficient to answer the query. Indeed, by Remark~\ref{remark:SegmentTree}, the lists are pairwise disjoint and their union is exactly $\B(q)$. Thus, $\sum_{i=1}^k \omega(L_i) = \omega(\B(q))$. The same remark guarantees that these lists are retrieved in time $2^{O(d)}\log^d N$.

    \item reporting: query $T(\B,\omega)$ on $q$ and retrieve all non-empty lists $L_1, \dots, L_k$. Build a linked list over $L_1, \dots, L_k$, then scan this list and output each element. Correctness follows from Remark~\ref{remark:SegmentTree}, since the union of these non-empty lists is exactly $\B(q)$, and they are pairwise disjoint. Moreover, the warm-up phase takes $2^{O(d)}\log^d N$ time, since retrieving and linking the non-empty lists takes the same asymptotic time as querying $T(\B,\omega)$. Once the linked list is constructed, each element can be reported with $O(1)$ delay.

    \item sampling: query $T(\B,\omega)$ on $q$ and retrieve the lists $L_1, \dots, L_k$. Choose a list $L_i$ with probability $\frac{\omega(L_i)}{\sum_{j=1}^k \omega(L_j)}$. Then, sample a box $B_j \in L_i$ with probability $\frac{\omega(B_j)}{\omega(L_i)}$. Again, from Remark~\ref{remark:SegmentTree} we have that $L_1 \uplus \cdots \uplus L_k = \B(q)$, thus the probability of sampling $B_j \in L_i$ is $\Pr(B_j) = \frac{\omega(L_i)}{\sum_{h=1}^k \omega(L_h)} \cdot \frac{\omega(B_j)}{\omega(L_i)} = \frac{\omega(B_j)}{\omega(\B(q))}$. Moreover, retrieving the lists takes $2^{O(d)}\log^d N$ time, choosing a list takes $O(k)$ time, and sampling a box within the chosen list takes $O(\log N)$ time by binary search on its prefix sums. Since $k \le 2^{O(d)}\log^d N$, the total query time is $2^{O(d)}\log^d N$.
\end{itemize}

For counting, both preprocessing and query time can be improved by a factor of $\log N$. In the two-dimensional case, we replace the second-dimensional segment tree associated with each node of the first-dimensional segment tree with a sorted list of signed interval endpoints, augmented with prefix sums of weights. For each box $B_i=[a,b]\times[c,d]$ assigned to a node $x$, we insert events $(c,+\omega(B_i))$ and $(d,-\omega(B_i))$ into the list associated with $x$, breaking ties so that starting events precede queries and ending events follow them. Each box contributes to $O(\log N)$ such lists, which can be constructed in $O(N\log N)$ total time by sorting the endpoints once. Fractional cascading then allows us to perform all predecessor searches along a first-dimensional segment-tree path in $O(\log N)$ total time, and the answer is obtained by summing the corresponding prefix sums. Applying this construction to the last two dimensions of the multidimensional segment tree yields preprocessing time $2^{O(d)}N\log^{d-1}N$ and query time $2^{O(d)}\log^{d-1}N$, for $d\geq 2$.

\subsection{Proof of \Cref{theorem:UnionStabbing}}\label{sec:UnionStabbingProof}
Let $T_m$ be a segment tree over $[m]$, and denote by $T_m^x$ the subtree rooted at a node $x$. Moreover, for every node $x$, store the weight
\begin{equation}
    \mu(x) = \omega\left(\bigcup \{J_y : y \in T_m^x \cap \Can(\cQ)\}\right).
\end{equation}
We call the resulting structure a \emph{union segment tree}, and denote it by $T_\mu(\cQ,\omega)$. Thus, $\mu(x)$ is the total weight covered by the canonical nodes in the subtree rooted at $x$. This is essentially the weighted version of the structure used by Bentley to solve \emph{Klee's measure problem}, which asks for the area of the union of axis-parallel rectangles (\cite[Section 8.4]{preparata2012computational}). Indeed, Bentley's algorithm can be adapted to answer counting union stabbing queries. We show that, by using the standard technique of \emph{path copying}, the same structure can also support listing and sampling queries. We first show how to answer these queries on a union segment tree for a fixed family $\cQ$. Then, we use path copying to obtain a persistent structure supporting all three query types for any $q \in [n]$.

\subsubsection{Querying the union tree}

Let $T_\mu(\cQ,\omega)$ be the union segment tree over $(\cQ,\omega)$. Recall that $\Can(\cQ)$ contains all canonical nodes of the intervals in $\cQ$. Although their intervals cover $\cQ^\cup$, they may overlap. We say that a node $x$ is a \emph{high canonical node} if $x\in\Can(\cQ)$ and no proper ancestor of $x$ belongs to $\Can(\cQ)$. In other words, high canonical nodes are the first canonical nodes encountered when descending from the root. We show that their intervals partition $\cQ^\cup$, and that $\mu(x)$ gives the weight covered within $J_x$ whenever no canonical node has been encountered above $x$.

\begin{claim}\label{claim:mux_Jx}
    Let $F$ be the set of high canonical nodes of $T_\mu(\cQ, \omega)$. Then the intervals $\{J_x : x \in F\}$ form a partition of $\cQ^{\cup}$. Moreover, for every node $x$ having no proper ancestor in $\Can(\cQ)$
    \begin{equation}
        \mu(x) = \omega(J_x \cap \cQ^{\cup}) = \sum_{y \in F \cap T_m^x}\omega(J_y).
    \end{equation}
\end{claim}
\begin{proof}
    By definition, $\cQ^\cup = \bigcup_{y \in \Can(\cQ)} J_y$. Every canonical node has a highest canonical ancestor, which belongs to $F$, thus $\cQ^\cup = \bigcup_{y \in F} J_y$. Moreover, by definition of $F$, two distinct nodes of $F$ cannot appear on the same path, and therefore  $\cQ^\cup = \biguplus_{y \in F} J_y$. This proves the first part of the claim.
    
    Now let $x$ have no proper ancestor in $\Can(\cQ)$. Then every canonical node intersecting $J_x$ must appear in $T^x_\mu(\cQ, \omega)$, therefore $J_x \cap Q^\cup = \bigcup_{y \in \Can(\cQ) \cap T_m^x} J_y$ and consequently $\mu(x) = \omega\big(\bigcup_{y \in \Can(\cQ) \cap T_m^x} J_y\big) = \omega(J_x \cap Q^\cup)$. Finally, since the intervals of the nodes in $F$ partition $\cQ^\cup$, we obtain $\omega(J_x \cap Q^\cup) = \sum_{y \in F \cap T_m^x} \omega(J_y)$.
\end{proof}

Claim~\ref{claim:mux_Jx} allows us to traverse $Q^\cup$ by descending from the root, skipping subtrees of weight zero and stopping at the first canonical node of each branch. Indeed, these are exactly the high canonical nodes, whose intervals partition $Q^\cup$. Moreover, at every node visited before stopping, $\mu(x)$ gives the weight of the portion of $J_x$ covered by $Q^\cup$, which we exploit for counting and sampling.

\begin{claim}\label{lemma:AugSegQueries}
Given the root $r$ of $T_\mu(\cQ, \omega)$, one can:
\begin{enumerate}
    \item return $\omega(\cQ^\cup)$ in time $O(1)$,
    \item list $\cQ^\cup$ in total time $O(|\cQ^\cup| \cdot \log m)$ and $O(\log m)$ delay,
    \item sample from the distribution $\left(\frac{\omega(u)}{\omega(\cQ^\cup)}\right)_{u \in \cQ^\cup}$ in time $O(\log m)$.
\end{enumerate}
\end{claim}

\begin{proof}
    For each of the three query types, we first introduce the procedure and then show that it is correct and runs within the claimed bounds. 

    \medskip\noindent\uline{Counting.} By Claim~\ref{claim:mux_Jx}, $\mu(r) = \omega(J_r \cap \cQ^\cup) = \omega(\cQ^\cup)$ since $J_r = [m]$. Thus, it suffices to return $\mu(r)$. This can be done in $O(1)$ time.

    \medskip\noindent\uline{Listing.} Starting from the root, recursively visit a node as follows:  if $\mu(x) = 0$ return. If $x \in \Can(\cQ)$, then list all the elements of $J_x$; otherwise, recurse on its children. The procedure stops at the first canonical node of each branch, thus visiting exactly the high canonical nodes of positive weight. By Claim~\ref{claim:mux_Jx}, these intervals partition $\cQ^\cup$, thus every element of $\cQ^\cup$ is listed exactly once. Since the tree has height $O(\log m)$ and $|F|\leq|\cQ^\cup|$, the total running time is $O(|\cQ^\cup|\log m)$. Finally, between two consecutive outputs, the traversal visits at most $O(\log m)$ nodes, yielding $O(\log m)$ delay.

    \medskip\noindent\uline{Sampling.} If $\mu(r)=0$, report $\cQ^\cup=\emptyset$. Otherwise, start at $x=r$ and, while $x\notin\Can(\cQ)$, move to its left child $x_l$ with probability $\mu(x_l)/\mu(x)$, and to its right child $x_r$ otherwise. Once a canonical node $y$ is reached, sample an element $u\in J_y$ with probability $\omega(u)/\omega(J_y)$, using the prefix sums. Let $y$ be the selected node, by Claim~\ref{claim:mux_Jx} every encountered node $x$ before $y$ satisfies $\mu(x) = \mu(x_l) + \mu(x_r)$. Hence, the probability of reaching $y$ telescopes to $\Pr[y] = \frac{\mu(y)}{\mu(r)} = \frac{\omega(J_y)}{\omega(\cQ^\cup)}$. Consequently, every $u \in J_y$ is sampled with probability
    \begin{equation*}
        \Pr[u] = \Pr[y] \cdot\Pr[u \, | \, y] = \frac{\omega(J_y)}{\omega(\cQ^\cup)} \cdot \frac{\omega(u)}{\omega(J_y)} = \frac{\omega(u)}{\omega(\cQ^\cup)}.
    \end{equation*}
    Since $F$ partitions $\cQ^\cup$, this is the required distribution. Finally, the procedure is a single traversal of a root-to-leaf path, and thus takes time at most $O(\log m)$. Let $x$ be a visited node, computing the probability of the next nodes takes $O(1)$ time. Finally, with a simple binary search Step 3 requires $O(\log m)$ time. Hence, $O(\log m)$ total time.
\end{proof}

\subsubsection{Making the tree persistent}\label{sec:StabbingQueries}


\begin{claim}
    Given a weighted union-stabbing instance $(\Pcal, \cQ, \omega)$, one can construct in time $O(n + m + N \log m)$ an implicit representation of $\{T_\mu(\cQ(q), \omega), q \in [n]\}$ such that the root of each tree can be accessed in $O(1)$ time. Moreover, for every node $x$ of $T_\mu(\cQ(q), \omega)$, membership in $\Can(\cQ(q))$ can be tested in $O(1)$ time.
\end{claim}

\begin{proof}
    The proof is organized as follows: we introduce the data structure in the first three paragraphs. Then, we prove its correctness and bounds.

    \medskip
    \noindent\uline{Construction.} Build a segment tree $T_m$ on $[m]$, with every node $x$ storing two integers, namely $C(x)$ and $\hat \mu(x)$, both initially set to zero. We also preprocess the prefix sums of $\omega$, so that $\omega(J_x)$ can be computed in $O(1)$ time for every node $x$

    \medskip
    \noindent\uline{Update.} We define a procedure $\textsc{Update}(x,I,\delta)$, where $x$ is a node, $I\subseteq[m]$ is an interval, and $\delta\in\{-1,+1\}$ indicates insertion or removal. If $J_x\cap I=\emptyset$, the procedure returns $x$ unchanged. Otherwise, it creates a copy $x'$ of $x$, initially retaining its children. If $J_x\subseteq I$, it sets $C(x')\gets C(x)+\delta$; otherwise, it recursively updates the children of $x'$ using $\textsc{Update}(x_l,I,\delta)$ and $\textsc{Update}(x_r,I,\delta)$. Finally, it sets
    \begin{equation}
    \hat\mu(x') =
    \begin{cases}
    \omega(J_x) & \text{if } C(x')>0,\\
    0 & \text{if } C(x')=0 \text{ and } x' \text{ is a leaf},\\
    \hat\mu(x'_l)+\hat\mu(x'_r) & \text{otherwise},
    \end{cases}
    \end{equation}
    and returns $x'$. Thus, $\textsc{Update}(x,I,+1)$ and $\textsc{Update}(x,I,-1)$ implement insertion and removal, respectively, while preserving all previous versions of the tree.

    \medskip
    \noindent\uline{Sweep.} For every $i \in [N]$, let $P_i=[a_i,b_i]$, create an insertion event of $(Q_i, +1)$ at $a_i$ and, if $b_i < n$, a removal event $(Q_i, -1)$ at $b_i + 1$. We group the events by position with a bucket sort on $[n]$. Let $r_0$ be the root of $T_m$. For $q = 1, \dots, n$, set $r_q \gets r_{q-1}$, and then $r_q \gets \textsc{Update}(r_q,I, \delta)$ for every event $(I, \delta)$.
    Finally, output $r_1, \dots, r_n$.

    \medskip
    \noindent\uline{Correctness.} Let $q \in [n]$. The root $r_q$ can be accessed in $O(1)$ time, and every node $x$ reachable from $r_q$ is either a node of $T_m$ or a copy of one, with the same interval $J_x$, whose children are reached through the pointers. 

    Let $x$ be a node reachable from $r_q$. We first show that $x \in \Can(\cQ(q))$ can be checked by testing $C(x) > 0$. Indeed note that, by definition, each insertion or removal changes the counter by $\delta$ exactly at the canonical nodes of the corresponding interval. Thus, since after the events at positions $1,\dots,q$ the intervals inserted are exactly those in $\cQ(q)$, the following equality holds
    \begin{equation}
        C(x) = |\{j \in [N]: q \in P_j, x \in \Can(Q_j)\}|.
    \end{equation}
    In words, $C(x)$ counts how many inserted intervals have $x$ as a canonical node. Thus, $C(x) > 0$ if and only if $x \in \Can(\cQ(q))$, giving the claimed constant time membership test.
    
    It remains to show that, for every $x$ reachable from $r_q$, the value stored in $x$ is its label in $T_{m}(\cQ, \omega)$; that is, $\hat\mu(x) = \mu(x)$. By construction, every node $x$ satisfies $\hat\mu(x) = \omega(J_x)$ if $C(x) > 0$, $\hat\mu(x) = 0$ if $C(x) = 0$ and $x$ is a leaf, and $\hat\mu(x) = \hat\mu(x_l) + \hat\mu(x_r)$. By definition, $\mu$ satisfies the same recursion with $x \in \Can(\cQ(q))$ in place of $C(x) > 0$. Indeed, if $x \in \Can(\cQ(q))$, then $\mu(x) = \omega(J_x)$, since $J_x$ itself appears in the union defining $\mu(x)$ and $J_y \subseteq J_x$ for every $y \in T_m^x$; if $x \notin \Can(\cQ(q))$ is a leaf, then the union is empty and $\mu(x) = 0$; if $x \notin \Can(\cQ(q))$ is internal, then every canonical node of $T_m^x$ lies in $T_m^{x_l}$ or in $T_m^{x_r}$, whose intervals are contained in the disjoint sets $J_{x_l}$ and $J_{x_r}$, hence $\mu(x) = \mu(x_l) + \mu(x_r)$. Since $C(x) > 0$ if and only if $x \in \Can(\cQ(q))$, by induction on the height of $x$ we conclude that $\hat\mu(x) = \mu(x)$ for every $x$ reachable from $r_q$. Thus, the tree rooted at $r_q$ is a representation of $T_\mu(\cQ(q), \omega)$.

    \medskip
    \noindent\uline{Time complexity.} Building $T_m$ on $[m]$ requires $O(m)$ time, while constructing and grouping the events take $O(n + N)$ time with a bucket sort on $[n]$.
    Each update visits $O(\log m)$ nodes, as in the standard segment tree insertion procedure (see \Cref{sec:SegmentTrees}), and performs $O(1)$ work per visited node. Since there are at most $2N$ events, the total update time is $O(N\log m)$. Moreover, each update creates $O(\log m)$ nodes, each storing $O(1)$ information. Together with the initial tree, the prefix sums, and the $n$ stored roots, this gives $O(n + m + N\log m)$ total time and space.
\end{proof}

\bibliographystyle{plain}
\bibliography{biblio,hypergraph,subgraph_counting_doi}

\end{document}